\documentclass[12pt]{report}
\usepackage{epubtk}
\usepackage{epsf}
\usepackage{indentfirst}
\usepackage{amsmath}
\usepackage{amssymb}
\usepackage{hyperref}

\def\defn#1{``#1''}
\def\edefn#1{\emph{#1}}
\def\program#1{\textsc{#1}}

\def\eg{e.g.\hbox{}}
\def\etal{\textit{et~al.\hbox{}}}
\def\ie{i.e.\hbox{}}

\def\del{\nabla}
\def\Lie{{\cal L}}
\def\Scri{{\cal J}}
\def\ltsim{\lesssim}

\def\ang{\text{ang}}
\def\coeff{\text{coeff}}
\def\Strahlkoerper{Strahlk\"{o}rper}

\def\i{\text{\footnotesize\textsc{i}}}
\def\j{\text{\footnotesize\textsc{j}}}
\def\k{\text{\footnotesize\textsc{k}}}
\def\J{{\mathbf{J}}}

\begin{document}

\title{Event and Apparent Horizon Finders for $3+1$ Numerical Relativity}

\author{%
\epubtkAuthorData{Jonathan Thornburg}{%
Max-Planck-Institut f\"ur Gravitationsphysik \\
Albert-Einstein-Institut}{%
jthorn@aei.mpg.de}{%
http//www.aei.mpg.de/~jthorn}%
}

\date{Version of  1 January 2006}
%%\preprint{AEI-2005-184}
\maketitle

%%%%%%%%%%%%%%%%%%%%%%%%%%%%%%%%%%%%%%%%

\begin{abstract}
Event and apparent horizons are key diagnostics for the presence and
properties of black holes.  In this article I review numerical algorithms
and codes for finding event and apparent horizons in numerically-computed
spacetimes, focusing on calculations done using the $3+1$ ADM formalism.

The event horizon of an asymptotically-flat spacetime is the boundary
between those events from which a future-pointing null geodesic can
reach future null infinity, and those events from which no such
geodesic exists.  The event horizon is a (continuous) null surface
in spacetime.  The event horizon is defined \emph{nonlocally in time}:
it's a global property of the entire spacetime, and must be found in
a separate post-processing phase \emph{after} (part of) the spacetime
has been numerically computed.

There are 3~basic algorithms for finding event horizons, based respectively
on integrating null geodesics \emph{forwards} in time, integrating
null geodesics \emph{backwards} in time, and integrating null
\emph{surfaces} backwards in time.  The last of these is generally
the most efficient and accurate.

In contrast to an event horizon, an apparent horizon is defined
locally in time in a spacelike slice, and so can be (and usually is)
found ``on the fly'' during the numerical computation of a spacetime.
A marginally outer trapped surface (MOTS) in a slice is a smooth closed
2-surface whose future-pointing outgoing null geodesics have zero
expansion~$\Theta$.  An apparent horizon is then defined as a MOTS
not contained in any other MOTS.  The MOTS condition is a nonlinear
elliptic partial differential equation (PDE) for the surface shape,
containing the ADM 3-metric, its spatial derivatives, and the extrinsic
curvature as coefficients.  Most ``apparent-horizon'' finders actually
find MOTSs.

There are a large number of apparent-horizon finding algorithms, with
differing trade-offs between speed, robustness, accuracy, and ease
of programming.  In axisymmetry, shooting algorithms work well and are
fairly easy to program.  In slices with no continuous symmetries,
Nakamura \etal{}'s algorithm and elliptic-PDE algorithms are fast and
accurate, but require good initial guesses to converge.  In many
cases, Schnetter's ``pretracking'' algorithm can greatly improve
an elliptic-PDE algorithm's robustness.  Flow algorithms are generally
quite slow, but can be very robust in their convergence.
\end{abstract}

%%%%%%%%%%%%%%%%%%%%%%%%%%%%%%%%%%%%%%%%

\epubtkKeywords{black holes,
		event horizons,
		apparent horizons,
		trapped surfaces,
		ADM formalism,
		null surfaces,
		numerical relativity,
		numerical algorithms,
		partial differential equations}

\newpage

%%%%%%%%%%%%%%%%%%%%%%%%%%%%%%%%%%%%%%%%%%%%%%%%%%%%%%%%%%%%%%%%%%%%%%%%%%%%%%%%
%%%%%%%%%%%%%%%%%%%%%%%%%%%%%%%%%%%%%%%%%%%%%%%%%%%%%%%%%%%%%%%%%%%%%%%%%%%%%%%%

\tableofcontents

%%%%%%%%%%%%%%%%%%%%%%%%%%%%%%%%%%%%%%%%%%%%%%%%%%%%%%%%%%%%%%%%%%%%%%%%%%%%%%%%
%%%%%%%%%%%%%%%%%%%%%%%%%%%%%%%%%%%%%%%%%%%%%%%%%%%%%%%%%%%%%%%%%%%%%%%%%%%%%%%%

\chapter{Introduction}
\label{chap-intro}

Compact objects -- ones which may contain event horizons and/or
apparent horizons -- are a major focus of numerical relativity.
The usual output of a numerical relativity simulation is some
(approximate, discrete) representation of the spacetime geometry
(the 4-metric and possibly its derivatives) and any matter fields,
but \emph{not} any explicit information about the existence, precise
location, or other properties of any event/apparent horizons.  To
gain this information, we must explicitly \emph{find} the horizons
from the numerically-computed spacetime geometry.  The subject of
this review is numerical algorithms and codes for doing this, focusing
on calculations done using the $3+1$ ADM formalism
(\cite{Arnowitt62, York79}).%%%
\footnote{%%%
	 There are many interesting uses of event and/or
	 apparent horizons in gaining physical understanding
	 of numerically-computed spacetimes.  However, a
	 discussion of these applications would encompass
	 much of strong-field numerical relativity, and
	 would be be far beyond the scope of this review.
	 }%%%
{}  Baumgarte and Shapiro~\cite[section~6]{Baumgarte:2002jm} have also
recently reviewed event and apparent-horizon finding algorithms.

In this review I distinguish between a numerical \edefn{algorithm}
(an abstract description of a mathematical computation; also often
known as a \defn{method} or \defn{scheme}), and a computer \edefn{code}
(a \defn{horizon finder}, a specific piece of computer software which
implements a horizon-finding algorithm or algorithms).  My main focus
is on the algorithms, but I also mention specific codes where they
are freely available to other researchers.

In this review I have tried to cover all the major horizon-finding
algorithms and codes, and to accurately credit the earliest publication
of important ideas.  However, in a field as large and active as numerical
relativity, it's inevitable that I have overlooked and/or misdescribed
some important research.  I apologise to anyone whose work I've slighted,
and I ask readers to help make this a truly ``living'' review by sending
me corrections, updates, and/or pointers to additional work (either
their own or others) which I should discuss in future revisions of
this review.

The general outline of this review is as follows:
In the remainder of this chapter I
define notation and terminology
	(section~\ref{sect-intro/notation,terminology}),
discuss how 2-surfaces should be parameterized
	(section~\ref{sect-intro/2-surface-parameterizations}),
and outline some of the software-engineering issues that arise in modern
numerical relativity codes
	(section~\ref{sect-intro/software-engineering}).
I then discuss numerical algorithms and codes for
finding event horizons
	(chapter~\ref{chap-EH})
and apparent horizons
	(chapter~\ref{chap-AH}).
Finally, in the appendices I briefly outline some of the excellent
numerical algorithms/codes available for two standard problems in
numerical analysis,
the solution of a single nonlinear algebraic equation
	(appendix~\ref{app-single-nonlinear-eqn})
and the time integration of a system of ordinary differential equations
	(appendix~\ref{app-ODEs}).

%%%%%%%%%%%%%%%%%%%%%%%%%%%%%%%%%%%%%%%%%%%%%%%%%%%%%%%%%%%%%%%%%%%%%%%%%%%%%%%%

\section{Notation and Terminology}
\label{sect-intro/notation,terminology}

I generally follow the sign and notation conventions of
Wald~\cite{Wald84}.  I assume that all spacetimes are globally hyperbolic,
and for event-horizon finding I further assume asymptotic flatness;
in this latter context $\Scri^+$ is future null infinity.  I use the
Penrose abstract-index notation, with summation over all repeated indices.
4-indices $abc$ range over all spacetime coordinates $\{x^a\}$,
and 3-indices $ijk$ range over the spatial coordinates $\{x^i\}$
in a spacelike slice $t = \text{constant}$.  The spacetime coordinates
are thus $x^a = (t,x^i)$.

$g_{ab}$ is the spacetime 4-metric,
and $g^{ab}$ the inverse spacetime 4-metric;
these are used to raise and lower 4-indices.
$\Gamma^c_{ab}$ are the 4-Christoffel symbols.
$\Lie_v$ is the Lie derivative along the 4-vector field $v^a$.

$g_{ij}$ is the 3-metric defined in a slice,
and $g^{ij}$ is the inverse 3-metric;
these are used to raise and lower 3-indices.
$\del_i$ is the associated 3-covariant derivative operator,
and $\Gamma^k_{ij}$ are the 3-Christoffel symbols.
$\alpha$ and $\beta^i$ are the $3+1$ lapse function and shift vector
respectively,%%%
\footnote{%%%
	 See York~\cite{York79} for a general overview
	 of the $3+1$ formalism as it's used in numerical
	 relativity.
	 }%%%
{} so the spacetime line element is
\begin{eqnarray}
ds^2	& = &	g_{ab} \, dx^a \, dx^b					\\
	& = &	- (\alpha^2 - \beta_i \beta^i) \, dt^2
		+ 2 \beta_i \, dx^i \, dt
		+ g_{ij} \, dx^i \, dx^j				%%%\\
\end{eqnarray}

As is common in $3+1$ numerical relativity, I follow the sign convention
of Misner, Thorne, and Wheeler~\cite{Misner73} and York~\cite{York79} in
defining the extrinsic curvature of the slice as
$K_{ij} = -\frac{1}{2} \Lie_n g_{ij} = -\del_i n_j$, where $n^a$ is the
future-pointing unit normal to the slice.  (In contrast,
Wald~\cite{Wald84} omits the minus signs from this definition.)
$K \equiv K_i{}^i$ is the trace of the extrinsic curvature $K_{ij}$.
$m_\text{ADM}$ is the ADM mass of an (asymptotically flat) slice.

Indices $uvw$ range over generic angular coordinates $(\theta,\phi)$
on $S^2$ or on a horizon surface.  Note that these coordinates are
conceptually completely distinct from the 3-dimensional spatial
coordinates $x^i$.  Depending on the context, $(\theta,\phi)$ may or
may not have the usual polar-spherical topology.

Indices $\i\j\k$ label angular grid points on $S^2$ or on a horizon
surface.  Notice that these are \emph{2-dimensional} indices: a
\emph{single} such index uniquely specifies an angular grid point.
$\delta_{\i\j}$ is the Kronecker delta on the space of these indices,
or equivalently on surface grid points.

I often write a differential operator as
$F = F(y, \partial_u y, \partial_{uv} y; g_{ij}, \partial_k g_{ij}, K_{ij})$.
The ``$;$'' notation means that $F$ is a (generally nonlinear) algebraic
function of the variable $y$ and its 1st and 2nd~angular derivatives,
and that $F$ also depends on the coefficients $g_{ij}$, $\partial_k g_{ij}$,
and $K_{ij}$ at the apparent horizon position.

There are 3~common types of spacetimes/slices where numerical event
or apparent horizon finding is of interest:  spherically-symmetric
spacetimes/slices, axisymmetric spacetimes/slices, and spacetimes/slices
with no continuous spatial symmetries (no spacelike Killing vectors).
I refer to the latter as \defn{fully generic} spacetimes/slices.

In this review I use the abbreviations
\defn{ODE} for ordinary differential equation,
\defn{PDE} for partial differential equation,
\defn{CE surface} for constant-expansion surface,
and \defn{MOTS} for marginally outer trapped surface.
Names in \textsc{Small Capitals} refer to horizon finders and other
computer software.

\begingroup
% local latex macros for this paragraph
\def\g{\mathsf{g}}
\def\s{\mathsf{s}}
\def\S{\mathsf{S}}
When discussing iterative numerical algorithms, it's often useful
to use the concept of an algorithm's \defn{radius of convergence}:
Suppose the solution space within which the algorithm is iterating
is $\S$.  Then given some norm $\| \cdot \|$ on $\S$, the algorithm's
radius of convergence about a solution $\s \in \S$ is defined as the
smallest $r > 0$ such that the algorithm will converge to the correct
solution $\s$ for any initial guess $\g$ with $\|\g - \s\| \le r$.
We only rarely know the exact radius of convergence of an algorithm,
but practical experience often provides a rough estimate.%%%
\footnote{%%%
	 An algorithm's actual \defn{convergence region}
	 (the set of all initial guesses for which the
	 algorithm converges to the correct solution)
	 may even be fractal in shape.  For example, the
	 Julia set is the convergence region of Newton's
	 method on a simple nonlinear algebraic equation.
	 }%%%
\endgroup	% local latex macros for this paragraph

%%%%%%%%%%%%%%%%%%%%%%%%%%%%%%%%%%%%%%%%%%%%%%%%%%%%%%%%%%%%%%%%%%%%%%%%%%%%%%%%

\section{2-Surface Parameterizations}
\label{sect-intro/2-surface-parameterizations}

%%%%%%%%%%%%%%%%%%%%%%%%%%%%%%%%%%%%%%%%

\subsection{Level-Set--Function Parameterizations}
\label{sect-intro/2-surface-parameterizations/level-set}

The most general way to parameterize a 2-surface in a slice is to
define a scalar \defn{level-set function}~$F$ on some neighborhood
of the surface, with the surface itself then being defined as the
level set
\begin{equation}
F = 0
	\qquad
	\text{on the surface}
						      \label{eqn-F=0-on-surface}
\end{equation}
\begin{sloppypar}
Assuming the surface to be orientable, it's conventional to choose
$F$ so that $F > 0$~($F < 0$) outside~(inside) the surface.
\end{sloppypar}

This parameterization is valid for any surface topology, including
time-dependent topologies.  The 2-surface itself can then be found
by a standard isosurface-finding algorithm such as the marching-cubes
algorithm~\cite{Lorensen-Cline-1987:marching-cubes}.  (This algorithm
is widely used in computer graphics, and is implemented in a number
of widely-available software libraries.)

%%%%%%%%%%%%%%%%%%%%%%%%%%%%%%%%%%%%%%%%

\subsection{\Strahlkoerper{} Parameterizations}
\label{sect-intro/2-surface-parameterizations/Strahlkoerper}

Most apparent-horizon finders, and many event-horizon finders,
assume that each connected component of the apparent (event) horizon
has $S^2$~topology.  With the exception of toroidal event horizons
(discussed in section~\ref{sect-EH/intro}), this is generally a
reasonable assumption.

To parameterize an $S^2$~surface's shape, it's common to further assume
that we are given (or can compute) some \defn{local coordinate origin}
point inside the surface such that the surface's 3-coordinate shape
relative to that point is a \defn{\Strahlkoerper},
(literally ``ray body'', or more commonly \defn{star-shaped region})
defined by Minkowski (\cite[p.~108]{Schroeder86}) as
\begin{quote}
a region in $n$-D Euclidean space containing
the origin and whose surface, as seen from the origin,
exhibits only one point in any direction.
\end{quote}

The \Strahlkoerper{} assumption is a significant restriction on the
horizon's coordinate shape (and the choice of the local coordinate
origin).  For example, it rules out the coordinate shape and local
coordinate origin illustrated in figure~\ref{fig-non-Strahlkoerper-shape}:
a horizon with such a coordinate shape about the local coordinate
origin couldn't be found by any horizon finder which assumes a
\Strahlkoerper{} surface parameterization.

%%%%%%%%%%
\epubtkImage{fig/non-Strahlkoerper-shape/shape.png}%%%
{%%%
\begin{figure}[bp]
\begin{center}
\centerline{%%%
  \epsfxsize=125mm
  \epsfbox{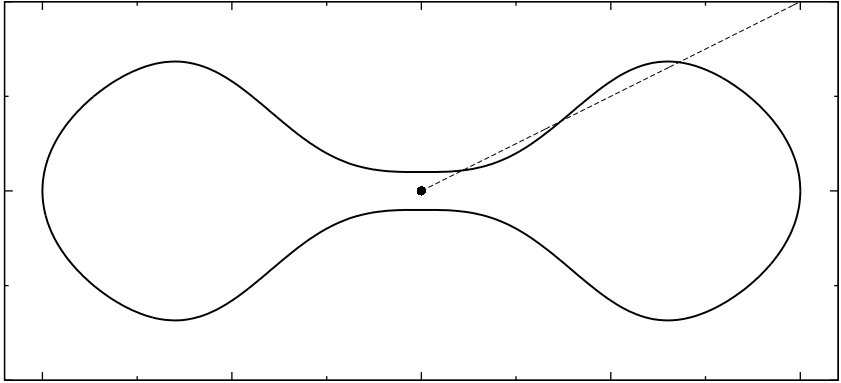}%%
}
\end{center}
\caption[Illustration of a Non-\Strahlkoerper{} Shape]
	{
	This figure shows a cross-section of a coordinate shape
	(the thick curve) which isn't a \Strahlkoerper{} about the
	local coordinate origin shown (the large dot).  The dashed
	line shows a ray from the local coordinate origin, which
	intersects the surface in more than one point.
	}
\label{fig-non-Strahlkoerper-shape}
\end{figure}%%%
}
%%%%%%%%%%

For event-horizon finding, algorithms and codes are now available
which allow an arbitrary horizon topology, with no \Strahlkoerper{}
assumption (see the discussion in
section~\ref{sect-EH/algorithms/null-surfaces-backwards/level-set}
for details).  For apparent-horizon finding, the flow algorithms discussed
in section~\ref{sect-AH/algorithms/flow} theoretically allow any surface
shape, although many implementations still make the \Strahlkoerper{}
assumption.  Removing this assumption for other apparent-horizon
finding algorithms might be a fruitful area for further research.

Given the \Strahlkoerper{} assumption, the surface can be explicitly
parameterized as
\begin{equation}
r = h(\theta,\phi)
						      \label{eqn-r=h(theta,phi)}
\end{equation}
where $r$~is the Euclidean distance from the local coordinate origin
to a surface point, $(\theta,\phi)$ are generic angular coordinates
on the horizon surface (or equivalently on $S^2$), and the
\defn{horizon shape function} $h: S^2 \to \Re^+$ is a positive
real-valued function on the domain of angular coordinates defining
the surface shape.

There are two common discretizations of this surface representation:
\begin{description}
\item[Spectral representation]\mbox{}\\
	Here we expand the horizon shape function $h$ in an infinite
	series in some (typically orthonormal) set of basis functions
	such as spherical harmonics $Y_{\ell m}$ or symmetric trace-free
	tensors,%%%
\footnote{%%%
	 For convenience of exposition I use spherical harmonics
	 here, but there are no essential differences if other
	 basis sets are used.
	 }%%%
	\begin{equation}
	h(\theta,\phi)
		= \sum_{\ell,m} a_{\ell m} Y_{\ell m}(\theta,\phi)
						     \label{eqn-h-expand-in-Ylm}
	\end{equation}

	This series can then be truncated at some finite order $\ell_{\max}$,
	and the $N_\coeff = \ell_{\max}(\ell_{\max}{+}1)$ coefficients
	$\{a_{\ell m}\}$ used to represent (discretely approximate)
	the horizon shape.  For reasonable accuracy, $\ell_{\max}$
	is typically on the order of 8 to 12.
\item[Finite difference representation]\mbox{}\\
	Here we choose some finite grid of angular coordinates
	$\{(\theta_\k,\phi_\k)\}$, $\k=1$, $2$, $3$, \dots, $N_\ang$
	on $S^2$ (or equivalently on the surface),%%%
\footnote{%%%
	 I discuss the choice of this angular grid in more detail in
	 section~\ref{sect-AH/algorithms/elliptic-PDE/angular-coords+grid+BCs}.
	 }%%%
{}	and represent (discretely approximate) the surface shape by
	the $N_\ang$
	values
	\begin{equation}
	\bigl\{ h(\theta_\k,\phi_\k) \bigr\}
	\qquad
	\text{$\k=1$, $2$, $3$, \dots, $N_\ang$}
						   \label{eqn-h-on-angular-grid}
	\end{equation}
	For reasonable accuracy, $N_\ang$ is typically on the order
	of a few thousand.
\end{description}

It's sometimes useful to explicitly construct a level-set function
describing a given \Strahlkoerper{}.  A common choice here is
\begin{equation}
F \equiv r - h(\theta,\phi)
						    \label{eqn-F=r-h(theta,phi)}
\end{equation}

%%%%%%%%%%%%%%%%%%%%%%%%%%%%%%%%%%%%%%%%

\subsection{Finite-Element Parameterizations}
\label{sect-intro/2-surface-parameterizations/finite-element}

Another way to parameterize a 2-surface is via finite elements,
where the surface is modelled as a triangulated mesh, \ie{} as a set
of interlinked \defn{vertices} (points in the slice, represented by
their spatial coordinates $\{x^i\}$), \defn{edges} (represented by
ordered pairs of vertices), and faces.  Typically only triangular faces
are used (represented as oriented triples of vertices).

A key benefit of this representation is that it allows an arbitrary
topology for the surface.  However, determining the actual surface
topology (\eg{} testing for whether or not the surface self-intersects)
is somewhat complicated.

This representation is similar to that of
Regge calculus~\cite{Regge:1961px, Gentle-2002:Regge-calculus-review},%%%
\footnote{%%%
	 There has been some controversy over whether, and if so
	 how quickly, Regge calculus converges to the continuum
	 Einstein equations.  (See, for example, the debate between
	 Brewin~\cite{Brewin-2000:convergence-of-Regge-calculus}
	 and Miller~\cite{Miller-1995:4th-order-Regge-calculus},
	 and the explicit convergence demonstration of
Gentle and Miller~\cite{Gentle-Miller-1998:Regge-calculus-model-of-Kasner}.)
	 However,
Brewin and Gentle~\cite{Brewin-Gentle-2001:convergence-of-Regge-calculus}
	 seem to have resolved this: Regge calculus does in fact
	 converge to the continuum solution, and this convergence
	 is generically 2nd~order in the resolution.
	 }%%%
{} and can similarly be expected to show 2nd~order convergence with
the surface resolution.

Finite element surface representations have been used for apparent-horizon
finding by Metzger~\cite{Metzger04}.

%%%%%%%%%%%%%%%%%%%%%%%%%%%%%%%%%%%%%%%%%%%%%%%%%%%%%%%%%%%%%%%%%%%%%%%%%%%%%%%%

\section{Software-Engineering Issues}
\label{sect-intro/software-engineering}

Historically, numerical relativists wrote their own codes from scratch.
As these became more complex, many researchers changed to working on
``group codes'' with multiple contributors.

%%%%%%%%%%%%%%%%%%%%%%%%%%%%%%%%%%%%%%%%

\subsection{Software Libraries and Toolkits}

More recently, particularly in work on fully generic spacetimes,
where all 3~spatial dimensions must be treated numerically, there has
been a strong trend towards the use of higher-level software libraries
and modular ``computational toolkits'' such as
\program{Cactus}~\cite{Goodale02a}~%%%
(\href{http://www.cactuscode.org}{\url{http://www.cactuscode.org}}).
These have a substantial learning overhead, but can allow researchers
to work much more productively by focusing more on numerical relativity,
instead of computer-science and software-engineering issues such as
parameter-file parsing, parallelization, I/O, etc.

A particularly important area for such software infrastructure is
mesh refinement.%%%
\footnote{%%%
	 See, for example, Choptuik~\protect\cite{Choptuik89},
	 Pretorius~\protect\cite{Pretorius:2003wc}, and
	 Pretorius and Choptuik~\protect\cite{Pretorius:2005amr}
	 for general surveys of the uses of, and methods
	 for, mesh refinement in numerical relativity.
	 }%%%
{}  This is essential to much current numerical-relativity research,
but is moderately difficult to implement even in only one spatial
dimension, and much harder in multiple spatial dimensions.  There are
now a number of software libraries providing multi-dimensional
mesh-refinement infrastructure (sometimes combined with parallelization),
such as
\program{DAGH}/\program{GrACE}~\cite{Parashar00a}
(\href{http://www.caip.rutgers.edu/~parashar/DAGH/}%%%
      {\url{http://www.caip.rutgers.edu/~parashar/DAGH/}})
and
\program{ParaMesh}~\cite{MacNeice00}
(\href{http://ct.gsfc.nasa.gov/paramesh/Users_manual/amr.html}%%%
      {\url{http://ct.gsfc.nasa.gov/paramesh/Users_manual/amr.html}}).
The \href{http://www.cactuscode.org}{\program{Cactus}} toolkit can
be used in either unigrid or mesh-refinement modes, the latter using
a ``mesh-refinement driver'' such as \program{PAGH} or
\program{Carpet}~\cite{Schnetter-etal-03b, carpet-author-and-web-site}
(\href{http://www.carpetcode.org}{\url{http://www.carpetcode.org}}).

In this review I point out event and apparent-horizon finders
which have been written in particular frameworks, and comment on
whether they work with mesh refinement.

%%%%%%%%%%%%%%%%%%%%%%%%%%%%%%%%%%%%%%%%

\subsection{Code Reuse and Sharing}

Another issue is that of code reuse and sharing.  It's common for
codes to be shared within a research group, but relatively uncommon
for them to be shared between different (competing) research groups.
Even apart from concerns about competitive advantage, without a
modular structure and clear documentation it's difficult to reuse
another group's code.  The use of a common computational toolkit
can greatly simplify such reuse.

If such reuse \emph{can} be accomplished, it becomes much easier for
other researchers to build on existing work, rather than having to
``reinvent the wheel''.  As well as the obvious ease of reusing
existing code that (hopefully!) already works and has been thoroughly
debugged and tested, there's another -- less obvious -- benefit of
code sharing: It greatly eases the replication of past work, which
is essential as a foundation for new development.  That is, without
access to another researcher's code, it can be surprisingly difficult
to replicate her results, because the success or failure of a numerical
algorithm frequently depends on subtle implementation details not
described in even the most complete of published papers.

Event and apparent-horizon finders are excellent candidates for
software reuse:  Many numerical-relativity researchers can benefit
from using them, and they have a relatively simple interface to an
underlying numerical-relativity simulation.  Even if a standard
computational toolkit isn't used, this makes it relatively easy to
port an event or apparent-horizon finder to a different code.

Throughout this review I note event and apparent-horizon finders
which are freely available to other researchers.

%%%%%%%%%%%%%%%%%%%%%%%%%%%%%%%%%%%%%%%%

\subsection{Using Multiple Event/Apparent Horizon Finders}

It's often useful to have multiple event or apparent-horizon finders
available: their strengths and weaknesses may complement each other,
and the extent of (dis)agreement between their results can give a
good measure of the numerical accuracy.  For example,
figure~\ref{fig-2BH-spiraling-collision/AH-masses} shows a comparison 
between the irreducible masses of apparent horizons in a binary
black hole coalescence simulation
(Alcubierre \etal{}~\cite[figure~4b]{Alcubierre2003:pre-ISCO-coalescence-times}),
as computed by two different apparent-horizon finders in the
\href{http://www.cactuscode.org}{\program{Cactus}} toolkit,
\program{AHFinder} and \program{AHFinderDirect}.  In this case the
two agree to within about $2\%$ for the individual horizons, and
$0.5\%$ for the common horizon.

%%%%%%%%%%%%%%%%%%%%%%%%%%%%%%%%%%%%%%%%%%%%%%%%%%%%%%%%%%%%%%%%%%%%%%%%%%%%%%%%
%%%%%%%%%%%%%%%%%%%%%%%%%%%%%%%%%%%%%%%%%%%%%%%%%%%%%%%%%%%%%%%%%%%%%%%%%%%%%%%%

\chapter{Finding Event Horizons}
\label{chap-EH}

%%%%%%%%%%%%%%%%%%%%%%%%%%%%%%%%%%%%%%%%%%%%%%%%%%%%%%%%%%%%%%%%%%%%%%%%%%%%%%%%

\section{Introduction}
\label{sect-EH/intro}

The black hole region of an asymptotically-flat spacetime is defined
(\cite{Hawking73, Hawking73a}) as the set of events from which no
future-point null geodesic can reach future null infinity ($\Scri^+$).
The event horizon is defined as the boundary of the black hole region.
The event horizon is a null surface in spacetime with
(in the words of Hawking and Ellis~\cite[p.~319]{Hawking73a})
``a number of nice properties'' for studying the causal stucture
of spacetime.

The event horizon is a \emph{global} property of an entire spacetime,
and is defined \emph{nonlocally in time}: the event horizon in a slice
is defined in terms of (and can't be computed without knowing) the full
\emph{future} development of that slice.

In practice, to find an event horizon in a numerically-computed spacetime,
we typically instrument a numerical evolution code to write out data
files of the 4-metric.  After the evolution has reached an
approximately-stationary final state, we then compute a numerical
approximation to the event horizon in a separate post-processing
pass, using the 4-metric data files as inputs.

As a null surface, the event horizon is necessarily continuous.
In theory it need not be \emph{anywhere} differentiable,%%%
\footnote{%%%
	 Chru{\'{s}}ciel and Galloway~%%%
	 \protect\cite{Chrusciel-Galloway-1998:non-differentiable-horizons}
	 showed that if a ``cloud of sand'' falls into a large
	 black hole, each ``sand grain'' generates a non-differentiable
	 caustic in the event horizon.
	 }%%%
{} but in practice this behavior rarely occurs:%%%
\footnote{%%%
	 This is a statement about the types of spacetimes
	 usually studied by numerical relativists, not a
	 statement about the mathematical properties of
	 the event horizon itself.
	 }%%%
{}  The event horizon is generally smooth except for possibly a
finite set of \defn{cusps} where new generators join the surface;
the surface normal has a jump discontinuity across each cusp.
(The classic example of such a cusp is the ``inseam'' of the
``pair of pants'' event horizon illustrated in
figures~\ref{fig-2BH-headon-collision/EH-and-generators-in-slice}
and~\ref{fig-2BH-headon-collision/EH-spacetime-view}.)

A black hole is defined as a connected component of the black hole
region in a $3+1$ slice.  The boundary of a black hole (the event horizon)
in a slice is a 2-dimensional set of events.  Usually this has 2-sphere
($S^2$) topology.  However, numerically simulating rotating dust
collapse, Abrahams \etal{}~\cite{Abrahams94a} found that in some
cases the event horizon in a slice may be \emph{toroidal} in topology.
Lehner \etal{}~\cite{Lehner-etal-1999:BH-coalescence-geometry} and
Husa and Winicour~\cite{Husa99a} have used null (characteristic) algorithms
to give a general analysis of the event horizon's topology in
black hole collisions; they find that there is generically a
(possibly brief) toroidal phase before the final 2-spherical state
is reached.  Lehner \etal{}~\cite{Lehner-etal-BH-bagels-movies} later
calculated movies showing this behavior for several asymmetric
black hole collisions.

%%%%%%%%%%%%%%%%%%%%%%%%%%%%%%%%%%%%%%%%%%%%%%%%%%%%%%%%%%%%%%%%%%%%%%%%%%%%%%%%

\section{Algorithms and Codes for Finding Event Horizons}
\label{sect-EH/algorithms}

There are 3~basic event-horizon finding algorithms:
\begin{itemize}
\item	Integrate null geodesics \emph{forwards} in time
	(section~\ref{sect-EH/algorithms/null-geodesics-forwards}).
\item	Integrate null geodesics \emph{backwards} in time
	(section~\ref{sect-EH/algorithms/null-geodesics-backwards}).
\item	Integrate null \emph{surfaces} backwards in time
	(section~\ref{sect-EH/algorithms/null-surfaces-backwards}).
\end{itemize}
I describe these in detail in the following subsections.

%%%%%%%%%%%%%%%%%%%%%%%%%%%%%%%%%%%%%%%%

\subsection{Integrating Null Geodesics Forwards in Time}
\label{sect-EH/algorithms/null-geodesics-forwards}

The first generation of event-horizon finders were based directly
on Hawking's original definition of an event horizon:
an event $\cal P$ is within the black hole region of spacetime if
and only if there is no future-pointing \defn{escape route} null geodesic
from $\cal P$ to $\Scri^+$; the event horizon is the boundary of
the black hole region.

That is, as described by Hughes \etal{}~\cite{Hughes94a}, we
numerically integrate the null geodesic equation
\begin{equation}
\frac{d^2 x^a}{d\lambda^2}
+ \Gamma^a_{bc} \frac{dx^a}{d\lambda} \frac{dx^b}{d\lambda}
	= 0
						       \label{eqn-null-geodesic}
\end{equation}
(where $\lambda$ is an affine parameter) forwards in time from a
set of starting events and check which events have ``escaping''
geodesics.  For analytical or semi-analytical studies like that of
Bishop~\cite{Bishop88}, this is an excellent algorithm.

For numerical work it's straightforward to rewrite the null geodesic
equation~\eqref{eqn-null-geodesic} as a coupled system of two
first-order equations, giving the time evolution of photon positions
and 3-momenta in terms of the $3+1$ geometry variables $\alpha$,
$\beta^i$, $g^{ij}$, and their spatial derivatives.  These can then
be time-integrated by standard numerical algorithms.%%%
\footnote{%%%
	 I briefly review ODE integration algorithms and
	 codes in appendix~\ref{app-ODEs}.
	 }%%%
{}  However, in practice several factors complicate this algorithm:

We typically only know the $3+1$~geometry variables on a discrete
lattice of spacetime grid points, and we only know the $3+1$~geometry
variables themselves, not their spatial derivatives.  Therefore we must
numerically differentiate the field variables, and interpolate the field
variables and their spacetime derivatives to each integration point
along each null geodesic.  This is straightforward to implement,%%%
\footnote{%%%
	 In practice the differentiation can usefully
	 be combined with the interpolation; I outline
	 how this can be done in
	 section ~\ref{sect-AH/intro/geometry-interp}.
	 }%%%
{} but the numerical differentiation tends to amplify any
numerical noise that may be present in the field variables.

Another complicating factor is that the numerically computations
generally only span a finite region of spacetime, so it's not entirely
obvious whether or not a given geodesic will eventually reach $\Scri^+$.
However, if the final numerically-generated slice contains an apparent
horizon, we can use this as an approximation: any geodesic which is
inside this apparent horizon will definitely \emph{not} reach $\Scri^+$,
while any other geodesic may be assumed to eventually reach $\Scri^+$
if its momentum is directed away from the apparent horizon.  If the
final slice is approximately stationary, the error from this
approximation should be small.  (I discuss the ``final slice is
approximately stationary'' assumption further in
section~\ref{sect-EH/algorithms/null-surfaces-backwards/error-bounds}.)

%%%%%%%%%%%%%%%%%%%%

\subsubsection{Spherically-Symmetric Spacetimes}

In spherical symmetry this algorithm works well, and has been used by a
number of researchers.  For example, Shapiro and Teukolsky~\cite{Shapiro79z,
Shapiro80, Shapiro85:partI, Shapiro85} used it to study event horizons
in a variety of dynamical evolutions of spherically symmetric collapse
systems.  Figure~\ref{fig-stellar-collapse/EH-and-AH} shows an example
of the event and apparent horizons in one of these simulations.

%%%%%%%%%%
\epubtkImage{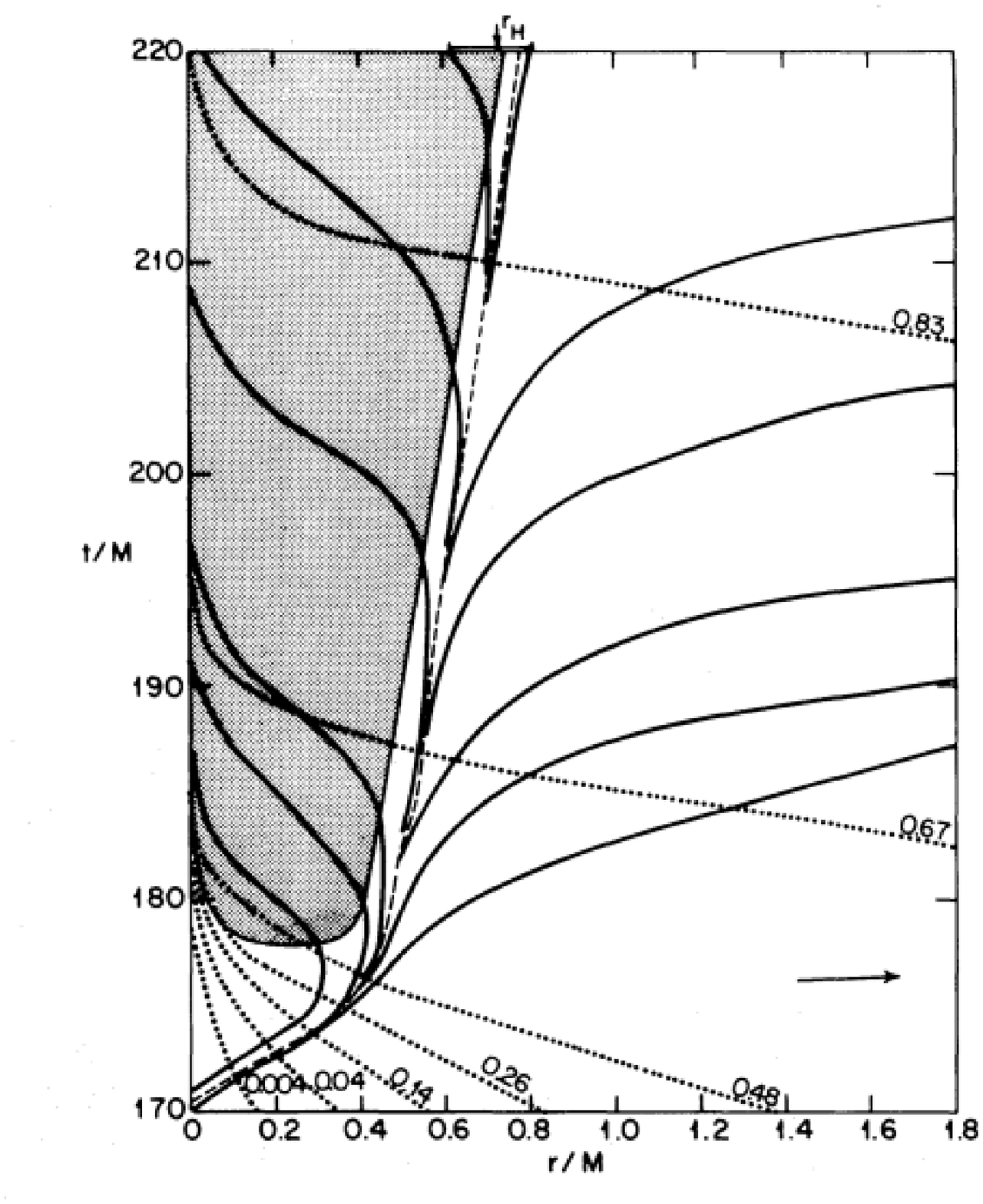}%%%
{%%%
\begin{figure}[bp]
\begin{center}
\centerline{%%%
  \epsfxsize=125mm
  \epsfbox{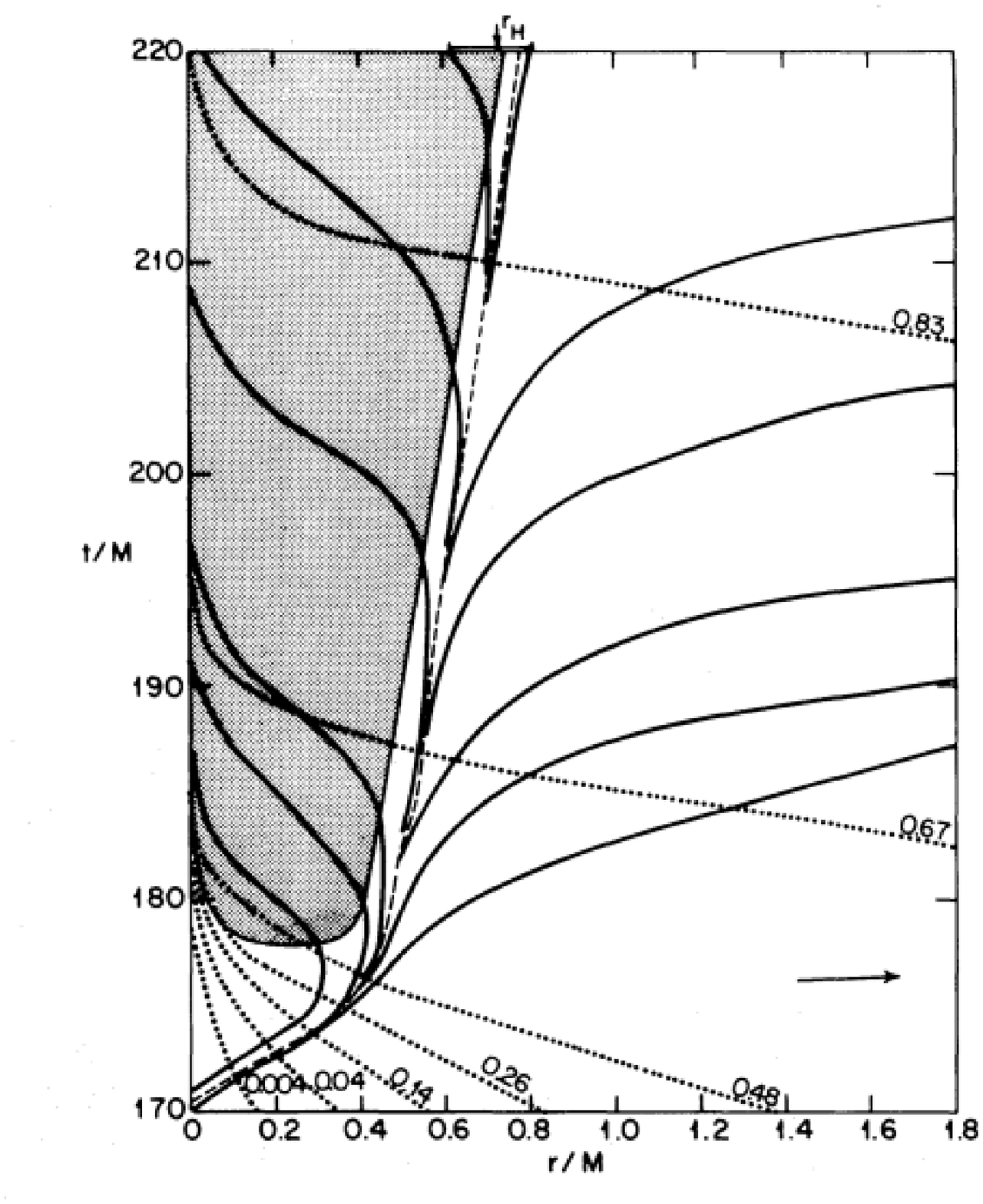}
}
\end{center}
\vspace{-10mm}
\caption[Event and Apparent Horizons in Spherically Symmetric Stellar Collapse]
	{
	This figure shows part of a simulation of the spherically symmetric
	collapse of a model stellar core (a $\Gamma = \frac{5}{3}$ polytrope)
	to a black hole.
	The event horizon (shown by the dashed line) was computed using the
	``integrate null geodesics forwards'' algorithm described in
	section~\ref{sect-EH/algorithms/null-geodesics-forwards};
	solid lines show outgoing null geodesics.
	The apparent horizon (the boundary of the trapped region, shown
	shaded) was computed using the zero-finding algorithm discussed in
	section~\ref{sect-AH/algorithms/zero-finding}.
	The dotted lines show the world lines of Lagrangian matter
	tracers, and are labeled by the fraction of baryons interior
	to them.
	Figure reprinted with permission from
	\href{http://adsabs.harvard.edu/cgi-bin/nph-bib_query?bibcode=1980ApJ...235..199S&amp;db_key=AST}%%%
	     {Shapiro and Teukolsky,
	     \textit{The Astrophysical Journal} \textbf{235}, 199--215 (1980)}.
	Copyright 1980 by the American Astronomical Society.
	}
\label{fig-stellar-collapse/EH-and-AH}
\end{figure}%%%
}
%%%%%%%%%%

%%%%%%%%%%%%%%%%%%%%

\subsubsection{Non--Spherically-Symmetric Spacetimes}

In a non--spherically-symmetric spacetime, several factors make this
algorithm very inefficient:
\begin{itemize}
\item	Many trial events must be tried to accurately resolve the
	event horizon's shape.  (Hughes \etal{}~\cite{Hughes94a}
	describe a 2-stage adaptive numerical algorithm for choosing
	the trial events so as to accurately locate the event horizon
	as efficiently as possible.)
\item	At each trial event we must try many different trial-geodesic
	starting directions to see if any of the geodesics escape to
	$\Scri^+$ (or our numerical approximation to it).
	Hughes \etal{}~\cite{Hughes94a} report needing only $48$~geodesics
	per trial event in several nonrotating axisymmetric spacetimes,
	but about $750$~geodesics per trial event in rotating axisymmetric
	spacetimes, with up to $3000$~geodesics per trial event
	in some regions of the spacetimes.
\item	Finally, each individual geodesic integration requires many
	(short) time steps for an accurate integration, particularly
	in the strong-field region near the event horizon.
\end{itemize}

Because of these limitations, for non--spherically-symmetric spacetimes
the ``integrate null geodesics forwards'' algorithm has generally been
supplanted by the more efficient algorithms I describe in the following
sections.

%%%%%%%%%%%%%%%%%%%%%%%%%%%%%%%%%%%%%%%%

\subsection{Integrating Null Geodesics Backwards in Time}
\label{sect-EH/algorithms/null-geodesics-backwards}

It's well-known that future-pointing outgoing null geodesics near
the event horizon tend to diverge exponentially in time away from
the event horizon.  Figure~\ref{fig-Schw/geodesics} illustrates this
behavior for Schwarzschild spacetime, but the behavior is actually
quite generic.

%%%%%%%%%%
\epubtkImage{fig/Schw--geodesics/geodesics.png}%%%
{%%%
\begin{figure}[bp]
\begin{center}
\centerline{%%%
  \epsfysize=160mm
  \epsfbox{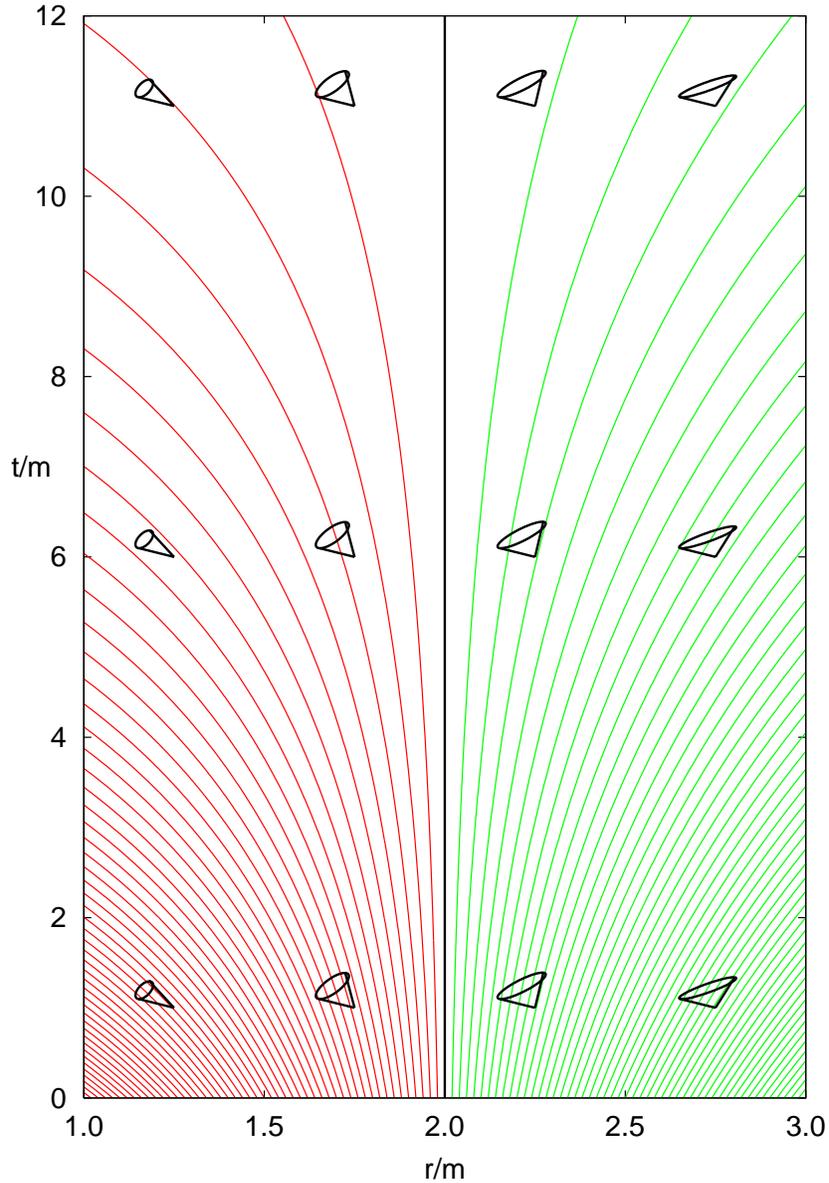}%%
}
\end{center}
\vspace{-10mm}
\caption[Illustration of Geodesics near the Event Horizon
	 in Schwarzschild Spacetime]
	{
	This figure shows a number of light cones and future-pointing
	outgoing null geodesics in a neighborhood of the event horizon
	in Schwarzschild spacetime, plotted in ingoing Eddington-Finkelstein
	coordinates $(t,r)$.  (These coordinates are defined by the
	conditions that $t\,{+}\,r$ is an ingoing null coordinate, while
	$r$ is an areal radial coordinate.)  The geodesics outside
	the event horizon are shown in green; those inside the event
	horizon are shown in red.  All the geodesics start out close
	together near the event horizon; they diverge away from each
	other exponentially in time (here with an $e$-folding time of $4m$
	near the horizon).  Equivalently, they \emph{converge} towards
	each other if integrated \emph{backwards} in time (downwards
	on the page).
	}
\label{fig-Schw/geodesics}
\end{figure}%%%
}
%%%%%%%%%%

Anninos \etal{}~\cite{Anninos94f} and Libson \etal{}~\cite{Libson94a}
observed that while this instability is a problem for the ``integrate
null geodesics forwards in time'' algorithm (it forces that algorithm
to take quite short time steps when integrating the geodesics), we can
turn it to our advantage by integrating the geodesics \emph{backwards}
in time: the geodesics will now \emph{converge} on to the horizon.%%%
\footnote{%%%
	 This convergence is only true in a global sense:
	 locally the event horizon has no special geometric
	 properties, and the Riemann tensor components which
	 govern geodesic convergence/divergence may have
	 either sign.
	 }%%%

This event-horizon finding algorithm is thus to integrate a large number
of such (future-pointing outgoing) null geodesics backwards in time,
starting on the final numerically-generated slice.  As the backwards
integration proceeds, even geodesics which started far from the event
horizon will quickly converge to it.  This can be seen, for example,
in figures~\ref{fig-stellar-collapse/EH-and-AH} and~\ref{fig-Schw/geodesics}.

Unfortunately, this convergence property holds only for \emph{outgoing}
geodesics.  In spherical symmetry the distinction between outgoing and
ingoing geodesics is trivial, but as described by
Libson \etal{}~\cite{Libson94a},
\begin{quote}
[\dots]
for the general 3D case,
when the two tangential directions of the EH are also considered,
the situation becomes more complicated.  Here normal and tangential
are meant in the 3D spatial, not spacetime, sense.  Whether or not a
trajectory can eventually be ``attracted'' to the EH, and how long
it takes for it to become ``attracted,'' depends on the photon's
starting direction of motion.  We note that even for a photon which
is already exactly on the EH at a certain instant, if its velocity at
that point has some component tangential to the EH surface as generated
by, say, numerical inaccuracy in integration, the photon will move
outside of the EH when traced backward in time.  For a small tangential
velocity, the photon will eventually return to the EH
[\dots{} but]
the position to which it returns will not be the original position.

This kind of tangential drifting is undesirable not just because it
introduces inaccuracy in the location of the EH, but more importantly,
because it can lead to spurious dynamics of the ``EH'' thus found.
Neighboring generators may cross, leading to numerically artificial
caustic points [\dots].
\end{quote}

Libson \etal{}~\cite{Libson94a} also observe that
\begin{quote}
Another consequence of the second order nature of the geodesic
equation is that not just the positions but also the directions
must be specified in starting the backward integration.  Neighboring
photons must have their starting direction well correlated in order
to avoid tangential drifting across one another.
\end{quote}

Libson \etal{}~\cite{Libson94a} give examples of the numerical difficulties
that can result from these difficulties, and conclude that this event-horizon
finding algorithm
\begin{quote}
[\dots]
is still quite demanding in finding an accurate history of the EH,
although the difficulties are much milder than those arising from
the instability of integrating forward in time.
\end{quote}
Because of this difficulty, this algorithm has generally been supplanted
by the ``backwards surface'' algorithm I describe in the next section.

%%%%%%%%%%%%%%%%%%%%%%%%%%%%%%%%%%%%%%%%

\subsection{Integrating Null Surfaces Backwards in Time}
\label{sect-EH/algorithms/null-surfaces-backwards}

Anninos \etal{}~\cite{Anninos94f}, Libson \etal{}~\cite{Libson94a}, and
Walker~\cite{Walker98a} introduced the important concept of explicitly
(numerically) finding the event horizon as a null \emph{surface} in
spacetime.  They observed that if we parameterize the event horizon
with a (any) level-set function~$F$ satisfying~\eqref{eqn-F=0-on-surface},
then the condition for the surface $F=0$ to be null is just
\begin{equation}
g^{ab} \partial_a F \partial_b F = 0
\end{equation}
Applying a $3+1$ decomposition to this then gives a quadratic equation
which can be solved to find the time evolution of the level-set function,
\begin{equation}
\partial_t F
	= \frac{
	       - g^{ti} \partial_i F
	       + \sqrt{
		      (g^{ti} \partial_i F)^2
		      - g^{tt} g^{ij} \partial_i f \partial_j f
		      }
	       }
	       {g^{tt}}
					    \label{eqn-null-surface-evolution/F}
\end{equation}

Alternatively, assuming the event horizon in each slice to be a
\Strahlkoerper{} in the manner of
section~\ref{sect-intro/2-surface-parameterizations/Strahlkoerper},
we can define a suitable level-set function~$F$
by~\eqref{eqn-F=r-h(theta,phi)}.  Substituting this definition
into~\eqref{eqn-null-surface-evolution/F} then gives an explicit
evolution equation for the horizon shape function,
\begin{equation}
\partial_t h
	= \frac{
	       - g^{tr}
	       + g^{ru} \partial_u h
	       + \sqrt{
		      (g^{tr} - g^{tu} \partial_u h)^2
		      - g^{tt} \left(
			       g^{rr}
			       - 2g^{ru} \partial_u h
			       + g^{uv} \partial_u h \partial_v h
			       \right)
		      }
	       }
	       {g^{tt}}
					    \label{eqn-null-surface-evolution/h}
\end{equation}

Surfaces near the event horizon share the same ``attraction'' property
discussed in section~\ref{sect-EH/algorithms/null-geodesics-backwards}
for geodesics near the event horizon.  Thus by integrating either
surface representation~\eqref{eqn-null-surface-evolution/F}
or~\eqref{eqn-null-surface-evolution/h} backwards in time, we can
refine an initial guess into a very accurate approximation to the
event horizon.

Notice that in contrast to the null geodesic
equation~\eqref{eqn-null-geodesic},
neither~\eqref{eqn-null-surface-evolution/F}
nor~\eqref{eqn-null-surface-evolution/h} contain any derivatives of
the 4-metric (or equivalently the $3+1$ geometry variables).  This
makes it much easier to integrate these latter equations accurately.%%%
\footnote{%%%
	 Diener~\protect\cite{Diener03a} describes how the
	 algorithm can be enhanced to also determine (integrate)
	 individual null generators of the event horizon.
	 This requires interpolating the 4-metric to the
	 generator positions, but (still) not taking any
	 derivatives of the 4-metric.
	 }%%%

This formulation of the event-horizon finding problem also
completely eliminates the tangential-drifting problem discussed in
section~\ref{sect-EH/algorithms/null-geodesics-backwards}, since the
level-set function only parameterizes motion normal to the surface.

%%%%%%%%%%%%%%%%%%%%

\subsubsection{Error Bounds: Integrating a Pair of Surfaces}
\label{sect-EH/algorithms/null-surfaces-backwards/error-bounds}

For a practical algorithm, it's useful to integrate a \emph{pair}
of trial null surfaces backwards, an \defn{inner-bound} one which
starts (and thus always is) inside the event horizon, and an
\defn{outer-bound} one which starts (and thus always is) outside
the event horizon.  If the final slice contains an apparent horizon,
then any 2-surface inside this can serve as our inner-bound surface.
However, choosing an outer-bound surface is more difficult.

It's this desire for a reliable outer bound on the event horizon
position that motivates our requirement for the final slice to be
approximately stationary, since (in the absence of time-dependent
equations of state or external perturbations entering the system)
this ensures that, for example, any surface substantially outside
the apparent horizon can serve as an outer-bound surface.

Assuming we have an inner- and an outer-bound surface on the final
slice, the spacing between these two surfaces after some period of
backwards integration then gives a reliable error bound for the
computed event horizon position.  Equivalently, a necessary (and,
if there are no other numerical problems, sufficient) condition
for the event-horizon finding algorithm to be accurate is that the
backwards integration must have proceeded far enough for the spacing
between the two trial surfaces to be ``small''.  For a reasonable
definition of ``small'', this typically takes at least $15m_\text{ADM}$
of backwards integration, with $20m_\text{ADM}$ or more providing
much higher accuracy.

In some cases it's difficult to obtain a long enough span of numerical
data for this backwards integration.  For example, in many simulations
of binary black hole collisions, the evolution becomes unstable and
crashes soon after a common apparent horizon forms.  This means that
we can't compute an accurate event horizon for the most interesting
region of the spacetime, that which is close to the black-hole merger.
There's no good solution to this problem except for the obvious one
of developing a stable (or less-unstable) simulation that can be
continued for a longer time.

%%%%%%%%%%%%%%%%%%%%

\subsubsection{Explicit \Strahlkoerper{} Surface Representation}
\label{sect-EH/algorithms/null-surfaces-backwards/Strahlkoerper}

The initial implementations of the ``integrate null surface backwards''
algorithm by Anninos \etal{}~\cite{Anninos94f}, Libson \etal{}~\cite{Libson94a},
and Walker~\cite{Walker98a} were based on the explicit \Strahlkoerper{}
surface integration formula~\eqref{eqn-null-surface-evolution/h},
further restricted to axisymmetry.%%%
\footnote{%%%
	 Walker~\protect\cite{Walker98a} mentions an
	 implementation for fully generic slices, but
	 only presents results for the axisymmetric case.
	 }%%%

For a single black hole the coordinate choice is straightforward.
For the two--black-hole case, the authors used topologically cylindrical
coordinates $(\rho,z,\phi)$, where the two black holes collide along
the axisymmetry ($z$)~axis.  Based on the symmetry of the problem,
they then assumed that the event horizon shape could be written in
the form
\begin{equation}
\rho = h(z)
\end{equation}
in each $t=\text{constant}$ slice.

This spacetime's event horizon has the now-classic ``pair of pants''
shape, with a non-differentiable cusp along the ``inseam'' (the $z$~axis
$\rho = 0$) where new generators join the surface.  The authors tried
two ways of treating this cusp numerically:
\begin{itemize}
\item	Since the cusp's location is known \textit{a priori},
	it can be treated as a special case in the angular finite
	differencing, using one-sided numerical derivatives as
	necessary.
\item	Alternatively, Thorne~\cite{Thorne94-EH-generators-suggestion}
	suggested calculating the \emph{union} of the event horizon
	and all its null generators (including those which haven't
	yet joined the surface).  This ``surface'' has a complicated
	topology (it self-intersects along the cusp), but it's smooth
	everywhere.  This is illustrated by
	figure~\ref{fig-2BH-headon-collision/EH-and-generators-in-slice},
	which shows a cross-section of this surface in a single slice,
	for a head-on binary black hole collision.  For comparison,
	figure~\ref{fig-2BH-headon-collision/EH-spacetime-view} shows a
	perspective view of part of the event horizon and some of its
	generators, for a similar head-on binary black hole collision.
\end{itemize}

%%%%%%%%%%
\epubtkImage{fig/2BH-headon-collision--EH-etal/EH-and-locus-of-generators.png}%%%
{%%%
\begin{figure}[bp]
\begin{center}
\centerline{%%%
  \epsfxsize=125mm
  \epsfbox{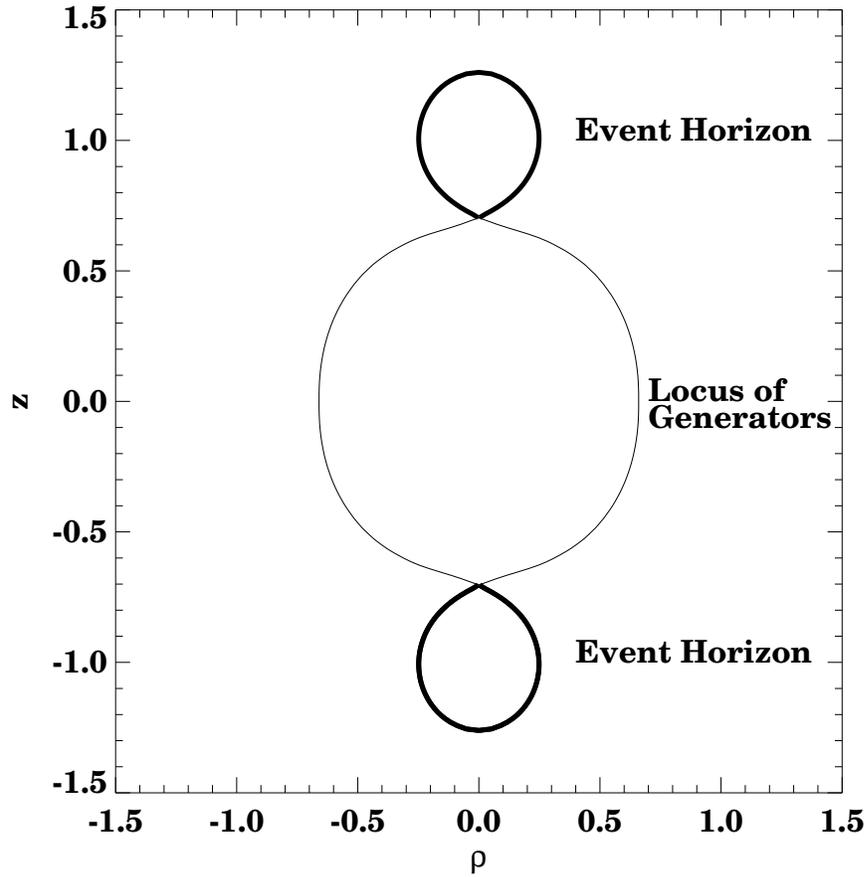}%%%
}
\end{center}
\caption[Event Horizon and Locus of Generators
	 for a Head-On Binary Black Hole Collision]
	{
	This figure shows a view of the numerically-computed
	event horizon in a single slice, together with the locus
	of the event horizon's generators that haven't yet joined
	the event horizon in this slice, for a head-on binary
	black hole collision.
	Notice how the event horizon is non-differentiable
	at the cusp where the new generators join it.
	Figure reprinted with permission from
	\href{http://link.aps.org/abstract/PRD/v53/p4335}%%%
	     {Libson \etal,
	      \textit{Physical Review~D} \textbf{53}, 4335--4350 (1996)}.
	Copyright 1996 by the American Physical Society.
	}
\label{fig-2BH-headon-collision/EH-and-generators-in-slice}
\end{figure}%%%
}
%%%%%%%%%%

%%%%%%%%%%
\epubtkImage{fig/2BH-headon-collision--EH-etal/EH.jpeg}%%%
{%%%
\begin{figure}[bp]
\begin{center}
\centerline{%%%
  \epsfxsize=140mm
  \epsfbox{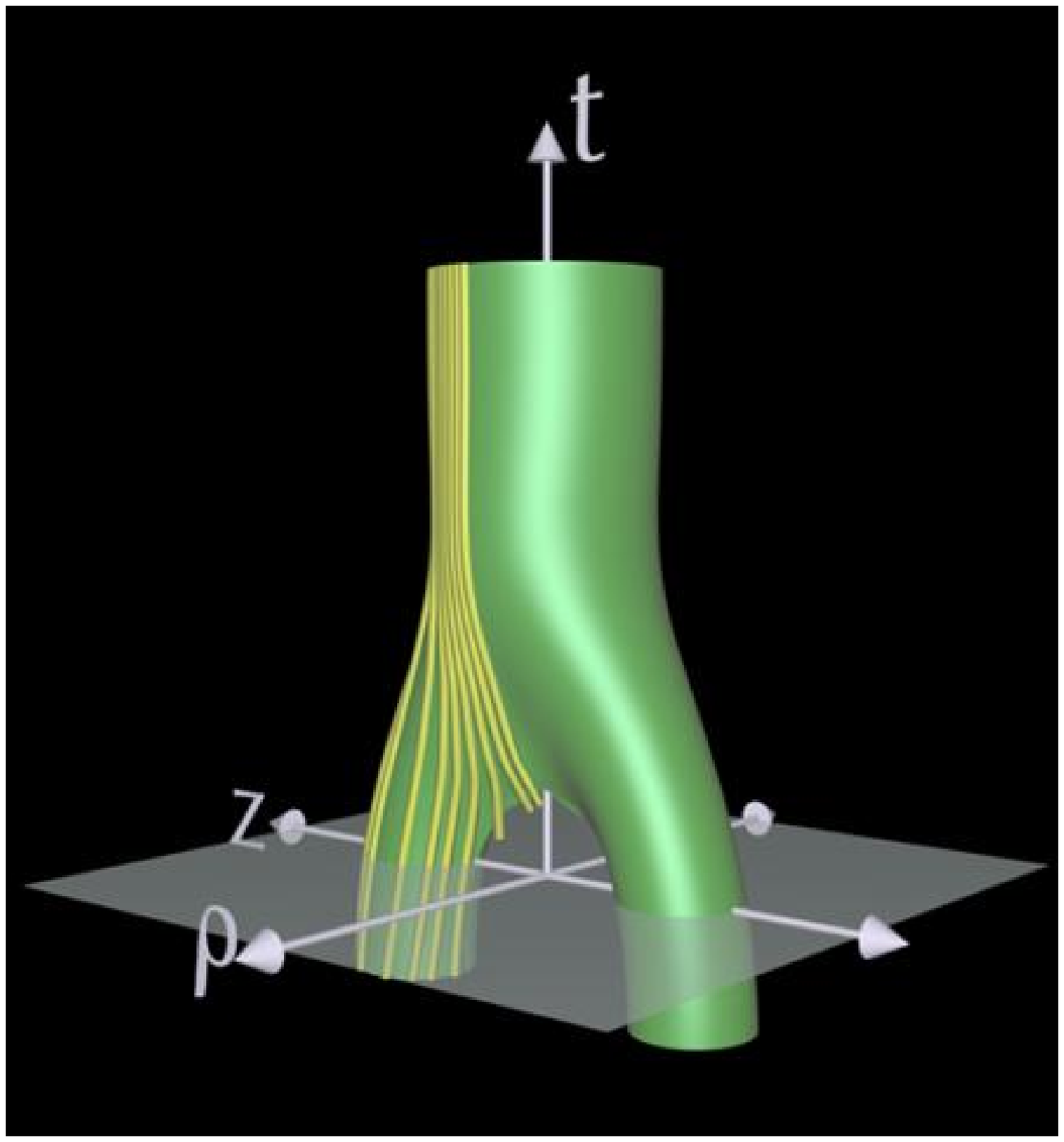}%%%
}
\end{center}
\caption[Event Horizon for a Head-On Binary Black Hole Collision]
	{
	This figure shows a perspective view of the numerically-computed
	event horizon, together with some of its generators, for the
	head-on binary black hole collision discussed in detail by
	\href{http://www.sciencemag.org/cgi/content/abstract/270/5238/941}%%%
	     {Matzner \etal{}~\protect\cite{Matzner95a}}.
	Figure courtesy of Edward Seidel.
	}
\label{fig-2BH-headon-collision/EH-spacetime-view}
\end{figure}%%%
}
%%%%%%%%%%

Caveny \etal{}~\cite{Caveny-2002, Caveny-Matzner-2003a} implemented
the ``integrate null surfaces backwards'' algorithm for fully generic
numerically-computed spacetimes, using the explicit \Strahlkoerper{}
surface integration formula~\eqref{eqn-null-surface-evolution/h}.  To
handle moving black holes, they recentered each black hole's
\Strahlkoerper{} parameterization~\eqref{eqn-r=h(theta,phi)} on the
black hole's coordinate centroid at each time step.

For single-black-hole test cases (Kerr spacetime in various coordinates),
they report typical accuracies of a few percent in determining the
event horizon position and area.

For binary-black-hole test cases (the Kastor-Traschen extremal-charge
black hole coalescence with a cosmological constant), they detect black
hole coalescence (which appears as a bifurcation in the backwards time
integration) by the ``necking off'' of the surface.
Figure~\ref{fig-Kastor-Traschen-2BH-headon-collision/EH}
shows an example of their results.

%%%%%%%%%%
\epubtkImage{fig/Kastor-Traschen-2BH-headon-collision--EH/EH.png}%%%
{%%%
\begin{figure}[bp]
\begin{center}
\centerline{%%%
  \epsfxsize=125mm
  \epsfbox{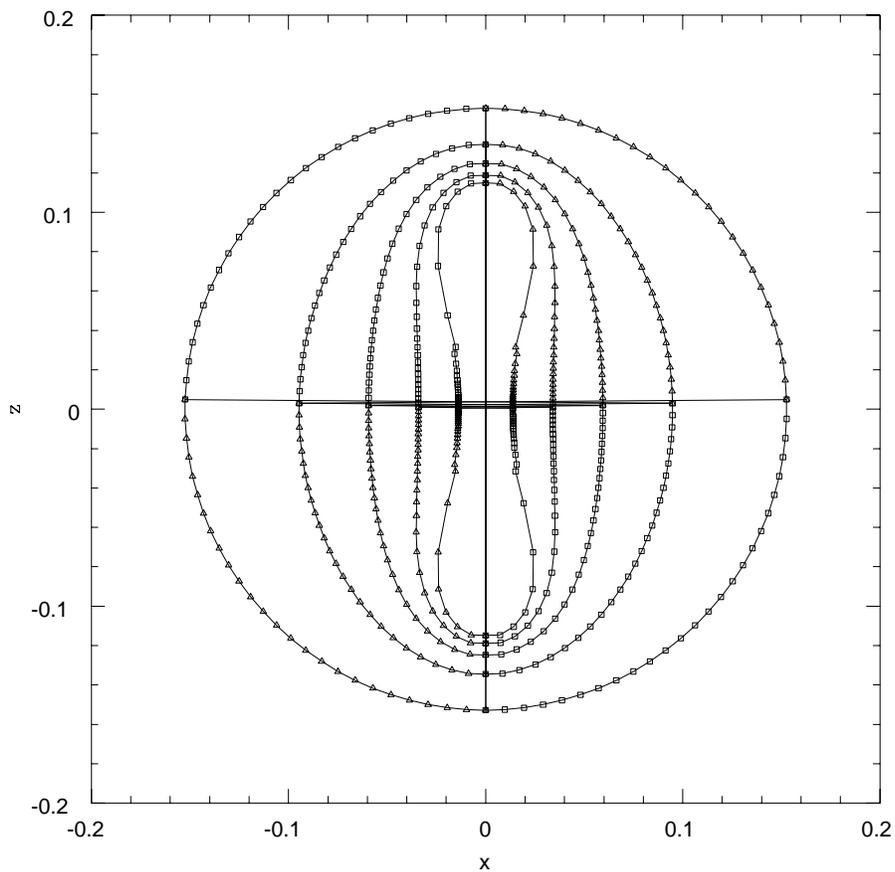}%%%
}
\end{center}
\caption[Event Horizon for a Head-On Kastor-Traschen Extremal Black Hole Collision]
	{
	This figure shows the cross-section of the numerically-computed
	event horizon in each of 5~different slices, for the head-on
	collision of two extremal Kastor-Traschen black holes.
	Figure reprinted with permission from
	\href{http://link.aps.org/abstract/PRD/v68/e104003}%%%
	     {Caveny and Matzner,
	      \textit{Physical Review~D} \textbf{68}, 104003 (2003)}.
	Copyright 2003 by the American Physical Society.
	}
\label{fig-Kastor-Traschen-2BH-headon-collision/EH}
\end{figure}%%%
}
%%%%%%%%%%

%%%%%%%%%%%%%%%%%%%%

\subsubsection{Level-Set Parameterization}
\label{sect-EH/algorithms/null-surfaces-backwards/level-set}

Caveny \etal{}~\cite{Caveny-2002, Caveny-Anderson-Matzner-2003a}
and Diener~\cite{Diener03a} (independently) implemented the
``integrate null surfaces backwards'' algorithm for fully generic
numerically-computed spacetimes, using the level-set function
integration formula~\eqref{eqn-null-surface-evolution/F}.
Here the level-set function~$F$ is initialized on the final
slice of the evolution, and evolved backwards in time
using~\eqref{eqn-null-surface-evolution/F} on (conceptually)
the entire numerical grid.  (In practice, only a smaller box
containing the event horizon need be evolved.)

This surface parameterization has the advantage that the event-horizon
topology and (non-)smoothness are completely unconstrained, allowing
the numerical study of configurations such as toroidal event horizons
(discussed in section~\ref{sect-EH/intro}).  It's also convenient that
the level-set function~$F$ is defined on the same numerical grid as the
spacetime geometry, so that no interpolation is needed for the evolution.

The major problem with this algorithm is that during the backwards
evolution, spatial gradients in~$F$ tend to steepen into a jump
discontinuity at the event horizon,%%%
\footnote{%%%
	 Equivalently, Diener~\protect\cite{Diener03a} observed
	 that the locus of any given nonzero value of the level-set
	 function~$F$ is itself a null surface, and tends to
	 move (exponentially) closer and closer to the event
	 horizon as the backwards evolution proceeds.
	 }%%%
{} eventually causing numerical difficulty.

Caveny \etal{}~\cite{Caveny-2002, Caveny-Anderson-Matzner-2003a} deal
with this problem by adding an artificial viscosity term to the level-set
function evolution equation, smoothing out the jump discontinuity in~$F$.
That is, instead of~\eqref{eqn-null-surface-evolution/F}, they actually
evolve~$F$ via
\begin{equation}
\partial_t F
	= \varepsilon^2 \del^2 F
	  + \text{RHS of \eqref{eqn-null-surface-evolution/F}}
\end{equation}
where $\del^2$ is a generic 2nd~order linear spatial differential
operator, and $\varepsilon > 0$ is a (small) dissipation constant.
This scheme works, but the numerical viscosity does seem to lead to
significant errors (several percent) in their computed event-horizon
positions and areas,%%%
\footnote{%%%
	 They describe how Richardson extrapolation can
	 be used to improve the accuracy of the solutions
	 from $O(\varepsilon)$ to $O(\varepsilon^2)$,
	 but it appears that this hasn't been done for
	 their published results.
	 }%%%
{} and even failure to converge to the correct solution for some
test cases (\eg{} rapidly-spinning Kerr black holes).

Alternatively, Diener~\cite{Diener03a} developed a technique of
periodically reinitializing the level-set function to approximately
the signed distance from the event horizon.  To do this, he periodically
evolves
\begin{equation}
\partial_\lambda F
	= - \frac{F}{\sqrt{F^2+1}} \bigl( |\del F| - 1 \bigr)
\end{equation}
in an unphysical ``pseudo-time'' $\lambda$ until an approximate
steady state has been achieved.  He reports that this works well in
most circumstances, but can significantly distort the computed
event horizon if the $F=0$ isosurface (the current approximation
to the event horizon) is only a few grid points thick in any direction,
as typically occurs just around the time of a topology change in
the isosurface.  He avoids this problem by estimating the minimum
thickness of this isosurface and, if it's below a threshold,
deferring the reinitialization.

In various tests on analytical data, Diener~\cite{Diener03a} found
this event-horizon finder, \program{EHFinder}, to be robust and
highly accurate, typically locating the event horizon to much less
than $1\%$~of the 3-dimensional grid spacing. Even with only 10~grid
points across the event horizon, this already corresponds to accuracies
on the order of~$0.1\%$, and this accuracy improves as expected
(2nd~order convergence) with increasing resolution.

As an example of the results obtained with \program{EHFinder},
figure~\ref{fig-2BH-spiraling-collision/EH-spacetime-view} shows two views
of the numerically-computed event horizon for a spiraling binary
black hole collision.  As another example,
figure~\ref{fig-rotating-NS-collapse/EH-and-AH} shows the
numerically-computed event and apparent horizons in the collapse
of a rapidly rotating neutron star to a Kerr black hole.
(The apparent horizons were computed using the \program{AHFinderDirect}
code described in section~\ref{sect-AH/algorithms/elliptic-PDE/summary}.)

\program{EHFinder} is implemented as a module (``thorn'') in the
\href{http://www.cactuscode.org}{\program{Cactus}} computational
toolkit.  It originally worked only with the \program{PUGH} unigrid driver,
but Diener~\cite{Diener-2005-modifying-EHFinder-to-work-with-Carpet}
is currently enhancing it to work with the
\href{http://www.carpetcode.org}{\program{Carpet}} mesh-refinement
driver.  \program{EHFinder} is freely available by anonymous~CVS,
and is now used by several research groups.

%%%%%%%%%%
%% the png file for the 2nd subfigure is
%%           fig/2BH-spiraling-collision--EH/EH-view1.png
\epubtkImage{fig/2BH-spiraling-collision--EH/EH-view1.png}%%%
{%%%
\begin{figure}[p]
\begin{center}
\centerline{%%%
  \epsfysize=160mm
  \epsfbox{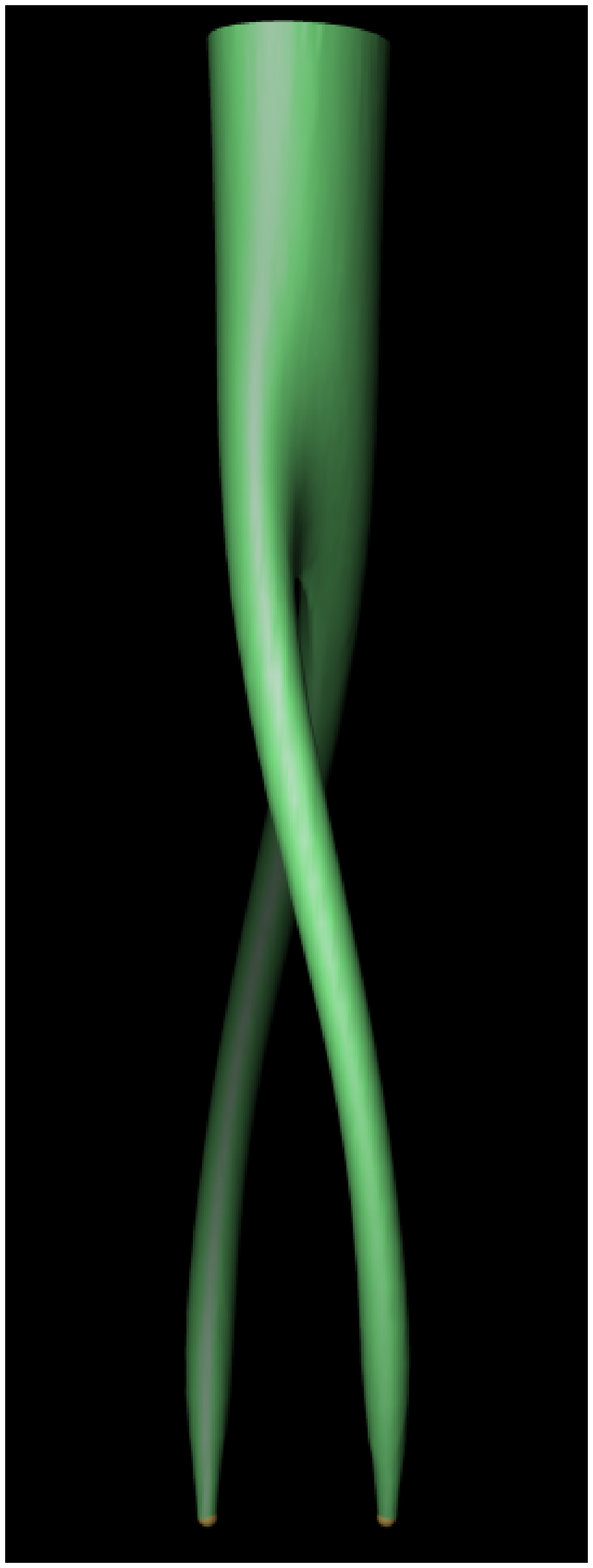}%%%
  \qquad%%%
  \epsfysize=160mm
  \epsfbox{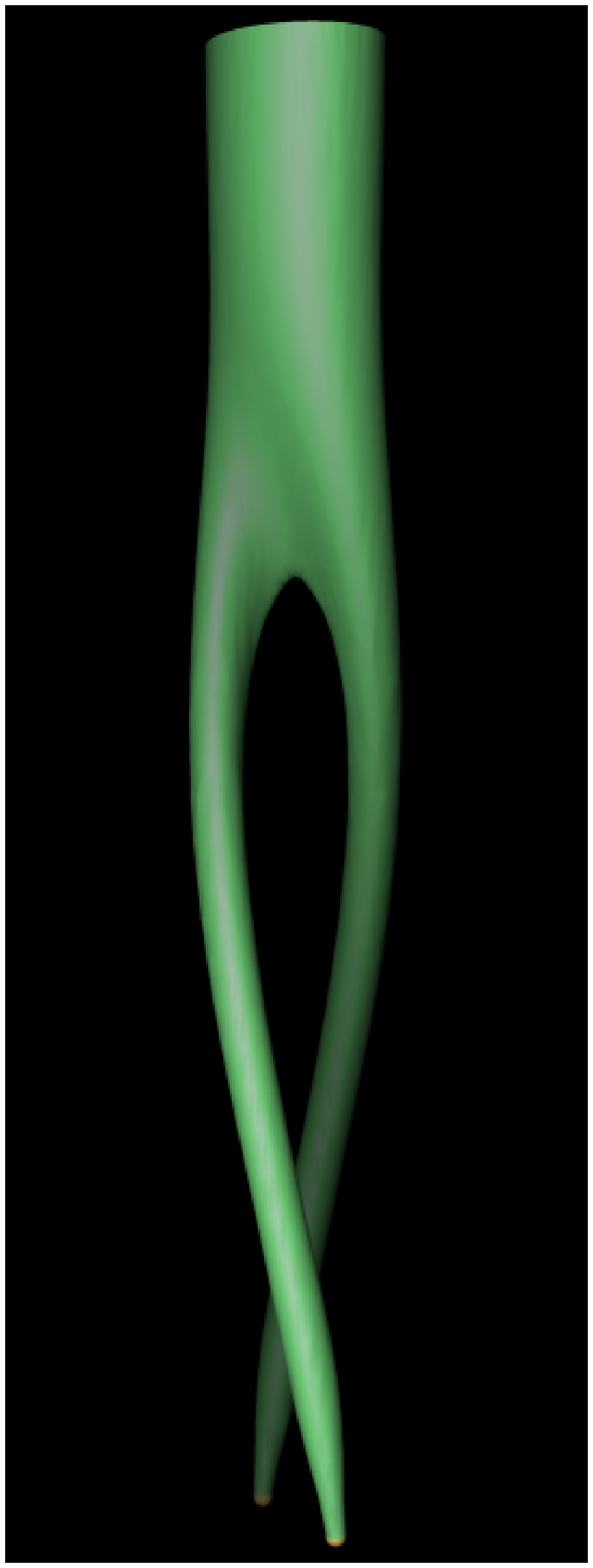}%%%
}
\end{center}
\vspace{-10mm}
\caption[Numerically-Computed Event Horizon in a Binary Black Hole Collision]
	{
	This figure shows two views of the numerically-computed
	event horizon for a spiraling binary black hole collision.
	The initial data was constructed to have an approximate
	helical Killing vector, corresponding to black holes
	in approximately circular orbits (the $D=18$ case of
	Grandcl\'{e}ment \etal{}~\protect\cite{Grandclement02}),
	with a proper separation of the apparent horizons of~$6.9m$.
	Figure courtesy of Peter Diener, visualization by Werner Benger.
	}
\label{fig-2BH-spiraling-collision/EH-spacetime-view}
\end{figure}%%%
}
%%%%%%%%%%

\epubtkImage{fig/rotating-NS-collapse--EH-and-AH/EH-and-AH.png}%%%
{%%%
\begin{figure}[p]
\begin{center}
\centerline{%%%
  \epsfxsize=125mm
  \epsfbox{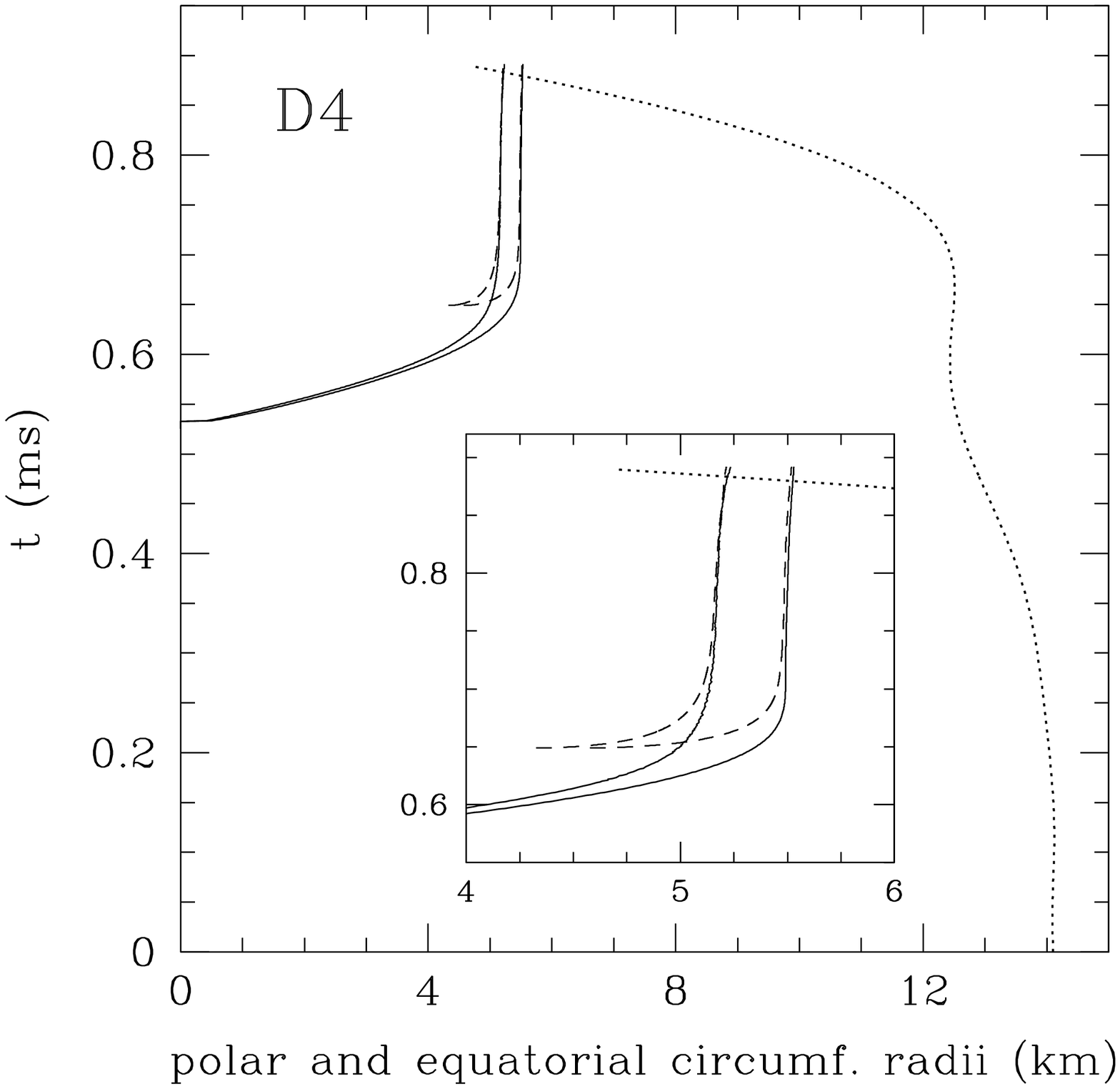}
}
\end{center}
\vspace{-10mm}
\caption[Numerically-Computed Event and Apparent Horizons
	 in Rotating Neutron-Star Collapse]
	{
	This figure shows the polar and equatorial radia
	of the event horizon (solid lines) and apparent horizon
	(dashed lines) in a numerical simulation of the collapse
	of a rapidly rotating neutron star to form a Kerr black hole.
	The dotted line shows the equatorial radius of the
	stellar surface.
	These results are from the D4 simulation of
	Baiotti \etal{}~\protect\cite{Baiotti04}.
	Notice how the event horizon grows from zero size,
	while the apparent horizon first appears at a finite
	size, and grows in a spacelike manner.  Notice
	also that both surfaces are flattened due to the
	rotation.
	Figure reprinted with permission from
	\href{http://link.aps.org/abstract/PRD/v71/e024035}%%%
	     {Baiotti \etal,
	      \textit{Physical Review~D} \textbf{71}, 024035 (2005)}.
	}
\label{fig-rotating-NS-collapse/EH-and-AH}
\end{figure}%%%
}
%%%%%%%%%%

%%%%%%%%%%%%%%%%%%%%%%%%%%%%%%%%%%%%%%%%%%%%%%%%%%%%%%%%%%%%%%%%%%%%%%%%%%%%%%%%

\section{Summary of Algorithms/Codes for Finding Event Horizons}
\label{sect-EH/algorithms/summary}

In spherical symmetry the ``integrate null geodesics forwards''
algorithm (section~\ref{sect-EH/algorithms/null-geodesics-forwards})
is reasonable, though the ``integrate null geodesics backwards''
algorithm (section~\ref{sect-EH/algorithms/null-geodesics-backwards})
is more efficient.

In non--spherically-symmetric spacetimes the
``integrate null surfaces backwards'' algorithm
(section~\ref{sect-EH/algorithms/null-surfaces-backwards}) is clearly
the best algorithm known: it's efficient, accurate, and fairly easy
to implement.  For generic spacetimes, Diener's event-horizon finder
\program{EHFinder}~\cite{Diener03a} is particularly notable as a
freely available implementation of this algorithm as a module (``thorn'')
in the widely-used \href{http://www.cactuscode.org}{\program{Cactus}}
computational toolkit.

%%%%%%%%%%%%%%%%%%%%%%%%%%%%%%%%%%%%%%%%%%%%%%%%%%%%%%%%%%%%%%%%%%%%%%%%%%%%%%%%
%%%%%%%%%%%%%%%%%%%%%%%%%%%%%%%%%%%%%%%%%%%%%%%%%%%%%%%%%%%%%%%%%%%%%%%%%%%%%%%%

\chapter{Finding Apparent Horizons}
\label{chap-AH}

%%%%%%%%%%%%%%%%%%%%%%%%%%%%%%%%%%%%%%%%%%%%%%%%%%%%%%%%%%%%%%%%%%%%%%%%%%%%%%%%

\section{Introduction}
\label{sect-AH/intro}

%%%%%%%%%%%%%%%%%%%%%%%%%%%%%%%%%%%%%%%%

\subsection{Definition}
\label{sect-AH/intro/defn}

Given a (spacelike) $3+1$ slice, a \defn{marginally outer trapped surface}
(MOTS) is defined as a smooth (differentiable) closed orientable
2-surface in the slice whose future-pointing outgoing null geodesics
have zero expansion $\Theta$.  There may be multiple MOTSs in a slice.
MOTSs may nest within each other, but they can't cross.  An
\defn{apparent horizon} is then defined as an outermost MOTS in a slice,
\ie{}, a MOTS not contained in any other MOTS.

Equivalently, a \defn{trapped surface} is defined as a smooth closed
2-surface in the slice whose future-pointing outgoing null geodesics
have \emph{negative} expansion.  The \defn{trapped region} in the
slice is then defined as the union of all trapped surfaces, and the
apparent horizon is defined as the outer boundary of the trapped region.

Notice that the apparent horizon is defined \emph{locally in time}
(it can be computed using only Cauchy data on a spacelike slice), but
(because of the requirement that it be closed) \emph{non-locally in space}.%%%
\footnote{%%%
	 As an indication of the importance of the ``closed''
	 requirement, Hawking~\cite{Hawking73} has observed
	 that if we consider two spacelike-separated events
	 in Minkowski spacetime, the intersection of their
	 backwards light cones satisfies all the conditions
	 of the MOTS definition, except that it's not closed.
	 }%%%
{}  Hawking and Ellis~\cite{Hawking73a} discuss the general properties
of MOTSs and apparent horizons in more detail.

%%%%%%%%%%%%%%%%%%%%%%%%%%%%%%%%%%%%%%%%

\subsection{General Properties}
\label{sect-AH/intro/properties}

Given certain technical assumptions (including energy conditions),
the existence of any MOTS (and hence any apparent horizon) implies
that the slice contains a black hole.%%%
\footnote{%%%
	 Note that the converse of this latter statement is
	 \emph{not} true:  An arbitrary (spacelike) slice through
	 a black hole need not contain any apparent horizon.
	 Notably, Wald and Iyer~\protect\cite{Wald91} have
	 constructed a family of angularly anisotropic slices
	 in Schwarzschild spacetime which approach arbitrarily
	 close to $r = 0$ yet contain no apparent horizons.  However,
	 Schnetter and Krishnan~\protect\cite{Schnetter-Krishnan-2005}
	 have recently studied the behavior of apparent horizons
	 in various anisotropic slices in Schwarzschild and Vaidya
	 spacetimes, finding that the Wald and Iyer behavior seems
	 to be rare.
	 }%%%
{}  Moreover, the apparent horizon necessarily coincides with, or is
contained in, an event horizon.  In a stationary spacetime the event
and apparent horizons coincide.

It's this relation to the event horizon which makes apparent horizons
valuable for numerical computation: an apparent horizon provides a
useful approximation to the event horizon in a slice, but unlike the
event horizon, an apparent horizon is defined locally in time and so
can be computed ``on the fly'' during a numerical evolution.

Given a family of spacelike $3+1$ slices which foliate part of
spacetime, the union of the slices' apparent horizons (assuming they exist)
forms a world-tube.  This world-tube is necessarily either null or
spacelike.  If it's null this world-tube is slicing-independent
(choosing a different family of slices gives the same world-tube,
at least so long as each slice still intersects the world-tube in
a surface with 2-sphere topology).  However, if the world-tube is
spacelike, it's \emph{slicing-dependent}: choosing a different
family of slices will in general give a different world-tube.%%%
\footnote{%%%
	 Ashtekar and Galloway~\protect\cite{Ashtekar05}
	 have recently proved ``a number of physically
	 interesting constraints'' on this slicing-dependence.
	 }%%%

Except for flow algorithms (section~\ref{sect-AH/algorithms/flow}),
all numerical ``apparent horizon'' finding algorithms and codes
actually find MOTSs, and hereinafter I generally follow the common
(albeit sloppy) practice in numerical relativity of blurring the
distinction between an MOTS and an apparent horizon.

%%%%%%%%%%%%%%%%%%%%%%%%%%%%%%%%%%%%%%%%

\subsection{Trapping, Isolated, and Dynamical Horizons}
\label{sect-AH/intro/trapping,isolated,dynamical-horizons}

Hayward~\cite{Hayward94a} introduced the important concept of a
\defn{trapping horizon}, roughly speaking an apparent-horizon world-tube
where the expansion becomes negative if the surface is deformed in
the inward null direction, along with several useful variants.
Ashtekar, Beetle, and Fairhurst~\cite{Ashtekar98a}
and Ashtekar and Krishnan~\cite{Ashtekar-etal-2002-dynamical-horizons}
later defined the related concepts of an \defn{isolated horizon},
essentially an apparent-horizon world-tube which is null, and a
\defn{dynamical horizon}, essentially an apparent-horizon world-tube
which is spacelike.

These world-tubes obey a variety of local and global conservation
laws, and have many applications in analyzing numerically-computed
spacetimes.  See the references cited above, and also
Dreyer \etal{}~\cite{Dreyer02a},
Ashtekar and Krishnan~\cite{Ashtekar03a, Ashtekar:2004cn},
Gourgoulhon and Jaramillo~\cite{Gourgoulhon-Jaramillo-Review}, and
Booth~\cite{Booth-review} for further discussions,
including applications to numerical relativity.

%%%%%%%%%%%%%%%%%%%%%%%%%%%%%%%%%%%%%%%%

\subsection{Description in Terms of the $3+1$ Variables}
\label{sect-AH/intro/3+1-description}

In terms of the $3+1$ variables, a marginally outer trapped surface
(and thus an apparent horizon) satisfies the condition
(\cite{York89}, \cite[section~IIA]{Gundlach98a})
\begin{equation}
\Theta \equiv
        \del_i s^i + K_{ij} s^i s^j - K = 0
							      \label{eqn-AH/s^i}
\end{equation}
where $s^i$ is the outward-pointing unit 3-vector normal to the surface.%%%
\footnote{%%%
	 Notice that in order for the 3-divergence
	 in~\eqref{eqn-AH/s^i} to be meaningful, $s^i$ (defined
	 only as a field on the marginally outer trapped surface)
	 must be smoothly continued off the surface, and extended
	 to a field in some 3-dimensional neighborhood of the
	 surface.  The off-surface continuation is non-unique,
	 but it's easy to see that this doesn't affect the value
	 of $\Theta$ on the surface.
	 }%%%

Assuming the \Strahlkoerper{} surface
parameterization~\eqref{eqn-r=h(theta,phi)}, \eqref{eqn-AH/s^i} can be
rewritten in terms of angular 1st and 2nd~derivatives of the horizon
shape function~$h$,
\begin{equation}
\Theta \equiv
	\Theta(h, \partial_u h, \partial_{uv} h;
	       g_{ij}, \partial_k g_{ij}, K_{ij})
		= 0
								\label{eqn-AH/h}
\end{equation}
where $\Theta$ is a complicated nonlinear algebraic function of the
arguments shown.  (Shibata~\cite{Shibata97a} and
Thornburg~\cite{Thornburg95, Thornburg2003:AH-finding} give the
$\Theta(h, \partial_u h, \partial_{uv} h)$ function explicitly.)

%%%%%%%%%%%%%%%%%%%%%%%%%%%%%%%%%%%%%%%%

\subsection{Geometry Interpolation}
\label{sect-AH/intro/geometry-interp}

$\Theta$ depends on the slice geometry variables $g_{ij}$,
$\partial_k g_{ij}$, and $K_{ij}$ at the horizon position.%%%
\footnote{%%%
	 Or, in the Huq \etal{}~\protect\cite{Huq96, Huq00}
	 algorithm described in
	 section~\ref{sect-AH/algorithms/elliptic-PDE/evaluating-Theta},
	 at the local Cartesian grid point positions.
	 }%%%
{}  In practice these variables are usually only known on the
(3-dimensional) numerical grid of the underlying numerical-relativity
simulation,%%%
\footnote{%%%
\label{footnote-spectral-methods}%%%
	 If the underlying simulation uses spectral methods
	 (see
	 Gottlieb and Orszag~\protect\cite{Gottlieb-Orszag:pseudospectral}
	 and Boyd~\protect\cite{Boyd00}
	 for general discussions of spectral methods,
	 and (for example)
	 Ansorg \etal{}~\cite{Ansorg:2003br, Ansorg:2004ds, Ansorg05},
	 Bonazzola \etal{}~\protect\cite{Bonazzola89,
	 Bonazzola-etal-1996:spectral-methods-in-gr,
	 Bonazzola-etal-1998:spectral-methods-in-gr-astrophysics},
	 Grandcl\'{e}ment \etal{}~\protect\cite{Grandclement-etal-2000:multi-domain-spectral-method},
	 Kidder \etal{}~\protect\cite{Kidder99a, Kidder00a, Kidder00b},
	 and Pfeiffer \etal{}~\cite{Pfeiffer:2002wt}
	 for applications to numerical relativity)
	 then the spectral series can be evaluated anywhere,
	 so no actual interpolation need be done.
	 }%%%
{} so they must be interpolated to the horizon position, and more
generally, to the position of each intermediate-iterate trial shape
the apparent-horizon finding algorithm tries in the process of
(hopefully) converging to the horizon position.

Moreover, usually the underlying simulation gives only $g_{ij}$ and
$K_{ij}$, so $g_{ij}$ must be numerically differentiated to obtain
$\partial_k g_{ij}$.
As discussed by Thornburg~\cite[section~6.1]{Thornburg2003:AH-finding},
it's somewhat more efficient to combine the numerical differentiation
and interpolation operations, essentially doing the differentiation
inside the interpolator.%%%
\footnote{%%%
	 An interpolator generally works by (conceptually)
	 locally fitting a fitting function (usually a
	 low-degree polynomial) to the data points in a
	 neighborhood of the interpolation point, then
	 evaluating the fitting function at the interpolation
	 point.  By evaluating the {\em derivative\/} of
	 the fitting function, the $\partial_k g_{ij}$
	 values can be obtained very cheaply at the same
	 time as the $g_{ij}$ values.
	 }%%%

Thornburg~\cite[section~6.1]{Thornburg2003:AH-finding} argues that for
an elliptic-PDE algorithm (section~\ref{sect-AH/algorithms/elliptic-PDE}),
for best convergence of the nonlinear elliptic solver, the interpolated
geometry variables should be smooth (differentiable) functions of the
trial horizon surface position.  He argues that that the usual Lagrange
polynomial interpolation doesn't suffice here (in some cases his
Newton's-method iteration failed to converge), because this interpolation
gives results which are only piecewise differentiable.%%%
\footnote{%%%
	 Thornburg~\protect\cite[appendix~F]{Thornburg98}
	 gives a more detailed discussion of this non-smoothness
	 of Lagrange-polynomial interpolation errors.
	 }%%%
{}  To avoid this problem,
Thornburg~\cite[section~6.1]{Thornburg2003:AH-finding} uses Hermite
polynomial interpolation; Cook and Abrahams~\cite{Cook92} use bicubic
spline interpolation.  Most other researchers either don't describe
their interpolation scheme, or use Lagrange polynomial interpolation,
and don't report serious non-convergence problems.

%%%%%%%%%%%%%%%%%%%%%%%%%%%%%%%%%%%%%%%%

\subsection{Criteria for Assessing Algorithms}
\label{sect-AH/intro/criteria}

Ideally, an apparent-horizon finder should have several attributes:
\begin{description}
\item[Robust:]
	The algorithm/code should find an (the) apparent horizon in
	a wide range of numerically-computed slices, without
	requiring extensive tuning of initial guesses, iteration
	parameters, etc.  This is often relatively easy to achieve
	for ``tracking'' the time evolution of an existing apparent
	horizon (where the most recent previously-found apparent horizon
	provides an excellent initial guess for the new apparent-horizon
	position), but may be difficult for detecting the appearance
	of a new (outermost) apparent horizon in an evolution, or
	for initial-data or other studies where there is no
	``previous time step''.
\item[Accurate:]
	The algorithm/code should find an (the) apparent horizon
	to high accuracy and shouldn't report spurious ``solutions''
	(``solutions'' which aren't actually apparent horizons
	or, at least, marginally outer trapped surfaces).
\item[Efficient:]
	The algorithm/code should be efficient in terms of its memory
	use and CPU time; in practice CPU time is generally the major
	constraint.  It's often desirable to find apparent horizons
	at each time step (or, at least, at frequent intervals) during
	a numerical evolution.  For this to be practical the
	apparent-horizon finder must be very fast.
\end{description}

In practice, no apparent-horizon finder is perfect in all these
dimensions, so trade-offs are inevitable, particularly when ease
of programming is considered.

As discussed in section~\ref{sect-intro/software-engineering}, there
are also significant advantages to having an apparent-horizon finder
that's freely available to other research groups, particularly if it's
designed and documented in such a way as to be relatively portable.

%%%%%%%%%%%%%%%%%%%%%%%%%%%%%%%%%%%%%%%%

\subsection{Local versus Global Algorithms}
\label{sect-AH/intro/local-vs-global}

Apparent-horizon finding algorithms can usefully be divided into two
broad classes:
\begin{description}
\item[Local] algorithms are those whose convergence is only guaranteed
	in some (functional) neighborhood of a solution.  These algorithms
	require a ``good'' initial guess in order to find the apparent
	horizon.  Most apparent-horizon finding algorithms are local.
\item[Global] algorithms are those which can (in theory, ignoring
	finite-step-size and other numerical effects) converge
	to the apparent horizon independent of any initial guess.
	Flow algorithms (section~\ref{sect-AH/algorithms/flow})
	are the only truely global algorithms.
	Zero-finding in spherical symmetry
	(section~\ref{sect-AH/algorithms/zero-finding})
	and shooting in axisymmetry (section~\ref{sect-AH/algorithms/shooting})
	are ``almost global'' algorithms: they require only
	1-dimensional searches, which (as discussed in
	appendix~\ref{app-single-nonlinear-eqn}) can be programmed
	to be very robust and efficient.  In many cases horizon
	pretracking (section~\ref{sect-AH/algorithms/pretracking})
	can semi-automatically find an initial guess for a local
	algorithm, essentially making the local algorithm behave
	like an ``almost global'' one.
\end{description}

One might wonder why local algorithms are ever used, given the apparently
superior robustness (guaranteed convergence independent of any initial
guess) of global algorithms.  There are two basic reasons:
\begin{itemize}
\item	In practice, local algorithms are much faster than global ones,
	particularly when ``tracking'' the time evolution of an existing
	apparent horizon.
\item	Due to finite-step-size and other numerical effects,
	in practice even ``global'' algorithms may fail to converge
	to an apparent horizon (that is, the algorithms may sometimes
	fail to find an apparent horizon even when one exists in the slice).
\end{itemize}

%%%%%%%%%%%%%%%%%%%%%%%%%%%%%%%%%%%%%%%%%%%%%%%%%%%%%%%%%%%%%%%%%%%%%%%%%%%%%%%%

\section{Algorithms and Codes for Finding Apparent Horizons}
\label{sect-AH/algorithms}

Many researchers have studied the apparent-horizon--finding problem,
and there are a large number of different apparent-horizon finding
algorithms and codes.  Almost all of these require (assume) that
any apparent horizon to be found is a \Strahlkoerper{}
(section~\ref{sect-intro/2-surface-parameterizations})
about some local coordinate origin; both finite-difference and
spectral parameterizations of the \Strahlkoerper{} are common.

For slices with continuous symmetries, special algorithms are sometimes
used:
\begin{description}
\item[Zero-Finding in Spherical Symmetry]
(section~\ref{sect-AH/algorithms/zero-finding})\\
	In spherical symmetry the apparent horizon
	equation~\eqref{eqn-AH/h} becomes a 1-dimensional
	nonlinear algebraic equation, which can be solved by
	zero-finding.
\item[The Shooting Algorithm in Axisymmetry]
(section~\ref{sect-AH/algorithms/shooting})\\
	In axisymmetry the apparent horizon equation~\eqref{eqn-AH/h}
	becomes a nonlinear 2-point boundary value ODE, which can
	be solved by a shooting algorithm.
\end{description}

Alternatively, all the algorithms described below for generic slices are
also applicable to axisymmetric slices, and can take advantage of the
axisymmetry to simplify the implementation and boost performance.

For fully generic slices, there are several broad categories of
apparent-horizon finding algorithms and codes:
\begin{description}
\item[Minimization Algorithms]
(section~\ref{sect-AH/algorithms/minimization})\\
	These algorithms define a scalar norm on $\Theta$ over
	the space of possible trial surfaces.  A general-purpose
	scalar-function-minimization routine is then used to search
	trial-surface-shape space for a minimum of this norm
	(which should give $\Theta = 0$).
\item[Nakamura \etal{}'s Spectral Integral-Iteration Algorithm]
(section~\ref{sect-AH/algorithms/Nakamura-etal})\\
	This algorithm expands the (\Strahlkoerper) apparent-horizon
	shape function in a spherical-harmonic basis, uses the orthogonality
	of spherical harmonics to write the apparent horizon equation
	as a set of integral equations for the spectral coefficients,
	and solves these equations using a functional-iteration algorithm.
\item[Elliptic-PDE Algorithms]
(section~\ref{sect-AH/algorithms/elliptic-PDE})\\
	These algorithms write the apparent horizon equation~\eqref{eqn-AH/h}
	as a nonlinear elliptic (boundary-value) PDE for the horizon shape,
	and solve this PDE using (typically) standard elliptic-PDE
	numerical algorithms.
\item[Horizon Pretracking]
(section~\ref{sect-AH/algorithms/pretracking})\\
	Horizon pretracking solves a slightly more general problem
	than apparent-horizon finding: roughly speaking, the determination
	of the smallest $E \ge 0$ such that the equation $\Theta = E$
	has a solution, and the determination of that solution.
	By monitoring the time evolution of $E$ and of the surfaces
	satisfying this condition, we can determine -- \emph{before}
	it appears -- approximately where (in space) and when (in time)
	a new marginally outer trapped surface \emph{will} appear in
	a dynamic numerically-evolving spacetime.
	Horizon pretracking is implemented as a 1-dimensional (binary)
	search using a slightly-modified elliptic-PDE apparent-horizon
	finding algorithm as a ``subroutine''.  
\item[Flow Algorithms]
(section~\ref{sect-AH/algorithms/flow})\\
	These algorithms start with a large 2-surface (larger than any
	possible apparent horizon in the slice), and shrink it inwards
	using an algorithm which ensures that the surface will stop
	shrinking when it coincides with the apparent horizon.
\end{description}

I describe the major algorithms and codes in these categories in detail
in the following subsections.

%%%%%%%%%%%%%%%%%%%%%%%%%%%%%%%%%%%%%%%%

\subsection{Zero-Finding in Spherical Symmetry}
\label{sect-AH/algorithms/zero-finding}

In a spherically symmetric slice, any apparent horizon must also be
spherically symmetric, so the apparent horizon equation~\eqref{eqn-AH/h}
becomes a 1-dimensional nonlinear algebraic equation $\Theta(h) = 0$
for the horizon radius~$h$.  For example, assuming the usual
polar-spherical spatial coordinates $x^i = (r,\theta,\phi)$, we have
(\cite[equation~(B7)]{Thornburg98})
\begin{equation}
\Theta \equiv
	\frac{\partial_r g_{\theta\theta}}{g_{\theta\theta} \sqrt{g_{rr}}}
	 - 2 \frac{K_{\theta\theta}}{g_{\theta\theta}}
		= 0
					    \label{eqn-AH/spherically-symmetric}
\end{equation}
Given the geometry variables $g_{rr}$, $g_{\theta\theta}$,
$\partial_r g_{\theta\theta}$, and $K_{\theta\theta}$, this equation
may be easily and accurately solved using one of the zero-finding
algorithms discussed in appendix~\ref{app-single-nonlinear-eqn}.%%%
\footnote{%%%
	 Note that $\partial_r g_{\theta\theta}$ is a known
	 coefficient field here, not an unknown (if necessary,
	 it can be obtained by numerically differentiating
	 $g_{\theta\theta}$).  Therefore, despite the appearance
	 of the derivative, \eqref{eqn-AH/spherically-symmetric}
	 is still an \emph{algebraic} equation for the horizon
	 radius~$h$, not a differential equation.
	 }%%%

Zero-finding has been used by many researchers, including
\cite{Shapiro79z, Shapiro80, Shapiro85:partI, Shapiro85,
Petrich-Shapiro-Teukolsky-1985, Choptuik86, Seidel92a, Anninos94e,
Thornburg98, Thornburg99}.%%%
\footnote{%%%
	 See also the work of Bizo\'{n}, Malec, and
	 \'{O}~Murchadha~\protect\cite{Bizon-Malec-OMurchadha-1988}
	 for an interesting analytical study giving necessary
	 and sufficient conditions for apparent horizons to form
	 in non-vacuum spherically symmetric spacetimes.
	 }%%%
{}  For example, the apparent horizons shown in
figure~\ref{fig-stellar-collapse/EH-and-AH}
were obtained using this algorithm.  As another example,
figure~\ref{fig-scalar-field-collapse/AH-and-Theta}
shows $\Theta(r)$ and $h$ at various times in a (different)
spherically symmetric collapse simulation.

%%%%%%%%%%
\epubtkImage{}%%%
{%%%
\begin{figure}[bp]
\begin{center}
\flushleft{\hspace{35mm}Part~(a)}\\[2mm]
\centerline{%%%
  \epsfxsize=75mm
  \epsfbox{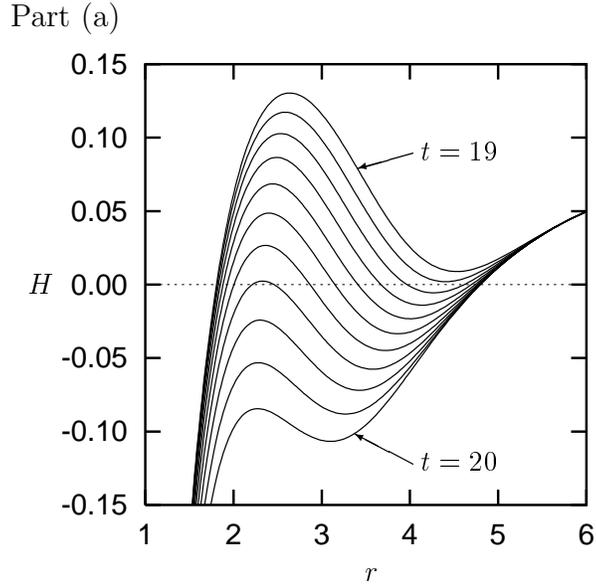}
}
\flushleft{\hspace{35mm}Part~(b)}\\[2mm]
\centerline{%%%
  \epsfxsize=75mm
  \epsfbox{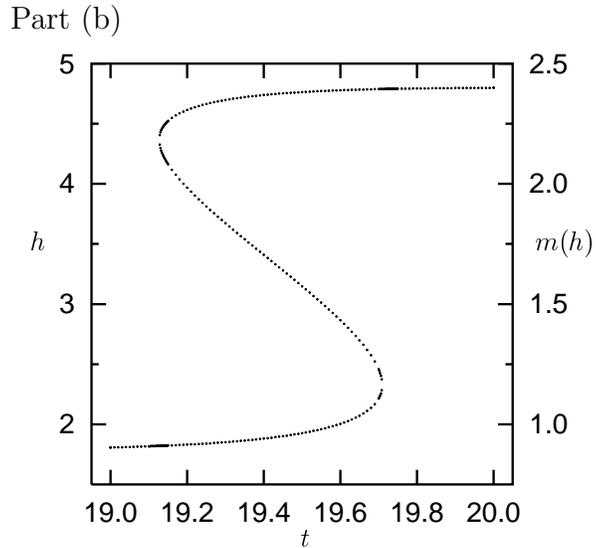}
}
\end{center}
\caption[$\Theta(r)$ and $h$ at various times in a scalar field collapse]
	{
	This figure shows results for a spherically symmetric
	numerical evolution of a black hole accreting a narrow shell
	of scalar field, the 800.pqw1 evolution of
	Thornburg~\cite{Thornburg99}.
	Part~(a) of this figure shows $\Theta(r)$ (here labelled $H$)
	for a set of equally-spaced times between $t{=}19$ and $t{=}20$,
	while part~(b) shows the corresponding horizon radius~$h(t)$
	and the Misner-Sharp~\protect\cite{Misner-Sharp-1964} mass
	$m(h)$ internal to each marginally outer trapped surface (MOTS).
	Notice how two new MOTSs appear when the local minimum in
	$\Theta(r)$ touches the $\Theta{=}0$~line, and two existing
	MOTS disappear when the local maximum in $\Theta(r)$ touches
	the $\Theta{=}0$ line.
	}
\label{fig-scalar-field-collapse/AH-and-Theta}
\end{figure}%%%
}
%%%%%%%%%%

%%%%%%%%%%%%%%%%%%%%%%%%%%%%%%%%%%%%%%%%

\subsection{The Shooting Algorithm in Axisymmetry}
\label{sect-AH/algorithms/shooting}

In an axisymmetric spacetime, the space of angular coordinates
$(\theta,\phi)$ is effectively 1-dimensional, and given the
\Strahlkoerper{} assumption, without further loss of generality
we can write the horizon shape function as $h = h(\theta)$, where
$\theta$ is the single nontrivial angular coordinate.  The apparent
horizon equation~\eqref{eqn-AH/h} then becomes a nonlinear 2-point
boundary-value ODE for the horizon shape function~$h$
(\cite[equation~(1.1)]{Shibata97a})
\begin{equation}
\Theta \equiv
	\Theta(h, \partial_\theta h, \partial_{\theta\theta} h;
	       g_{ij}, \partial_k g_{ij}, K_{ij})
		= 0
						    \label{eqn-AH/h/axisymmetry}
\end{equation}
where $\Theta(h)$ is a nonlinear 2nd~order (ordinary) differential
operator in $h$ as shown.

Taking the angular coordinate $\theta$ to have the usual polar-spherical
topology, local smoothness of the apparent horizon gives the boundary
conditions
\begin{equation}
\partial_\theta h = 0
	\qquad
	\text{at $\theta{=}0$ and $\theta{=}\theta_{\max}$}
						\label{eqn-AH/h/axisymmetry/BCs}
\end{equation}
where $\theta_{\max}$ is $\pi/2$ if there is \defn{bitant} reflection
symmetry about the $z=0$~plane, or $\pi$ otherwise.

As well as the more general algorithms described in the following
subsections, this may be solved by a shooting algorithm:
\begin{enumerate}
\item	Guess the value of $h$ at one endpoint, say
	$h(\theta{=}0) \equiv h_\ast$.
\item	Use this guessed value of $h(\theta{=}0)$ together with the
	boundary condition there~\eqref{eqn-AH/h/axisymmetry/BCs}
	as initial data to integrate (``shoot'') the
	ODE~\eqref{eqn-AH/h/axisymmetry} from $\theta{=}0$ to
	the other endpoint $\theta{=}\theta_{\max}$.%%%
\footnote{%%%
	 I briefly review ODE integration algorithms and
	 codes in appendix~\ref{app-ODEs}.
	 }%%%
\item	If the numerically computed solution satisfies the other
	boundary condition~\eqref{eqn-AH/h/axisymmetry/BCs} at
	$\theta{=}\theta_{\max}$ to within some tolerance, then
	the just-computed $h(\theta)$ describes the (an) apparent
	horizon, and the algorithm is finished.
\item	Otherwise, adjust the guessed value $h(\theta{=}0) \equiv h_\ast$
	and try again.  Because there's only a single parameter ($h_\ast$)
	to be adjusted, this can be done easily and efficiently using
	one of the 1-dimensional zero-finding algorithms discussed in
	 appendix~\ref{app-single-nonlinear-eqn}.
\end{enumerate}

This algorithm is fairly efficient and easy to program.  By trying a
sufficiently wide range of initial guesses~$h_\ast$ this algorithm can
give a high degree of confidence that all apparent horizons have been
located, although this, of course, increases the cost.

Shooting algorithms of this type have been used by many researchers,
for example \cite{Cadez74, Dykema80, Abrahams92b, Bishop82, Bishop84,
Shapiro92a, Abrahams92c, Alcubierre98b}.

%%%%%%%%%%%%%%%%%%%%%%%%%%%%%%%%%%%%%%%%

\subsection{Minimization Algorithms}
\label{sect-AH/algorithms/minimization}

This class of algorithms defines a scalar norm $\| \cdot \|$ on the
expansion~$\Theta$ over the space of possible trial surfaces, typically
the mean-squared norm
\begin{equation}
\bigl\|\Theta\bigr\|
	\equiv \int \Theta^2 \, d\Omega
						       \label{eqn-AH-error-norm}
\end{equation}
where the integral is over all solid angles on a trial surface.

Assuming the horizon surface to be a \Strahlkoerper{} and adopting
the spectral representation~\eqref{eqn-h-expand-in-Ylm}
for the horizon surface, we can view the norm~\eqref{eqn-AH-error-norm}
as being defined on the space of spectral coefficients~$\{a_{\ell m}\}$.

This norm clearly has a global minimum $\|\Theta\| = 0$ for each
solution of the apparent horizon equation~\eqref{eqn-AH/h}.  To find
the apparent horizon we numerically search the spectral-coefficient
space for this (a) minimum, using a general-purpose ``function-minimization''
algorithm (code) such as Powell's algorithm.%%%
\footnote{%%%
	 See, for example,
	 Dennis and Schnabel~\cite{Dennis-Schnabel-1983}
	 or Brent~\cite{Brent2002} for general surveys
	 of general-purposes function-minimization
	 algorithms and codes.
	 }%%%

Evaluating the norm~\eqref{eqn-AH-error-norm} requires a numerical
integration over the horizon surface:  We choose some grid of
$N_\ang$~points on the surface, interpolate the slice geometry
fields ($g_{ij}$, $\partial_k g_{ij}$, and $K_{ij}$) to this grid
(see section~\ref{sect-AH/intro/geometry-interp}), and use numerical
quadrature to approximate the integral.
In practice this must be done for many different trial surface shapes
(see section~\ref{sect-AH/algorithms/minimization/poor-performance}),
so it's important that it be as efficient as possible.
Anninos \etal{}~\cite{Anninos98b} and Baumgarte \etal{}~\cite{Baumgarte96}
discuss various ways to optimize and/or parallelize this calculation.

Unfortunately, minimization algorithms have two serious disadvantages
for apparent-horizon finding: they are susceptible to spurious local
minima, and they're very slow.  I discuss these disadvantages further
in the following two subsections.

%%%%%%%%%%%%%%%%%%%%

\subsubsection{Spurious Local Minima}

While the norm~\eqref{eqn-AH-error-norm} clearly has a single
\emph{global} minimum $\|\Theta\| = 0$ for each marginally outer
trapped surface $\Theta = 0$, it typically also has a large number
of other \emph{local} minima with $\Theta \ne 0$, which are \defn{spurious}
in the sense that they don't correspond (even approximately) to
marginally outer trapped surfaces.%%%
\footnote{%%%
	 There's a simple heuristic argument
	 (see, for example, Press \etal{}~\cite[section~9.6]{Press92})
	 that at least some spurious local minima should be expected:
	 We are trying to solve a system of $N_\ang$ nonlinear
	 equations $\Theta_\i = 0$ (one equation for each
	 horizon-surface grid point).  Equivalently, we are trying
	 to find the intersection of the $N_\ang$ codimension-one
	 hypersurfaces $\Theta_\i = 0$ in surface-shape space.
	 The problem is that anywhere two or more of these
	 hypersurfaces closely approach, but don't actually touch,
	 there is a spurious local minimum in $\|\Theta\|$.
	 }%%%
{}  Unfortunately, general-purpose ``function-minimization'' routines
only locate local minima, and thus may easily converge to one of the
spurious $\Theta \ne 0$ minima.

What this problem means in practice is that a minimization algorithm
needs quite a good (accurate) initial guess for the horizon shape in
order to ensure that the algorithm converges to the true global minimum
$\Theta = 0$ rather than to one of the spurious $\Theta \ne 0$ local
minima.

To view this problem from a different perspective, once the
function-minimization algorithm does converge, we must somehow determine
whether the ``solution'' found is the true one $\Theta = 0$ or a spurious
one $\Theta \ne 0$.  Due to numerical errors in the geometry interpolation
and the evaluation of the integral~\eqref{eqn-AH-error-norm}, $\|\Theta\|$
will almost never evaluate to \emph{exactly} zero; rather, we must
set a tolerance level for how large $\|\Theta\|$ may be.  Unfortunately,
in practice it's hard to choose this tolerance: if it's too small, the
genuine solution may be falsely rejected, while if it's too large, we
may accept a spurious solution (which may be very different from any
of the true solutions).

Anninos \etal{}~\cite{Anninos98b} and Baumgarte \etal{}~\cite{Baumgarte96}
suggest screening out spurious solutions by repeating the algorithm
with varying resolutions of the horizon-surface grid, and checking
that $\|\Theta\|$ shows the proper convergence towards zero.  This
seems like a good strategy, but it's tricky to automate and, again, it
may be difficult to choose the necessary error tolerances in advance.

%%%%%%%%%%%%%%%%%%%%

\subsubsection{Performance}
\label{sect-AH/algorithms/minimization/poor-performance}

For convenience of exposition, suppose the spectral
representation~\eqref{eqn-h-expand-in-Ylm}
of the horizon-shape function~$h$ uses spherical harmonics $Y_{\ell m}$.
(Symmetric trace-free tensors or other basis sets don't change the
argument in any important way.)  Then if we keep harmonics up to some
maximum degree $\ell_{\max}$, the number of coefficients is then
$N_\coeff = (\ell_{\max}{+}1)^2$.  $\ell_{\max}$ is set by the desired
accuracy (angular resolution) of the algorithm, and is typically
on the order of~$6$ to~$12$.

To find a minimum in an $N_\coeff$-dimensional space (here the
space of surface-shape coefficients $\{a_{\ell m}\}$), a general-purpose
function-minimization algorithm typically needs on the order of
$5 N_\coeff^2$ to $10 N_\coeff^2$ iterations.%%%
\footnote{%%%
	 % local latex macros for this footnote
	 \def\a{\mathsf{a}}%%%
	 \def\B{\mathsf{B}}%%%
	 \def\x{\mathsf{x}}%%%
	 \def\v{\mathsf{v}}%%%
	 A simple counting argument suffices to show that any
	 general-purpose function-minimization algorithm in
	 $n$~dimensions must involve at least $O(n^2)$ function
	 evaluations (see, for example,
	 Press \etal{}~\protect\cite[section~10.6]{Press92}):
	 Suppose the function to be minimized is $f: \Re^n \to \Re$,
	 and suppose $f$ has a local minimum near some point
	 $\x_0 \in \Re^n$.  Taylor-expanding $f$ in a neighborhood
	 of $x_0$ gives
	 $
	 f(\x) = f(\x_0)
		 + \a^T (\x{-}\x_0)
		 + (\x{-}\x_0)^T \B (\x{-}\x_0)
		 + O(\|\x{-}\x_0\|^3)
	 $,
	 where $\a \in \Re^n$, $\B \in \Re^{n{\times}n}$ is symmetric,
	 and $\v^T$ denotes the transpose of the column vector $\v \in \Re^n$.

	 Neglecting the higher order terms (\ie{}, approximating $f$
	 as a quadratic form in $\x$ in a neighborhood of $x_0$),
	 and ignoring $f(\x_0)$ (which doesn't affect the position of
	 the minimum), there are a total of $N = n + \frac{1}{2}n(n+1)$
	 coefficients in this expression.  Changing any of these coefficients
	 may change the position of the minimum, and at each function
	 evaluation the algorithm ``learns'' only a single number
	 (the value of $f$ at the selected evaluation point), so
	 the algorithm must make at least $N = O(n^2)$ function
	 evaluations to (implicitly) determine all the coefficients.

	 Actual functions aren't exact quadratic forms, so in practice
	 there are additional $O(1)$ multiplicative factors in the
	 number of function evaluations.  Minimization algorithms
	 may also make additional performance and/or space-versus-time
	 trade-offs to improve numerical robustness or to avoid
	 explicitly manipulating $n \,{\times}\, n$ Jacobian matrices.
	 }%%%
{}  Thus the number of iterations grows as $\ell_{\max}^4$.

Each iteration requires an evaluation of the norm~\eqref{eqn-AH-error-norm}
for some trial set of surface-shape coefficients $\{a_{\ell m}\}$,
which requires $O(N_\coeff) = O(\ell_{\max}^2)$ work to compute the
surface positions, together with $O(N_\ang)$ work to interpolate the
geometry fields to the surface points and compute the numerical
quadrature of the integral~\eqref{eqn-AH-error-norm}.

The result is that minimization horizon-finders tend to be quite slow,
particularly if high accuracy is required (large $\ell_{\max}$ and $N_\ang$).
The one exception is in axisymmetry, where only spherical harmonics
$Y_{\ell m}$ with $m{=}0$~need be considered.  In this case minimization
algorithms are much faster, though probably still slower than shooting
or elliptic-PDE algorithms.

%%%%%%%%%%%%%%%%%%%%

\subsubsection{Summary of Minimization Algorithms/Codes}
\label{sect-AH/algorithms/minimization/summary}

Minimization algorithms are fairly easy to program and have been used
by many researchers, for example \cite{Brill63, Eppley77, Libson95a,
Anninos98b, Baumgarte96, Alcubierre98b}.  However, they're susceptible
to spurious local minima, have relatively poor accuracy, and tend to
be quite slow.  I believe that the other algorithms discussed in the
following sections are generally preferable.

Alcubierre's apparent-horizon finder
\program{AHFinder}~\cite{Alcubierre98b} includes a minimization
algorithm based on the work of Anninos \etal{}~\cite{Anninos98b}.%%%
\footnote{%%%
	 \program{AHFinder} also includes a ``fast flow''
	 algorithm (section~\ref{sect-AH/algorithms/flow}).
	 }%%%
{}  It's implemented as a module (``thorn'') in the
\href{http://www.cactuscode.org}{\program{Cactus}} computational
toolkit, and is freely available by anonymous~CVS (it's part of the
\program{CactusEinstein} set of thorns included with the standard
\href{http://www.cactuscode.org}{\program{Cactus}} distribution).
It has been used by a number of research groups.

%%%%%%%%%%%%%%%%%%%%%%%%%%%%%%%%%%%%%%%%

\subsection{Nakamura \etal{}'s Spectral Integral-Iteration Algorithm}
\label{sect-AH/algorithms/Nakamura-etal}

Nakamura, Kojima, and Oohara~\cite{Nakamura84} developed a
functional-iteration algorithm for solving the apparent horizon
equation~\eqref{eqn-AH/h}.

This algorithm begins by choosing the usual polar-spherical topology
for the angular coordinates $(\theta,\phi)$, and rewriting the
apparent horizon equation~\eqref{eqn-AH/h} in the form
\begin{equation}
L \equiv
	\partial_{\theta\theta} h
	+ \frac{\partial_\theta h}{\tan \theta}
	+ \frac{\partial_{\phi\phi} h}{\sin^2 \theta}
		= F(\partial_{\theta\phi} h, \partial_{\phi\phi} h,
		    \partial_\theta h, \partial_\phi h;
		   g_{ij}, K_{ij}, \Gamma^k_{ij})
					     \label{eqn-AH/h-with-Laplacian-LHS}
\end{equation}
where $F$ is a complicated nonlinear algebraic function of the arguments
shown, which remains regular even at $\theta{=}0$ and $\theta{=}\pi$,
and where for future use we define $L$ to be the left hand side
of~\eqref{eqn-AH/h-with-Laplacian-LHS}.

Next we expand $h$ in spherical harmonics~\eqref{eqn-h-expand-in-Ylm}.
Because the left hand side $L$ of~\eqref{eqn-AH/h-with-Laplacian-LHS}
is just the flat-space angular Laplacian of $h$, which has the
$Y_{\ell m}$ as orthogonal eigenfunctions, multiplying both sides
of~\eqref{eqn-AH/h-with-Laplacian-LHS} by $Y_{\ell m}^\ast$
(the complex conjugate of $Y_{\ell m}$) and integrating over all
solid angles gives
\begin{equation}
a_{\ell m} = - \frac{1}{\ell(\ell{+}1)} \int Y_{\ell m}^\ast F \, d\Omega
						   \label{eqn-Nakamura-etal/alm}
\end{equation}
for each $\ell$ and $m$ except $\ell=m=0$.

Based on this, Nakamura \etal{}~\cite{Nakamura84} propose the following
functional-iteration algorithm for solving~\eqref{eqn-AH/h-with-Laplacian-LHS}:
\begin{enumerate}
\item	Start with some (initial-guess) set of horizon-shape
	coefficients $\{a_{\ell m}\}$.  These determine a surface
	shape via~\eqref{eqn-h-expand-in-Ylm}.
\item	Interpolate the geometry variables to this surface shape
	(see section~\ref{sect-AH/intro/geometry-interp}).
\item	For each $\ell$ and $m$ except $\ell = m = 0$, evaluate the
	integral~\eqref{eqn-Nakamura-etal/alm} by numerical quadrature
	to obtain a next-iteration coefficient $a_{\ell m}$.
\item	Determine a next-iteration coefficient $a_{00}$ by
	numerically solving (finding a root of) the equation
	\begin{equation}
	\int Y^\ast_{00} F \, d\Omega = 0
						   \label{eqn-Nakamura-etal/a00}
	\end{equation}
	This can be done using any of the 1-dimensional zero-finding
	algorithms discussed in appendix~\ref{app-single-nonlinear-eqn}.
\item	Iterate until all the coefficients $\{a_{\ell m}\}$ converge.
\end{enumerate}

Gundlach~\cite{Gundlach98a} observed that the subtraction and inversion
of the flat-space angular Laplacian operator in this algorithm is actually
a standard technique for solving nonlinear elliptic PDEs by spectral
methods.  I discuss this observation and its implications further
in section~\ref{sect-AH/algorithms/flow/fast-flow}.

Nakamura \etal{}~\cite{Nakamura84} report that their algorithm works
well, but Nakao~\cite{Nakao97:problems-of-Nakamura-etal-AH-finder}
has argued that it tends to become inefficient (and possibly inaccurate)
for large $\ell$ (high angular resolution) because the $Y_{\ell m}$
fail to be numerically orthogonal due to the finite resolution of the
numerical grid.  I know of no other published work addressing Nakao's
criticism.

Kemball and Bishop~\cite{Kemball91a} investigated the behavior of
Nakamura \etal{}'s algorithm, and found that its (only) major weakness
seems to be that the $a_{00}$-update equation~\eqref{eqn-Nakamura-etal/a00}
``may have multiple roots or minima even in the presence of a
single marginally outer trapped surface, and all should be tried
for convergence''.

Kemball and Bishop~\cite{Kemball91a} suggested and tested several
modifications to improve the algorithm's convergence behavior.  They
verified that (either in its original form or with their modifications)
the algorithm's convergence speed (number of iterations to a given error
level) is roughly independent of the degree $\ell_{\max}$ of
spherical-harmonic expansion used.  They also give an analysis that
the algorithm's cost is $O(\ell_{\max}^4)$, and its accuracy
$\varepsilon = O(1/\ell_{\max})$, \ie{} the cost is $O(1/\varepsilon^4)$.

Despite what appears to be fairly good numerical behavior and
reasonable ease of implementation, this algorithm has not been widely
used apart from later work by its original developers
(see, for example, \cite{Oohara85, Oohara86}).

%%%%%%%%%%%%%%%%%%%%%%%%%%%%%%%%%%%%%%%%

\subsection{Elliptic-PDE Algorithms}
\label{sect-AH/algorithms/elliptic-PDE}

The basic concept of elliptic-PDE algorithms is simple: we view the
apparent horizon equation~\eqref{eqn-AH/h} as a nonlinear elliptic PDE
for the horizon shape function $h$ on the angular-coordinate space
and solve this equation by standard finite-differencing techniques,%%%
\footnote{%%%
	 In theory this equation could also be solved by a
	 spectral method on $S^2$, using spectral differentiation
	 to evaluate the angular derivatives.  (See the references
	 cited in footnote~\ref{footnote-spectral-methods}
	 on page~\pageref{footnote-spectral-methods}
	 for further discussion of spectral methods.)
	 This should yield a highly efficient apparent-horizon
	 finder.  However, I know of no published work taking
	 this approach.
	 }%%%
{} generally using Newton's method to solve the resulting set of
nonlinear algebraic (finite-difference) equations.  Algorithms of
this type have been widely used both in axisymmetry and in fully
generic slices.

%%%%%%%%%%%%%%%%%%%%

\subsubsection{Angular Coordinates, Grid, and Boundary Conditions}
\label{sect-AH/algorithms/elliptic-PDE/angular-coords+grid+BCs}

In more detail, elliptic-PDE algorithms assume that the horizon is
a \Strahlkoerper{} about some local coordinate origin, and choose
an angular coordinate system and a finite-difference grid of
$N_\ang$~points on $S^2$ in the manner discussed in
section~\ref{sect-intro/2-surface-parameterizations/Strahlkoerper}.

The most common choices are the usual polar-spherical coordinates
$(\theta,\phi)$ and a uniform ``latitude/longitude'' grid in these
coordinates.  Since these coordinates are ``unwrapped'' relative to
the actual $S^2$ trial-horizon-surface topology, the horizon shape
function~$h$ satisfies periodic boundary conditions across the artificial
grid boundary at $\phi = 0$ and $\phi = 2\pi$.  The north and south
poles $\theta = 0$ and $\theta = \pi$ are trickier, but
Huq \etal{}~\cite{Huq96, Huq00}, Shibata and Ury\={u}~\cite{Shibata-Uryu-2000b},
and Schnetter~\cite{Schnetter02a, Schnetter03a} all describe suitable
``reaching across the pole'' boundary conditions for these
artificial grid boundaries.

Alternatively, Thornburg~\cite{Thornburg2003:AH-finding} avoids
the $z$~axis coordinate singularity of polar-spherical coordinates
by using an ``inflated-cube'' system of 6~angular patches to cover
$S^2$.  Here each patch's nominal grid is surrounded by a
\defn{ghost zone} of additional grid points where $h$ is determined
by interpolation from the neighboring patches.  The interpatch
interpolation thus serves to tie the patches together, enforcing
the continuity and differentiability of~$h$ across patch boundaries.
Thornburg reports that this scheme works well but was quite complicated
to program.

Overall, the latitude/longitude grid seems to be the superior choice:
it works well, is simple to program, and eases interoperation with
other software.

%%%%%%%%%%%%%%%%%%%%

\subsubsection{Evaluating the Expansion $\Theta$}
\label{sect-AH/algorithms/elliptic-PDE/evaluating-Theta}

The next step in the algorithm is to evaluate the expansion~$\Theta$
given by~\eqref{eqn-AH/h} on the angular grid given a trial horizon
surface shape function~$h$ on this same grid~\eqref{eqn-h-on-angular-grid}.

Most researchers compute~$\Theta$ via 2-dimensional angular finite
differencing of~\eqref{eqn-AH/h} on the trial horizon surface.
2nd~order angular finite differencing is most common, but
Thornburg~\cite{Thornburg2003:AH-finding} uses 4th~order angular
finite differencing for increased accuracy.

With a $(\theta,\phi)$ latitude/longitude grid the
$\Theta(h, \partial_u h, \partial_{uv} h)$ function
in~\eqref{eqn-AH/h} is singular on the $z$~axis (at the north and
south poles $\theta = 0$ and $\theta = \pi$), but can be regularized
by applying L'Hopital's rule.  Schnetter~\cite{Schnetter02a, Schnetter03a}
observes that using a \emph{Cartesian} basis for all tensors greatly
aids in this regularization.

Huq \etal{}~\cite{Huq96, Huq00} choose, instead, to use a completely
different computation technique for~$\Theta$, based on \emph{3-dimensional}
\emph{Cartesian} finite differencing:
\begin{enumerate}
\item	\label{Huq-etal/item-F-defn}
	They observe that the scalar field~$F$ defined
	by~\eqref{eqn-F=r-h(theta,phi)} can be evaluated at any
	(3-dimensional) position in the slice by computing the
	corresponding $(r,\theta,\phi)$ using the usual flat-space
	formulas, then interpolating~$h$ in the 2-dimensional
	$(\theta,\phi)$ surface grid.
\item	Rewrite the apparent horizon condition~\eqref{eqn-AH/s^i}
	in terms of $F$ and it's (3-dimensional) \emph{Cartesian}
	derivatives,
	\begin{equation}
	\Theta \equiv
		\Theta(F, \partial_i F, \partial_{ij} F;
		       g_{ij}, \partial_k g_{ij}, K_{ij})
			= 0
								\label{eqn-AH/F}
	\end{equation}
	Huq \etal{}~\cite{Huq96, Huq00} give the
	$\Theta(F, \partial_i F, \partial_{ij} F)$
	function explicitly.
\item	For each (latitude/longitude) grid point on the trial
	horizon surface, define a $3{\times}3{\times}3$-point
	\emph{local Cartesian grid} centered at that point.
	The spacing of this grid should be such as to allow
	accurate finite differencing, \ie{} in practice it
	should probably be roughly comparable to that of the
	underlying numerical-relativity simulation's grid.
\item	Evaluate $F$ on the local Cartesian grid as described in
	step~\ref{Huq-etal/item-F-defn} above.
\item	Evaluate the Cartesian derivatives in~\eqref{eqn-AH/F}
	by centered 2nd~order Cartesian finite differencing of
	the $F$~values on the local Cartesian grid.
\end{enumerate}

Comparing the different ways of evaluating~$\Theta$, 2-dimensional
angular finite differencing of~\eqref{eqn-AH/h} seems to me to be both
simpler (easier to program) and likely more efficient than 3-dimensional
Cartesian finite differencing of~\eqref{eqn-AH/F}.

%%%%%%%%%%%%%%%%%%%%

\subsubsection{Solving the Nonlinear Elliptic PDE}
\label{sect-AH/algorithms/elliptic-PDE/solving-nonlinear-elliptic-PDE}

A variety of algorithms are possible for actually solving the nonlinear
elliptic PDE~\eqref{eqn-AH/h} (or~\eqref{eqn-AH/F} for the
Huq \etal{}~\cite{Huq96, Huq00} horizon finder).

The most common choice is to use some variant of Newton's method.
That is, finite differencing~\eqref{eqn-AH/h} or~\eqref{eqn-AH/F}
(as appropriate) gives a system of $N_\ang$ nonlinear algebraic
equations for the horizon shape function~$h$ at the $N_\ang$
angular grid points; these can be solved by Newton's method in
$N_\ang$~dimensions.  (As explained by
Thornburg~\cite[section~VIII.C]{Thornburg95}, this is
usually equivalent to applying the Newton-Kantorovich algorithm
(\cite[appendix~C]{Boyd00}) to the original nonlinear elliptic
PDE~\eqref{eqn-AH/h} or~\eqref{eqn-AH/F}.)

Newton's method converges very quickly once the trial horizon surface
is sufficiently close to a solution (a marginally outer trapped surface).
However, for a less accurate initial guess, Newton's method may converge
very slowly or even fail to converge at all.  There's no usable way of
determining \textit{a priori} just how large the radius of convergence
of the iteration will be, but in practice $\frac{1}{4}$~to $\frac{1}{3}$
of the horizon radius is often a reasonable estimate.%%%
\footnote{%%%
	 Thornburg~\protect\cite{Thornburg95} used a
	 Monte-Carlo survey of horizon-shape perturbations
	 to quantify the radius of convergence of Newton's
	 method for apparent-horizon finding.  He found that
	 if strong high-spatial-frequency perturbations are
	 present in the slice's geometry then the radius
	 of convergence may be very small.  Fortunately,
	 this problem rarely occurs in practice.
	 }%%%

Thornburg~\cite{Thornburg95} described the use of various
\defn{line search} modifications to Newton's method to improve its
radius and robustness of convergence, and reported that even fairly
simple modifications of this sort roughly doubled the radius of
convergence.

Schnetter~\cite{Schnetter02a, Schnetter03a} used the \program{PETSc}
general-purpose elliptic-solver library~\cite{petsc-home-page,
petsc-manual, petsc-efficient} to solve the equations.  This offers
a wide variety of Newton-like algorithms already implemented in a
highly optimized form.

Rather than Newton's method or one of its variants,
Shibata \etal{}~\cite{Shibata97a, Shibata-Uryu-2000b} use a functional
iteration algorithm directly on the nonlinear elliptic PDE~\eqref{eqn-AH/h}.
This seems likely to be less efficient than Newton's method but
avoids having to compute and manipulate the Jacobian matrix.

%%%%%%%%%%%%%%%%%%%%

\subsubsection{The Jacobian Matrix}
\label{sect-AH/algorithms/elliptic-PDE/Jacobian-matrix}

Newton's method, and all its variants, require an explicit computation
of the Jacobian matrix
\begin{equation}
\J_{\i\j} = \frac{\partial \Theta_\i}{\partial h_\j}
							 \label{eqn-AH-Jacobian}
\end{equation}
where the indices $\i$ and $\j$ label angular grid points on the
horizon surface (or equivalently on $S^2$).

Notice that $\J$ includes contributions both from the direct dependence
of $\Theta$ on $h$, $\partial_u h$, and $\partial_{uv} h$, and also
from the indirect dependence of $\Theta$ on $h$ through the position-dependence
of the geometry variables $g_{ij}$, $\partial_k g_{ij}$, and $K_{ij}$
(since $\Theta$ depends on the geometry variables
\emph{at the horizon surface position}, and this position
is determined by $h$).  Thornburg~\cite{Thornburg95} discusses this
indirect dependence in detail.

There are two basic ways to compute the Jacobian matrix.

%%%%%%%%%%

\paragraph{Numerical Perturbation:}

The simplest way to determine the Jacobian matrix is by
\defn{numerical perturbation}, where for each horizon-surface grid
point~$\j$, $h$ is perturbed by some (small) amount $\varepsilon$ at the
$\j$\,th grid point (that is, $h_\i \to h_\i + \varepsilon \delta_{\i\j}$),
and the expansion~$\Theta$ is recomputed.%%%
\footnote{%%%
	 A very important optimization here is that
	 $\Theta$ only needs to be recomputed within
	 the finite difference domain of dependence
	 of the $\j$\,th grid point.
	 }%%%
{}  The $\j$\,th column of the Jacobian matrix~\eqref{eqn-AH-Jacobian}
is then estimated as
\begin{equation}
\J_{\i\j} \approx
	\frac{\Theta_\i(h + \varepsilon \delta_{\i\j}) - \Theta_\i(h)}
	     {\varepsilon}
				     \label{eqn-Jacobian/numerical-perturbation}
\end{equation}
Curtis and Reid~\cite{Curtis-Reid-1974} and
Stoer and Bulirsch~\cite[section~5.4.3]{Stoer-Bulirsch-1980}
discuss the optimum choice of $\varepsilon$ in this algorithm.%%%
\footnote{%%%
	 Because of the one-sided finite differencing, the
	 approximation~\eqref{eqn-Jacobian/numerical-perturbation}
	 is only $O(\varepsilon)$ accurate.  However, in
	 practice this doesn't seriously impair the convergence
	 of a horizon finder, and the extra cost of a
	 centered--finite-differencing $O(\varepsilon^2)$
	 approximation isn't warranted.
	 }%%%

This algorithm is easy to program but somewhat inefficient.
It's used by a number of researchers, including
Schnetter~\cite{Schnetter02a, Schnetter03a}
and Huq \etal{}~\cite{Huq96, Huq00}.

%%%%%%%%%%

\paragraph{Symbolic Differentiation:}

A more efficient, although somewhat more complicated, way to
determine the Jacobian matrix is the \defn{symbolic differentiation}
algorithm described by Thornburg~\cite{Thornburg95}, and also used by
Pasch~\cite{Pasch-1997:level-sets-and-curvature-flow},
Shibata \etal{}~\cite{Shibata97a, Shibata-Uryu-2000b},
and Thornburg~\cite{Thornburg2003:AH-finding}.  Here the internal
structure of the finite differenced $\Theta(h)$ function is used
to directly determine the Jacobian matrix elements.

This algorithm is best illustrated by an example which is simpler than
the full apparent horizon equation: Suppose we discretize the left hand
side $L$ of the apparent horizon equation~\eqref{eqn-AH/h-with-Laplacian-LHS}
with centered 2nd~order finite differences in $\theta$ and $\phi$.
Then neglecting finite-differencing trunation errors, and temporarily
adopting the usual notation for 2-dimensional grid functions,
$h_{i,j} \equiv h(\theta{=}\theta_i, \phi{=}\phi_j)$, $L$ is given by
\begin{eqnarray}
L_{i,j}
	& = &	\frac{h_{i-1,j} - 2h_{i,j} + h_{i+1,j}}{(\Delta\theta)^2}
		+ \frac{1}{\tan\theta}
		  \frac{h_{i+1,j} - h_{i-1,j}}{2 \, \Delta\theta}
							\nonumber	\\
	&   &	\hbox{}
		+ \frac{1}{\sin^2\theta}
		  \frac{h_{i,j-1} - 2h_{i,j} + h_{i,j+1}}{(\Delta\phi)^2}
									%%%\\
\end{eqnarray}
The Jacobian of $L$ is thus given by
\begin{equation}
\frac{\partial L_{(i,j)}}{\partial h_{(k,\ell)}}
	= \begin{cases}
	  \displaystyle
	  \frac{1}{(\Delta\theta)^2}
	  \pm \frac{1}{2\, \tan\theta \, \Delta\theta}
					& \text{if $(k,\ell)=(i{\pm}1,j)$}
									\\[2ex]
	  \displaystyle
	  \frac{1}{\sin^2\theta \, (\Delta\phi)^2}
					& \text{if $(k,\ell)=(i,j{\pm}1)$}
									\\[2ex]
	  \displaystyle
	  - \frac{2}{(\Delta\theta)^2}
	  - \frac{2}{\sin^2\theta\, (\Delta\phi)^2}
					& \text{if $(k,\ell)=(i,j)$}
									\\[2ex]
	  \displaystyle
	  0
						& \text{otherwise}
									%%%\\
	  \end{cases}
\end{equation}
Thornburg~\cite{Thornburg95} describes how to generalize this to
nonlinear differential operators without having to explicitly
manipulate the nonlinear finite difference equations.

%%%%%%%%%%%%%%%%%%%%

\subsubsection{Solving the Linear Equations}
\label{sect-AH/algorithms/elliptic-PDE/solving-linear-eqns}

All the algorithms described in
section~\ref{sect-AH/algorithms/elliptic-PDE/solving-nonlinear-elliptic-PDE}
for treating nonlinear elliptic PDEs require solving a sequence of
linear systems of $N_\ang$~equations in $N_\ang$~unknowns.
$N_\ang$ is typically on the order of a few thousand, and the
Jacobian matrices in question are sparse due to the locality of
the angular finite differencing
(see section~\ref{sect-AH/algorithms/elliptic-PDE/Jacobian-matrix}).
Thus, for reasonable efficiency, it's essential to use linear solvers
that exploit this sparsity.  Unfortunately, many such algorithms/codes
only handle symmetric positive-definite matrices while, due to the
angular boundary conditions%%%
\footnote{%%%
	 Or the interpatch interpolation conditions
	 in Thornburg's multiple-grid-patch
	 scheme~\protect\cite{Thornburg2003:AH-finding}.
	 }%%%
{} (see section~\ref{sect-AH/algorithms/elliptic-PDE/angular-coords+grid+BCs}),
the Jacobian matrices that arise in apparent-horizon finding are
generally neither of these.

The numerical solution of large sparse linear systems is a whole
subfield of numerical analysis.  See, for example,
Duff, Erisman, and Reid~\cite{Duff-Erisman-Reid-1986} and
Saad~\cite{Saad-2003:iterative-methods-2nd-ed} for extensive discussions.%%%
\footnote{%%%
\label{footnote-multigrid-refs}%%%
	 Multigrid algorithms are also important here;
	 these exploit the geometric structure of the
	 underlying elliptic PDE.  See
	 Briggs, Henson, and
	 McCormick~\protect\cite{Briggs-Henson-McCormick-2000:multigrid}
	 and Trottenberg, Oosterlee, and
 Sch\"{u}ller~\protect\cite{Trottenberg-Oosterlee-Schueller-2001:multigrid}
	 for general introductions to multigrid algorithms.
	 }%%%
{}  In practice, a numerical relativist is unlikely to write her own
linear solver but, rather, will use an existing subroutine (library).

Kershaw's~\cite{Kershaw1978:ilucg} \program{ILUCG} iterative solver
is often used; this is only moderately efficient, but is quite easy
to program.%%%
\footnote{%%%
	 Madderom's Fortran
	 subroutine~\program{DILUCG}~\protect\cite{Madderom-1984:ILUCG}
	 has been used by a number of numerical relativists
	 for both this and other purposes.
	 }%%%
{}  Schnetter~\cite{Schnetter02a, Schnetter03a} reports good results
with an ILU-preconditioned GMRES solver from the \program{PETSc} library.
Thornburg~\cite{Thornburg2003:AH-finding} experimented with both an
\program{ILUCG} solver and a direct sparse $\sf LU$~decomposition solver
(Davis's \program{UMFPACK} library~\cite{Davis-Duff-1997-UMFPACK,
Davis-Duff-1999-UMFPACK, Davis-2002a-UMFPACK-report,
Davis-2002b-UMFPACK-report}), and found each to be more efficient
in some situations; overall, he found the \program{UMFPACK} solver
to be the best choice.

%%%%%%%%%%%%%%%%%%%%

\subsubsection{Sample Results}
\label{sect-AH/algorithms/elliptic-PDE/sample-results}

As an example of the results obtained with this type of apparent-horizon
finder, figure~\ref{fig-2BH-headon-collision/AH-perspective-view} shows the
numerically-computed apparent horizons (actually, marginally outer
trapped surfaces) at two times in a head-on binary black hole collision.
(The physical system being simulated here is very similar to that
simulated by Matzner \etal{}~\protect\cite{Matzner95a}, a view of
whose event horizon is shown in
figure~\ref{fig-2BH-headon-collision/EH-spacetime-view}.)

%%%%%%%%%%
\epubtkImage{}%%%
{%%%
\begin{figure}[bp]
\begin{center}
\flushleft{%%%
  \hspace{7mm}$t=3.93m_\text{ADM}$\hspace{50.6mm}$t=6.28m_\text{ADM}$%%%
}\\
\vspace{-40mm}
\centerline{%%%
  \epsfxsize=100mm
  \epsfbox{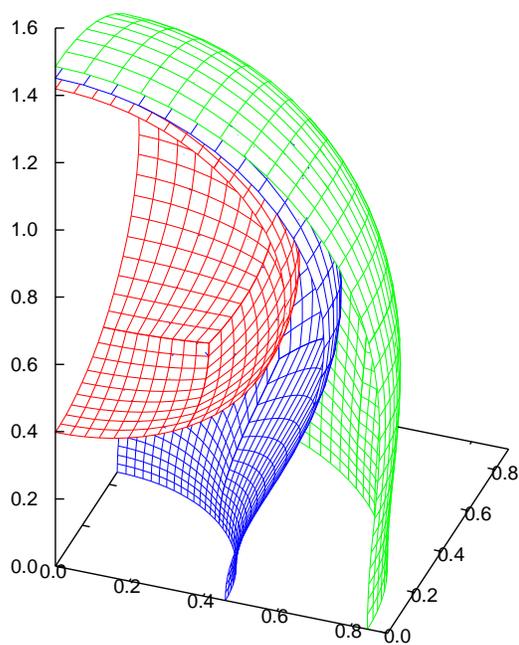}%%%
  \hspace{-25mm}
  \epsfxsize=100mm
  \epsfbox{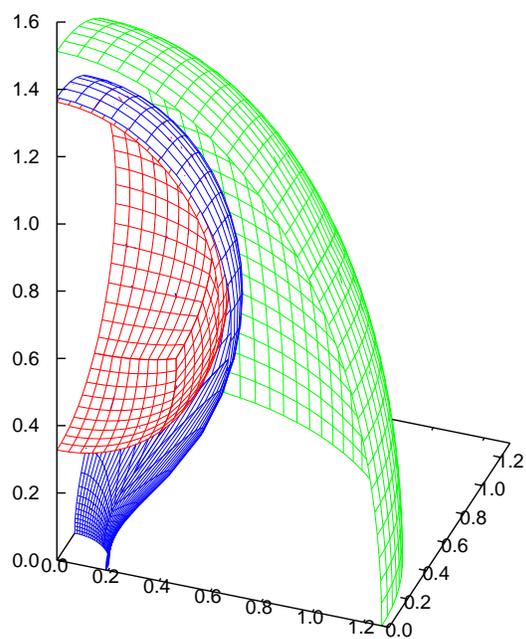}%%%
}
\vspace{-20mm}
\end{center}
\caption[Numerically-Computed Apparent Horizons in a Binary Black Hole Collision]
	{
	This figure shows the numerically-computed apparent horizons
	(actually marginally outer trapped surfaces) at two times in a
	head-on binary black hole collision.
	Figure reprinted with permission from
	\href{http://stacks.iop.org/0264-9381/21/743}%%%
	     {Thornburg,
	      \textit{Classical and Quantum Gravity} \textbf{21}, 743--766}.
	Copyright 2004 by IOP Publishing Ltd.
	}
\label{fig-2BH-headon-collision/AH-perspective-view}
\end{figure}%%%
}
%%%%%%%%%%

As another example, figure~\ref{fig-2BH-spiraling-collision/AH-masses}
shows the time dependence of the irreducible masses of apparent horizons
found in a (spiraling) binary black hole collision, simulated at several
different grid resolutions, as found by both \program{AHFinderDirect}
and another \href{http://www.cactuscode.org}{\program{Cactus}}
apparent-horizon finder, \program{AHFinder}.%%%
\footnote{%%%
	 \program{AHFinder} incorporates both a minimization
	 algorithm (section~\ref{sect-AH/algorithms/minimization})
	 and a fast-flow algorithm
	 (section~\ref{sect-AH/algorithms/flow/fast-flow});
	 these tests used the fast-flow algorithm.
	 }%%%
{}  For this evolution, the two apparent-horizon finders give irreducible
masses which agree to within about $2\%$ for the individual horizons
and $0.5\%$ for the common horizon.

%%%%%%%%%%
\epubtkImage{fig/2BH-spiraling-collision--AH-masses/AH-masses.png}%%%
{%%%
\begin{figure}[bp]
\begin{center}
\centerline{%%%
  \epsfxsize=125mm
  \epsfbox{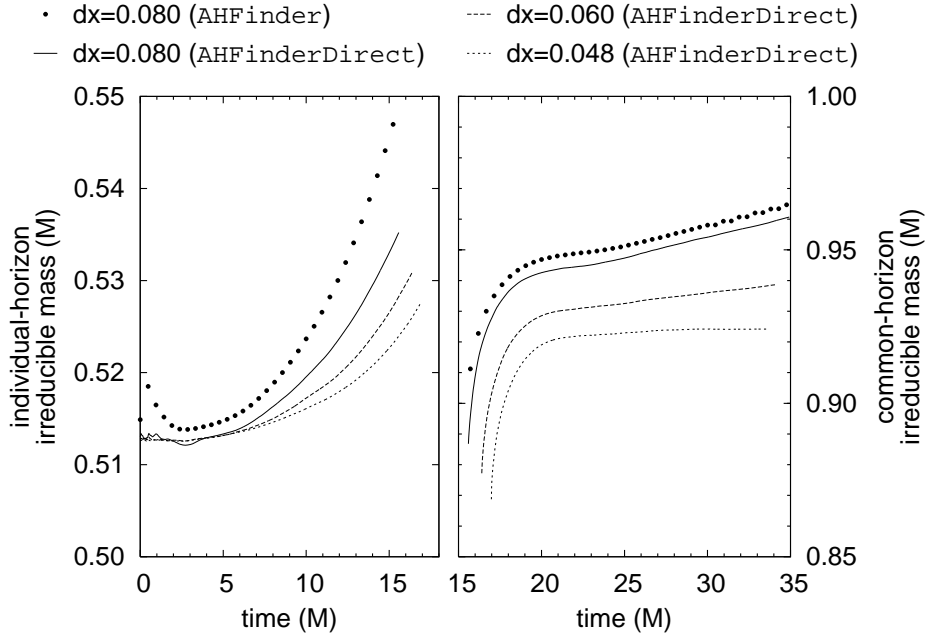}
}
\end{center}
\caption[Apparent Horizon Masses Computed by Two Different Apparent-Horizon Finders in a Binary Black Hole Collision]
	{
	This figure shows the irreducible masses ($\sqrt{\text{area}}/16\pi$)
	of individual and common apparent horizons in a binary
	black hole collision, as calculated by
	two different apparent-horizon finders in the
	\href{http://www.cactuscode.org}{\program{Cactus}} toolkit,
	\program{AHFinder} and \program{AHFinderDirect}.
	(\program{AHFinderDirect} was also run in simulations
	at several different resolutions.)
	Notice that when both apparent-horizon finders are
	run in the same simulation (resolution $dx{=}0.080$),
	there are only small differences between their results.
	Figure reprinted with permission from
	\href{http://link.aps.org/abstract/PRD/v72/e044004}%%%
	     {Alcubierre \etal,
	      \textit{Physical Review~D} \textbf{72}, 044004 (2005)}.
	Copyright 2005 by the American Physical Society.
	}
\label{fig-2BH-spiraling-collision/AH-masses}
\end{figure}%%%
}
%%%%%%%%%%

As a final example,
figure~\ref{fig-rotating-NS-collapse/EH-and-AH} shows the
numerically-computed event and apparent horizons in the collapse
of a rapidly rotating neutron star to a Kerr black hole.  (The event
horizons were computed using the \program{EHFinder} code described in
section~\ref{sect-EH/algorithms/null-surfaces-backwards/level-set}.)

%%%%%%%%%%%%%%%%%%%%

\subsubsection{Summary of Elliptic-PDE Algorithms/Codes}
\label{sect-AH/algorithms/elliptic-PDE/summary}

Elliptic-PDE apparent-horizon finders have been developed by
many researchers, including Eardley~\cite{Eardley75},
Cook~\cite{Cook90, Cook90a, Cook92}, and Thornburg~\cite{Thornburg95}
in axisymmetry, and Shibata \etal{}~\cite{Shibata97a, Shibata-Uryu-2000b},
Huq \etal{}~\cite{Huq96, Huq00}, Schnetter~\cite{Schnetter02a, Schnetter03a},
and Thornburg~\cite{Thornburg2003:AH-finding} in fully generic slices.

Elliptic-PDE algorithms are (or can be implemented to be) generally
the fastest horizon-finding algorithms.  For example,
Thornburg~\cite{Thornburg2003:AH-finding} reports that the production
version of his \program{AHFinderDirect} elliptic-PDE apparent-horizon
finder, when run at each time step of a binary black hole evolution,
averaged 1.7~seconds per time step, as compared with 61~seconds for
an alternate ``fast-flow'' apparent-horizon finder \program{AHFinder}
(discussed in more detail in section~\ref{sect-AH/algorithms/flow}).
However, achieving maximum performance comes at some cost in implementation
effort (\eg{}~the ``symbolic differentiation'' Jacobian computation
discussed in section~\ref{sect-AH/algorithms/elliptic-PDE/Jacobian-matrix}).

Elliptic-PDE algorithms are probably somewhat more robust in their
convergence (\ie{} they have a slightly larger radius of convergence)
than other types of local algorithms, particularly if the ``line search''
modifications of Newton's method described by Thornburg~\cite{Thornburg95}
are implemented.%%%
\footnote{%%%
	 The convergence problems Thornburg~\protect\cite{Thornburg95}
	 noted when high-spatial-frequency perturbations are
	 present in the slice's geometry, seem to be rare in
	 practice.
	 }%%%
{}  Their typical radius of convergence is on the order of $30\%$
of the horizon radius, but cases are known where it's much smaller.
For example, Schnetter, Herrmann, and Pollney~\cite{Schnetter04}
report that (with no ``line search'' modifications) it's only about
$10\%$ for some slices in a binary black hole coalescence simulation.

Schnetter's \program{TGRapparentHorizon2D}~\cite{Schnetter02a, Schnetter03a}
and Thornburg's \program{AHFinderDirect}~\cite{Thornburg2003:AH-finding}
are both elliptic-PDE apparent-horizon finders implemented as modules
(``thorns'') in the \href{http://www.cactuscode.org}{\program{Cactus}}
computational toolkit.  Both are freely available by anonymous~CVS,
and work with either the \program{PUGH} unigrid driver or the
\href{http://www.carpetcode.org}{\program{Carpet}} mesh-refinement
driver for \href{http://www.cactuscode.org}{\program{Cactus}}.
\program{TGRapparentHorizon2D} is no longer maintained, but
\program{AHFinderDirect} is actively supported and is now used by
many different research groups.%%%
\footnote{%%%
	 In addition, at least two different research
	 groups have now ported, or are in the process
	 of porting, \program{AHFinderDirect} to their own
	 (non-\href{http://www.cactuscode.org}{\program{Cactus}})
	 numerical relativity codes.
	 }%%%

%%%%%%%%%%%%%%%%%%%%%%%%%%%%%%%%%%%%%%%%

\subsection{Horizon Pretracking}
\label{sect-AH/algorithms/pretracking}

Schnetter \etal{}~\cite{Schnetter03a, Schnetter04} introduced the
important concept of \defn{horizon pretracking}.  They focus on the
case where we want to find a common apparent horizon as soon as
it appears in a binary black-hole (or neutron-star) simulation.
While a global (flow) algorithm (section~\ref{sect-AH/algorithms/flow})
could be used to find this common apparent horizon, these algorithms
tend to be very slow.  They observe that the use of a local (elliptic-PDE)
algorithm for this purpose is somewhat problematic:
\begin{quote}
The common [apparent] horizon [\dots] appears instantaneously at
some late time and without a previous good guess for its location.
In practice, an estimate of the surface location and shape can be
put in by hand.  The quality of this guess will determine the rate
of convergence of the finder and, more seriously, also determines
whether a horizon is found at all.  Gauge effects in the strong field
region can induce distortions that have a large influence on the
shape of the common horizon, making them difficult to predict,
particularly after a long evolution using dynamical coordinate
conditions.  As such, it can be a matter of some expensive trial
and error to find the common apparent horizon at the earliest
possible time.  Further, if a common apparent horizon is not found,
it is not clear whether this is because there is none, or whether
there exists one which has only been missed due to unsuitable
initial guesses -- for a fast apparent horizon finder, a good
initial guess is crucial.
\end{quote}

Pretracking tries (usually successfully) to eliminate these
difficulties by determining -- \emph{before} it appears --
approximately where (in space) and when (in time) the common apparent
horizon will appear.

%%%%%%%%%%%%%%%%%%%%

\subsubsection{Constant-Expansion Surfaces}
\label{sect-AH/algorithms/pretracking/CE-surfaces}

The basic idea of horizon pretracking is to consider surfaces
of constant expansion (\defn{CE surfaces}), \ie{} smooth closed
orientable 2-surfaces in a slice satisfying the condition
\begin{equation}
\Theta = E
							  \label{eqn-CE-surface}
\end{equation}
where the expansion $E$~is a specified real number.  Each marginally
outer trapped surface (including the apparent horizon) is thus a CE
surface with expansion~$E=0$; more generally \eqref{eqn-CE-surface}
defines a 1-parameter family of 2-surfaces in the slice.  As discussed
by Schnetter \etal{}~\cite{Schnetter03a, Schnetter04}, for asymptotically
flat slices containing a compact strong-field region, some of the
$E > 0$ members of this family typically foliate the weak-field region.

In the binary-coalescence context, for each $t=\text{constant}$ slice
we define $E_\ast$ to be the smallest $E \ge 0$ for which a CE surface
(containing both strong-field regions) exists with expansion~$E$.
If $E_\ast = 0$ then this \defn{minimum-expansion CE surface} is
the common apparent horizon, while if $E_\ast > 0$ this surface is
an approximation to where the common apparent horizon \emph{will} appear.
We expect the minimum-expansion CE surface to change continuously during
the evolution, and its expansion~$E_\ast$ to decrease towards~$0$.
Essentially, horizon pretracking follows the time evolution of the
minimum-expansion CE surface and uses it as an initial guess for
(searching for) the common apparent horizon.

%%%%%%%%%%%%%%%%%%%%

\subsubsection{Generalized Constant-Expansion Surfaces}
\label{sect-AH/algorithms/pretracking/generalized-CE-surfaces}

Schnetter~\cite{Schnetter03a} implemented an early form of horizon
pretracking, which followed the evolution of the minimum-expansion
constant-expansion surface, and found that it worked well for simple
test problems.  However, Schnetter \etal{}~\cite{Schnetter04} found that
for more realistic binary--black-hole coalescence systems the algorithm
needs to be extended:
\begin{itemize}
\item	While the expansion is zero for a common apparent horizon,
	it's also zero for a 2-sphere at spatial infinity.  
	Figure~\ref{fig-Schw/expansions} illustrates this
	for Schwarzschild spacetime.  Notice that for small positive
	$E_\ast$ there will generally be two distinct CE surfaces
	with $E = E_\ast$, an inner surface just outside the
	horizon, and an outer one far out in the weak-field region.
	The inner CE surface converges to the common apparent
	horizon as $E_\ast$ decreases towards~$0$, and is the surface
	we would like the pretracking algorithm to follow.  Unfortunately,
	without measures such as those described below, there's
	nothing to prevent the algorithm from following the outer
	surface, which does \emph{not} converge to the common
	apparent horizon as $E_\ast$ decreases towards~$0$.

%%%%%%%%%%
\epubtkImage{fig/Schw--expansions/expansion.png}%%%
{%%%
\begin{figure}[bp]
\begin{center}
\centerline{%%%
  \epsfxsize=125mm
  \epsfbox{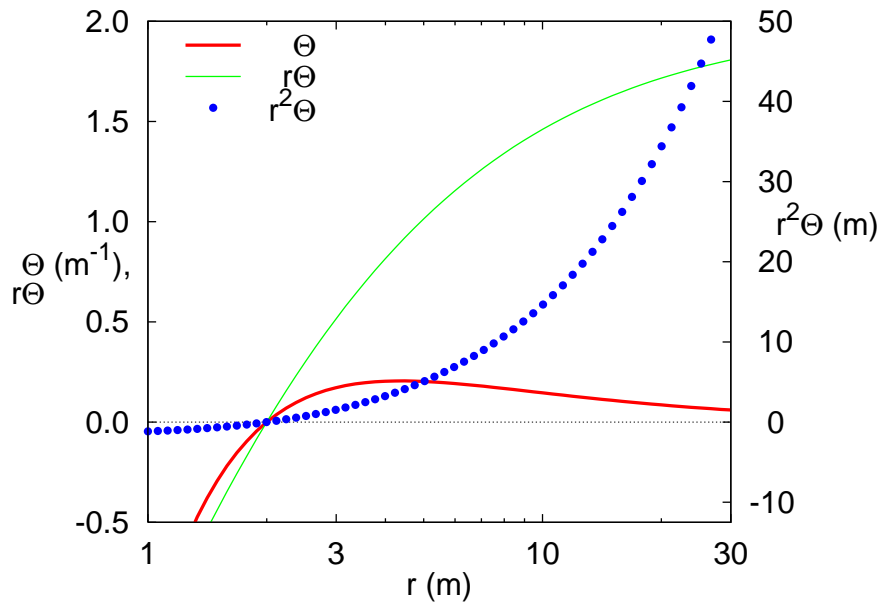}%%
}
\end{center}
\vspace{-10mm}%%%
\caption[Expansion and Related Quantities for Spherical Surfaces
	 in Schwarzschild Spacetime]
	{
	This figure shows the expansion~$\Theta$ (left scale),
	and the \defn{generalized expansions}~$r\Theta$ (left scale)
	and $r^2 \Theta$ (right scale), for various $r=\text{constant}$
	surfaces in an Eddington-Finkelstein slice of Schwarzschild
	spacetime.  Notice that all three functions
	have zeros at the horizon~$r=2m$, and that while $\Theta$
	has a maximum at $r \approx 4.4m$, both $r\Theta$ and
	$r^2\Theta$ increase monotonically with~$r$.
	}
\label{fig-Schw/expansions}
\end{figure}%%%
}
%%%%%%%%%%

\item	In a realistic binary-coalescence simulation, the actual
	minimum-expansion CE surface may be highly distorted,
	which makes it hard to represent accurately with a
	finite-resolution angular grid.
\end{itemize}

Schnetter \etal{}~\cite{Schnetter04} discuss these problems in more
detail, arguing that to solve them, the expansion~$\Theta$ should
be generalized to a \defn{shape function}~$H$ given by one of
\begin{subequations}
					 \label{eqn-pretracking-shape-functions}
\begin{eqnarray}
H_1	& = &	\Theta							\\
H_r	& = &	h \Theta						\\
H_{r^2}	& = &	h^2 \Theta						%%%\\
\end{eqnarray}
\end{subequations}
CE surfaces are then generalized to surfaces satisfying
\begin{equation}
H = E
					      \label{eqn-generalized-CE-surface}
\end{equation}
for some specified~$E \ge 0$.

Note that unlike $H_1$, both $H_r$ and $H_{r^2}$ are typically
monotonic with radius.  Neither $H_r$ nor $H_{r^2}$ are 3-covariantly
defined, but they both still have the property that $E = 0$
in~\eqref{eqn-generalized-CE-surface} implies the surface is a
marginally outer trapped surface, and in practice
they work better for horizon pretracking.

%%%%%%%%%%%%%%%%%%%%

\subsubsection{Goal Functions}
\label{sect-AH/algorithms/pretracking/goal-fns}

To select a single ``smallest'' surface at each time,
Schnetter \etal{}~\cite{Schnetter04} introduce a second generalization,
that of a \defn{goal function}~$G$, which maps surfaces to real numbers.
The pretracking search then attempts, on each time slice, to find the
surface (shape) satisfying $H = E$ with the minimum value of~$G$.
They experimented with several different goal functions,
\begin{subequations}
						\label{eqn-pretracking-goal-fns}
\begin{eqnarray}
G_H	& = &	\overline{H}						\\
G_{rH}	& = &	\overline{h} \, \overline{H}				\\
G_r	& = &	\overline{h}						%%%\\
\end{eqnarray}
\end{subequations}
where in each case the overline (\,$\overline{\phantom{h}}$\,) denotes
an average over the surface.%%%
\footnote{%%%
	 Schnetter \etal{}~\protect\cite{Schnetter04}
	 use a simple arithmetic mean over all surface
	 grid points.  In theory this average could be
	 defined 3-covariantly by taking the induced
	 metric on the surface into account, but in
	 practice they found that this wasn't worth
	 the added complexity.
	 }%%%

%%%%%%%%%%%%%%%%%%%%

\subsubsection{The Pretracking Search}
\label{sect-AH/algorithms/pretracking/pretracking-search}

Schnetter's~\cite{Schnetter03a} original implementation of horizon
pretracking (which followed the evolution of the minimum-expansion
CE surface) used a binary search on the desired expansion~$E$.
Because~$E$ appears only on the right hand side of the generalized
CE condition~\eqref{eqn-generalized-CE-surface}, it's trivial to
modify any apparent-horizon finder to search for a surface of specified
expansion~$E$.  (Schnetter used his \program{TGRapparentHorizon2D}
elliptic-PDE apparent-horizon finder described in
section~\ref{sect-AH/algorithms/elliptic-PDE/summary} for this.)
A binary search on~$E$ can then be used to find the minimum
value~$E_\ast$.%%%
\footnote{%%%
	 There is one complication here:  Any local
	 apparent-horizon finding algorithm may fail
	 if the initial guess isn't good enough,
	 \emph{even if the desired surface is actually present}.
	 The solution is to use the constant-expansion
	 surface for a slightly larger expansion~$E$ as
	 an initial guess, gradually ``walking down'' the
	 value of~$E$ to find the minimum value~$E_\ast$.
	 Thornburg~\protect\cite[appendix~C]{Thornburg2003:AH-finding}
	 describes such a
	 \defn{continuation-algorithm binary search}
	 algorithm in detail.
	 }%%%

Implementing a horizon-pretracking search on any of the generalized
goal functions~\eqref{eqn-pretracking-goal-fns} is conceptually
similar but somewhat more involved:  As described by
Schnetter \etal{}~\cite{Schnetter04} for the case of an elliptic-PDE
apparent-horizon finder,%%%
\footnote{%%%
	 So far as I know this is the only case that
	 has so far been considered for horizon pretracking.
	 Extension to other types of apparent-horizon
	 finders might be a fruitful area for further
	 research.
	 }%%%
{} we first write the equation defining a desired pretracking surface as
\begin{equation}
H - \overline{H} + G - p = 0
						 \label{eqn-pretracking-surface}
\end{equation}
where $p$ is the desired value of the goal function~$G$.  Since $H$
is the only term in~\eqref{eqn-pretracking-surface} which varies over
the surface, it must be constant for the equation to be satisfied.
In this case $H - \overline{H}$ vanishes, so the equation just
gives $G = p$, as desired.

Because $\overline{H}$ depends on $H$ at \emph{all} surface points,
directly finite differencing~\eqref{eqn-pretracking-surface} would
give a non-sparse Jacobian matrix, which would greatly slow the
linear-solver phase of the elliptic-PDE apparent-horizon finder
(section~\ref{sect-AH/algorithms/elliptic-PDE/solving-linear-eqns}).
Schnetter \etal{}~\cite[section~III.B]{Schnetter04} show how this problem
can be solved by introducing a single extra unknown into the discrete
system.  This gives a Jacobian which has a single non-sparse row and
column, but is otherwise sparse, so the linear equations can still
be solved efficiently.

When doing the pretracking search, the cost of a single binary-search
iteration is approximately the same as that of finding an apparent
horizon.  Schnetter \etal{}~\cite[figure~5]{Schnetter04} report that
their pretracking implementation (a modified version of
Thornburg's \program{AHFinderDirect}~\cite{Thornburg2003:AH-finding}
elliptic-PDE apparent-horizon finder described in
section~\ref{sect-AH/algorithms/elliptic-PDE/summary})
typically takes on the order of 5 to 10 binary-search iterations.%%%
\footnote{%%%
	 This refers to the period before a common apparent
	 horizon is found.  Once a common apparent horizon
	 is found, then pretracking can be disabled as the
	 apparent-horizon finder can easily ``track'' the
	 apparent horizon's motion from one time step to
	 the next.
	 }%%%
$^,$%%%
\footnote{%%%
	 With a binary search the number of iterations depends
	 only weakly (logarithmically) on the pretracking
	 algorithm's accuracy tolerance.  It might be possible
	 to replace the binary search by a more sophisticated
	 1-dimensional search algorithm (I discuss such algorithms
	 in appendix~\ref{app-single-nonlinear-eqn}), potentially
	 cutting the number of iterations substantially.  This
	 might be a fruitful area for further research.
	 }%%%
{}  The cost of pretracking is thus on the order of 5 to 10 times that
of finding a single apparent horizon.  This is substantial, but not
prohibitive, particularly if the pretracking algorithm isn't run at
every time step.

%%%%%%%%%%%%%%%%%%%%

\subsubsection{Sample Results}
\label{sect-AH/algorithms/pretracking/sample-results}

As an example of the results obtained from horizon pretracking,
figure~\ref{fig-2BH-headon-collision/pretracking-expansions} shows
the expansion~$\Theta$ for various pretracking surfaces (\ie{} various
choices for the shape function~$H$ in a head-on binary black hole
collision.  Notice how all three of the shape
functions~\eqref{eqn-pretracking-shape-functions} result in
pretracking surfaces whose expansions converge smoothly to zero
just when the apparent horizon appears (at about $t=1.1$).

%%%%%%%%%%
\epubtkImage{fig/2BH-headon-collision--pretracking-expansions/Theta-etal.png}%%%
{%%%
\begin{figure}[bp]
\begin{center}
\centerline{%%%
  \epsfxsize=150mm
  \epsfbox{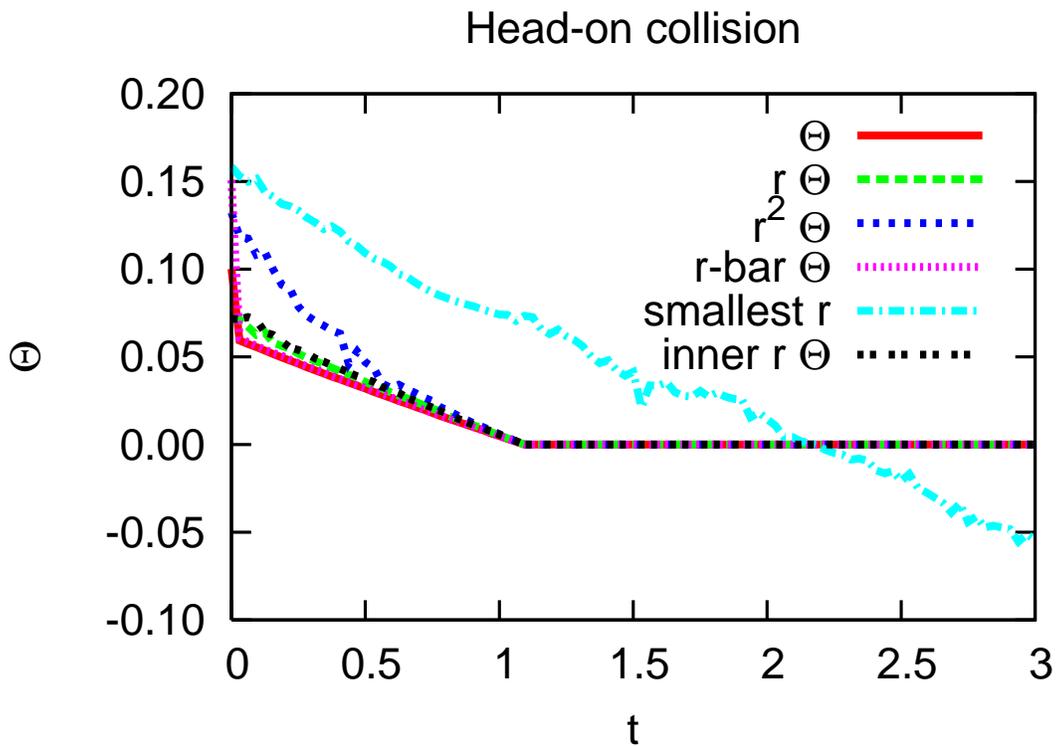}%%
}
\end{center}
\caption[Expansion and Related Quantities for Various Pretracking Surfaces
	 in a Head-On Black Hole Collision]
	{
	This figure shows the expansion~$\Theta$ for various pretracking
	surfaces, \ie{}, for various choices for the shape function~$H$,
	in a head-on binary black hole collision.  Notice how the three
	shape functions~\eqref{eqn-pretracking-shape-functions}
	(here labelled $\Theta$, $r\Theta$, and $r^2 \Theta$)
	result in pretracking surfaces whose expansions converge
	smoothly to zero just when the apparent horizon appears
	(at about $t=1.1$).  Notice also that these three expansions
	have all converged to each other somewhat before the common
	apparent horizon appears.
	Figure reprinted with permission from
	\href{http://link.aps.org/abstract/PRD/v71/e044033}%%%
	     {Schnetter, Herrmann, and Pollney,
	      \textit{Physical Review~D} \textbf{71}, 044033 (2005)}.
	Copyright 2005 by the American Physical Society.
	}
\label{fig-2BH-headon-collision/pretracking-expansions}
\end{figure}%%%
}
%%%%%%%%%%

As a further example,
figure~\ref{fig-2BH-spiraling-collision/pretracking-surfaces} shows
the pretracking surfaces at various times in a spiraling binary black
hole collision, projected into the black hole's orbital plane.

%%%%%%%%%%
%% the png files for other subfigures are
%%           fig/2BH-spiraling-collision--pretracking-surfaces/surfaces.t=5.png
%%           fig/2BH-spiraling-collision--pretracking-surfaces/surfaces.t=10.png
%%           fig/2BH-spiraling-collision--pretracking-surfaces/surfaces.t=16.png
\epubtkImage{fig/2BH-spiraling-collision--pretracking-surfaces/surfaces.t=0.png}%%%
{%%%
\begin{figure}[bp]
\begin{center}
\centerline{%%%
  \hspace{5mm}%%%
  \epsfxsize=75mm
  \epsfbox{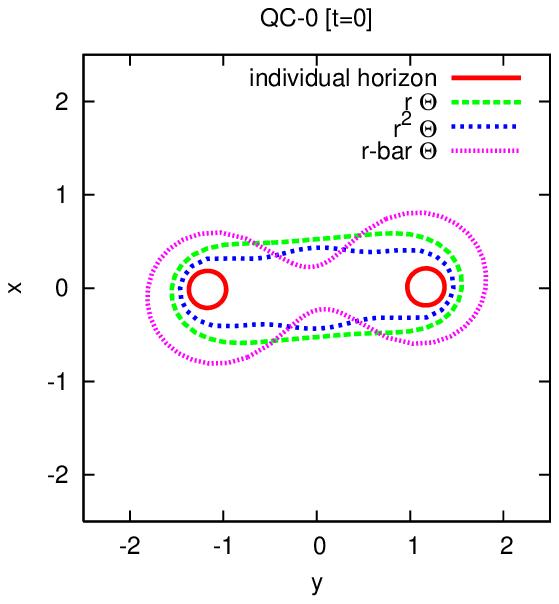}%%%
  \epsfxsize=75mm
  \epsfbox{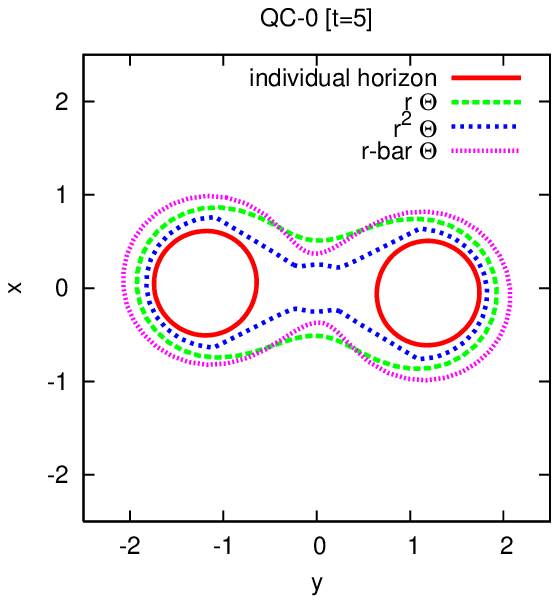}%%%
\vspace{5mm}%%%
}%%%
\centerline{%%%
  \hspace{5mm}%%%
  \epsfxsize=75mm
  \epsfbox{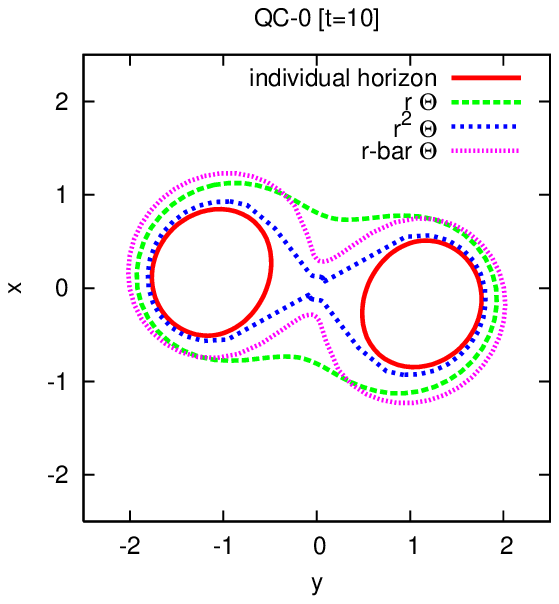}%%%
  \epsfxsize=75mm
  \epsfbox{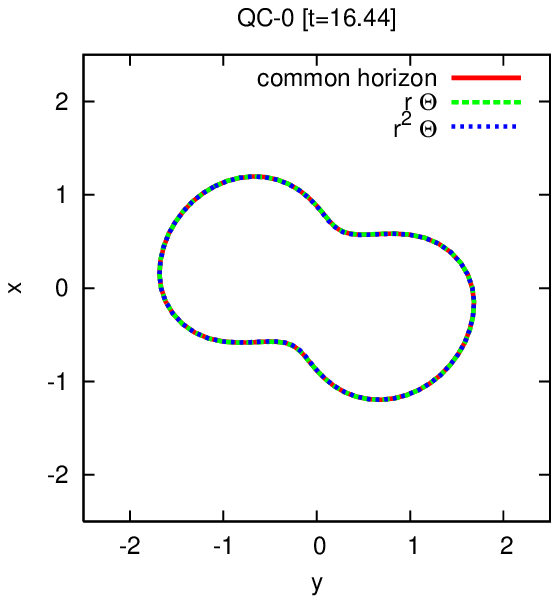}%%%
}%%%
\vspace{-10mm}%%%
\end{center}
\caption[Pretracking Surfaces in a Spiraling Black Hole Collision]
	{
	This figure shows the pretracking surfaces at various times
	in a spiraling binary black hole collision, projected into
	the black holes' orbital plane.  Notice how, even well before
	the common apparent horizon first appears ($t=16.44m_\text{ADM}$,
	bottom right plot), the $r\Theta$ pretracking surface is already
	a reasonable approximation to the eventual common apparent-horizon's
	shape.
	Figure reprinted with permission from
	\href{http://link.aps.org/abstract/PRD/v71/e044033}%%%
	     {Schnetter, Herrmann, and Pollney,
	      \textit{Physical Review~D} \textbf{71}, 044033 (2005)}.
	Copyright 2005 by the American Physical Society.
	}
\label{fig-2BH-spiraling-collision/pretracking-surfaces}
\end{figure}%%%
}
%%%%%%%%%%

%%%%%%%%%%%%%%%%%%%%

\subsubsection{Summary of Horizon Pretracking}
\label{sect-AH/algorithms/pretracking/summary}

Pretracking is a very valuable addition to the horizon-finding
repertoire: it essentially gives a local algorithm (in this case an
elliptic-PDE algorithm) most of the robustness of a global algorithm
in terms of finding a common apparent horizon as soon as it appears.
It's implemented as a higher-level algorithm which uses a slightly-modified
elliptic-PDE apparent-horizon finding algorithm as a ``subroutine''.

The one significant disadvantage of pretracking is its cost: each
pretracking search typically takes 5 to 10 times as long as finding
an apparent horizon.  Further research to reduce the cost of pretracking
would be useful.

Schnetter \etal's pretracking implementation~\cite{Schnetter04} is
implemented as a set of modifications to Thornburg's
\program{AHFinderDirect}~\cite{Thornburg2003:AH-finding}
apparent-horizon finder.  Like the original \program{AHFinderDirect},
the modified version is a ``thorn'' in the
\href{http://www.cactuscode.org}{\program{Cactus}} toolkit
and is freely available by anonymous CVS.

%%%%%%%%%%%%%%%%%%%%%%%%%%%%%%%%%%%%%%%%

\subsection{Flow Algorithms}
\label{sect-AH/algorithms/flow}

Flow algorithms define a ``flow'' on 2-surfaces, \ie{}, they define
an evolution of 2-surfaces in some pseudo-time~$\lambda$, such that
the apparent horizon is the $\lambda \to \infty$ limit of a (any)
suitable starting surface.  Flow algorithms are different from other
apparent-horizon finding algorithms (except for zero-finding in
spherical symmetry), in that their convergence doesn't depend on
having a good initial guess.  In other words, flow algorithms are
global algorithms (section~\ref{sect-AH/intro/local-vs-global}).

To find the (an) apparent horizon, \ie{}, an \emph{outermost} MOTS,
the starting surface should be outside the largest possible MOTS in
the slice.  In practice, it generally suffices to start with a 2-sphere
of areal radius substantially greater than $2m_\text{ADM}$.

The global convergence property requires that a flow algorithm always
flow from a large starting surface into the apparent horizon.  This
means that the algorithm gains no particular benefit from already
knowing the approximate position of the apparent horizon.  In particular,
flow algorithms are no faster when ``tracking'' the apparent horizon
(repeatedly finding it at frequent intervals) in a numerical time
evolution.  (In contrast, in this situation a local apparent-horizon
finding algorithm can use the most recent previously-found apparent
horizon as an initial guess, greatly speeding the algorithm's convergence.)

Flow algorithms were first proposed for apparent-horizon finding by
Tod~\cite{Tod91}.  He initially considered the case of a time-symmetric
slice (one where $K_{ij} = 0$).  In this case a marginally outer
trapped surface (and thus an apparent horizon) is a surface of
minimal area, and may be found by a \defn{mean curvature flow}
\begin{equation}
\partial_\lambda x^i
	= - \kappa s^i
						 \label{eqn-mean-curvature-flow}
\end{equation}
where $x^i$ are the spatial coordinates of a horizon-surface point,
$s^i$ is as before the outward-pointing unit 3-vector normal to
the surface, and $\kappa \equiv \del_k s^k$ is the mean curvature of
the surface as embedded in the slice.  This is a gradient flow for the
surface area, and Grayson~\cite{Grayson-1987} has proven that if the
slice contains a minimum-area surface, this will in fact be the stable
$\lambda \to \infty$ limit of this flow.  Unfortunately, this proof is
valid only for the time-symmetric case.

For non--time-symmetric slices, Tod~\cite{Tod91} proposed generalizing
the mean curvature flow to the \defn{expansion flow}
\begin{equation}
\partial_\lambda x^i
	= - \Theta s^i
						      \label{eqn-expansion-flow}
\end{equation}
There is no theoretical proof that this flow will converge to the (an)
apparent horizon, but since the flow velocity is zero there, and the
flow is identical to the mean curvature flow~\eqref{eqn-mean-curvature-flow}
in the principle part, convergence is at least theoretically plausible.
Numerical experiments by Bernstein~\cite{Bernstein-1993:mean-curvature-flow},
Shoemaker \etal{}~\cite{Shoemaker99, Shoemaker-Huq-Matzner-2000}, and
others show that that the expansion flow~\eqref{eqn-expansion-flow}
does in fact converge robustly to the apparent horizon.

In the following subsections I discuss a number of important
implementation details for, and refinements of, this basic algorithm.

%%%%%%%%%%%%%%%%%%%%

\subsubsection{Implicit Pseudo-Time Stepping}
\label{sect-AH/algorithms/flow/implicit-pseudo-time-stepping}

Assuming the \Strahlkoerper{} surface
parameterization~\eqref{eqn-r=h(theta,phi)}, the expansion
flow~\eqref{eqn-expansion-flow} is a \emph{parabolic} equation for
the horizon shape function~$h$.%%%
\footnote{%%%
	 Linearizing the $\Theta(h)$ function~\eqref{eqn-AH/h}
	 gives a negative Laplacian in $h$ as the principal part.
	 }%%%
{}  This means that any fully explicit
scheme to integrate it (in the pseudo-time~$\lambda$) must severely
restrict its pseudo-time step $\Delta \lambda$ for stability, and
this restriction grows (quadratically) worse at higher spatial resolutions.%%%
\footnote{%%%
	 For a spatial resolution $\Delta x$, an explicit
	 scheme is generally limited to a pseudo-time step
	 $\Delta \lambda \ltsim (\Delta x)^2$.
	 }%%%
{}  This makes the horizon-finding process very slow.

To avoid this restriction, practical implementations of flow algorithms
use implicit pseudo-time integration schemes; these can have large
pseudo-time steps and still be stable.  Because we only care about the
$\lambda \to \infty$ limit, a highly accurate pseudo-time integration
isn't important; only the accuracy of approximating the spatial derivatives
matters.  Bernstein~\cite{Bernstein-1993:mean-curvature-flow} used a
modified Du~Fort-Frankel scheme~\cite{DuFort-Frankel-1953},%%%
\footnote{%%%
 Richtmyer and Morton~\protect\cite[section~7.5]{Richtmyer-Morton-2nd-edition}
	 give a very clear presentation and analysis
	 of the Du~Fort-Frankel scheme.
	 }%%%
{} but found some problems with the surface shape gradually
developing high-spatial-frequency noise.
Pasch~\cite{Pasch-1997:level-sets-and-curvature-flow} reports that
an \defn{exponential} integrator
(Hochbruck \etal{}~\cite{Hochbruck-etal-1998:exponential-integrator})
works well, provided the flow's Jacobian matrix is computed accurately.%%%
\footnote{%%%
	 More precisely,
	 Pasch~\protect\cite{Pasch-1997:level-sets-and-curvature-flow}
	 found that that an exponential integrator worked well when the
	 flow's Jacobian matrix was computed exactly (using the
	 symbolic-differentiation technique described in
	 section~\ref{sect-AH/algorithms/elliptic-PDE/Jacobian-matrix}).
	 However, when the Jacobian matrix was approximated using
	 the numerical-perturbation technique described in
	 section~\ref{sect-AH/algorithms/elliptic-PDE/Jacobian-matrix},
	 Pasch found that the pseudo-time integration became
	 unstable at high numerical resolutions.
	 }%%%
$^,$%%%
\footnote{%%%
	 Pasch~\protect\cite{Pasch-1997:level-sets-and-curvature-flow}
	 also notes that the exponential integrator uses a very large
	 amount of memory.
	 }%%%
{}  The most common choice is probably that of
Shoemaker \etal{}~\cite{Shoemaker99, Shoemaker-Huq-Matzner-2000},
who use the iterated Crank-Nicholson (\defn{ICN}) scheme.%%%
\footnote{%%%
	 Teukolsky~\cite{Teukolsky00} and
	 Leiler and Rezzolla~\cite{Leiler_Rezzolla06} have
	 analyzed ICN's stability under various conditions.
	 }%%%
{}  They report that this works very well; in particular, they don't
report any noise problems.

By refining his finite-element grid
(section~\ref{sect-intro/2-surface-parameterizations/finite-element})
in a hierarchical manner, Metzger~\cite{Metzger04} is able to use
standard conjugate-gradient elliptic solvers in a multigrid-like
fashion,%%%
\footnote{%%%
	 See the references cited in
	 footnote~\ref{footnote-multigrid-refs}
	 on page~\pageref{footnote-multigrid-refs}
	 for general introductions to multigrid algorithms
	 for elliptic PDEs.
	 }%%%
{} using each refinement level's solution as an initial guess for
the next higher refinement level's iterative solution.  This greatly
speeds the flow integration: Metzger reports that the performance of
the overall surface-finding algorithm is ``of the same order of magnitude''
as that of Thornburg's \program{AHFinderDirect}~\cite{Thornburg2003:AH-finding}
elliptic-PDE apparent-horizon finder (described in
section~\ref{sect-AH/algorithms/elliptic-PDE/summary}).

In a more general context than numerical relativity,
Osher and Sethian~\cite{Osher-Sethian-1988} have discused a general
class of numerical algorithms for integrating ``fronts propagating
with curvature-dependent speed''.  These flow a level-set function
(section~\ref{sect-intro/2-surface-parameterizations/level-set})
which implicitly locates the actual ``front''.

%%%%%%%%%%%%%%%%%%%%

\subsubsection{Varying the Flow Speed}
\label{sect-AH/algorithms/flow/varying-flow-speed}

Another important performance optimization of the standard expansion
flow~\eqref{eqn-expansion-flow} is to replace $\Theta$ in the right-hand
side by a suitable nonlinear function of $\Theta$, chosen so the
surface shrinks faster when it's far from the apparent horizon.
For example, Shoemaker \etal{}~\cite{Shoemaker99, Shoemaker-Huq-Matzner-2000}
use the flow
\begin{equation}
\partial_\lambda x^i
	= - \left[
	    (\Theta - c)
	    \arctan^2 \left( \frac{\Theta - c}{\Theta_0} \right)
	    \right] s^i
\end{equation}
for this purpose, where $\Theta_0$ is the value of $\Theta$ on the
initial-guess surface, and $c$ (which is gradually decreased towards~$0$
as the iteration proceeds) is a ``goal'' value for $\Theta$.

%%%%%%%%%%%%%%%%%%%%

\subsubsection{Surface Representation and the Handling of Bifurcations}
\label{sect-AH/algorithms/flow/representation,bifurcations}

Since a flow algorithm starts with (topologically) a single large
2-sphere, if there are multiple apparent horizons present the
surface must change topology (bifurcate) at some point in the flow.
Depending on how the surface is represented, this may be easy or
difficult.

Pasch~\cite{Pasch-1997:level-sets-and-curvature-flow}
and Shoemaker \etal{}~\cite{Shoemaker99, Shoemaker-Huq-Matzner-2000}
use a level-set function approach
(section~\ref{sect-intro/2-surface-parameterizations/level-set}).
This automatically handles any topology or topology change.  However,
it has the drawback of requiring the flow to be integrated throughout
the entire volume of the slice (or at least in some neighborhood of
each surface).  This is likely to be much more expensive than only
integrating the flow on the surface itself.  Shoemaker \etal{} also
generate an explicit \Strahlkoerper{} surface representation
(section~\ref{sect-intro/2-surface-parameterizations/Strahlkoerper}),
monitoring the surface shape to detect an imminent bifurcation
and reparameterizing the shape into 2~separate surfaces if a
bifurcation happens.

Metzger~\cite{Metzger04} uses a finite-element surface representation
(section~\ref{sect-intro/2-surface-parameterizations/finite-element}),
which can represent any topology.  However, if the flow bifurcates,
then to explicitly represent each apparent horizon the code must
detect that the surface self-intersects, which may be expensive.

%%%%%%%%%%%%%%%%%%%%

\subsubsection{Gundlach's ``Fast Flow''}
\label{sect-AH/algorithms/flow/fast-flow}

Gundlach~\cite{Gundlach98a} introduced the important concept of a
\defn{fast flow}.  He observed that the subtraction and inversion
of the flat-space Laplacian in Nakamura \etal{}'s spectral
integral-iteration algorithm (section~\ref{sect-AH/algorithms/Nakamura-etal})
is an example of
``a standard way of solving nonlinear elliptic problems numerically,
namely subtracting a simple linear elliptic operator from the     
nonlinear one, inverting it by pseudo-spectral algorithms and iterating''.
Gundlach then interpreted Nakamura \etal{}'s algorithm as a type of
flow algorithm where each pseudo-time step of the flow corresponds
to a single functional-iteration step of the Nakamura \etal{} algorithm.

In this framework, Gundlach defines a 2-parameter family of flows
interpolating between Nakamura \etal{}'s algorithm and
Tod's~\cite{Tod91} expansion flow~\eqref{eqn-expansion-flow},
\begin{equation}
\partial_\lambda h = -A(1 - B\del^2)^{-1} \rho \Theta
\end{equation}
where $A \ge 0$ and $B \ge 0$ are parameters, $\rho > 0$ is a weight
functional which depends on $h$ through at most 1st~derivatives,
$\del^2$ is the flat-space Laplacian operator, and
$(1 - B\del^2)^{-1}$ denotes inverting the operator $(1 - B\del^2)$.%%%
\footnote{%%%
	 The inversion is only formal, because
	 Nakamura \etal{}'s algorithm treats the
	 $a_{00}$ spectral coefficient specially.
	 Gundlach~\protect\cite{Gundlach98a} discusses
	 this in more detail.
	 }%%%

Gundlach then argues that intermediate \defn{fast flow} members
of this family should be a useful compromises between the speed of
Nakamura \etal{}'s algorithm, and the robustness of Tod's expansion flow.
Based on numerical experiments, Gundlach suggests a particular choice
for the weight functional $\rho$ and the parameters $A$ and $B$.
The resulting algorithm updates high-spatial-frequency components
of $h$ essentially the same as Nakamura \etal{}'s algorithm, but
should reduce low-spatial-frequency error components faster.

Alcubierre's \program{AHFinder}~\cite{Alcubierre98b} horizon finder
includes an implementation of Gundlach's fast flow algorithm.%%%
\footnote{%%%
\label{footnote-AHFinder-also-includes-minimization}
	 \program{AHFinder} also includes a minimization
	 algorithm (section~\ref{sect-AH/algorithms/minimization}).
	 }%%%
{}  \program{AHFinder} is implemented as a module (``thorn'') in the
\href{http://www.cactuscode.org}{\program{Cactus}} computational
toolkit, and is freely available by anonymous~CVS (it's part of the
\program{CactusEinstein} set of thorns included with the standard
\href{http://www.cactuscode.org}{\program{Cactus}} distribution).
\program{AHFinder} has been used by a large number of research groups.

%%%%%%%%%%%%%%%%%%%%

\subsubsection{Summary of Flow Algorithms/Codes}
\label{sect-AH/algorithms/flow/summary}

Flow algorithms are the only truely global apparent-horizon finding
algorithms, and as such can be much more robust than local algorithms.
In particular, flow algorithms can guarantee convergence to the
\emph{outermost} MOTS in a slice.  Unfortunately, these convergence
guarantees hold only for time-symmetric slices.

In the forms which have strong convergence guarantees, flow algorithms
tend to be very slow.  (Metzger's algorithm~\cite{Metzger04} is a
notable exception: it's very fast.)  There are modifications which
can make flow algorithms much faster, but then their convergence is
no longer guaranteed.  In particular, practical experience has
shown that in some binary black hole coalescence simulations
(Alcubierre \etal{}~\cite{Alcubierre2003:pre-ISCO-coalescence-times},
Diener \etal{}~\cite{Diener-etal-2006a}), ``fast flow'' algorithms
(section~\ref{sect-AH/algorithms/flow/fast-flow}) can miss common
apparent horizons which are found by other (local) algorithms.

Alcubierre's apparent-horizon finder
\program{AHFinder}~\cite{Alcubierre98b} includes a ``fast flow''
algorithm based on the work of Gundlach~\cite{Gundlach98a}.%%%
$^\textrm{\ref{footnote-AHFinder-also-includes-minimization}}$%%%
{}  It's implemented as a module (``thorn'') in the
\href{http://www.cactuscode.org}{\program{Cactus}} computational
toolkit, and is freely available by anonymous~CVS (it's part of the
\program{CactusEinstein} set of thorns included with the standard
\href{http://www.cactuscode.org}{\program{Cactus}} distribution).
It has been used by a number of research groups.

%%%%%%%%%%%%%%%%%%%%%%%%%%%%%%%%%%%%%%%%%%%%%%%%%%%%%%%%%%%%%%%%%%%%%%%%%%%%%%%%

\section{Summary of Algorithms/Codes for Finding Apparent Horizons}
\label{sect-AH/summary}

%%%%%%%%%%%%%%%%%%%%%%%%%%%%%%%%%%%%%%%%

\subsection{Summary of Apparent-Horizon Finding Algorithms}
\label{sect-AH/summary/algorithms}

There are a large number of apparent-horizon finding algorithms, with
differing trade-offs between speed, robustness of convergence, accuracy,
and ease of programming.

In spherical symmetry, zero-finding
	(section~\ref{sect-AH/algorithms/zero-finding})
is fast, robust, and easy to program.
In axisymmetry, shooting algorithms
	(section~\ref{sect-AH/algorithms/shooting})
work well and are fairly easy to program.  Alternatively, any of
the algorithms for generic slices (summarized below) can be used
with implementations tailored to the axisymmetry.

Minimization algorithms
	(section~\ref{sect-AH/algorithms/minimization})
are fairly easy to program, but are susceptible to spurious local
minima, have relatively poor accuracy, and tend to be very slow
unless axisymmetry is assumed.

Nakamura \etal{}'s spectral integral-iteration algorithm
	(section~\ref{sect-AH/algorithms/Nakamura-etal})
and elliptic-PDE algorithms
	(section~\ref{sect-AH/algorithms/elliptic-PDE})
are both fast and accurate, but are moderately difficult to program.
Their main disadvantage is the need for a fairly good initial guess
for the horizon position/shape.

In many cases Schnetter's ``pretracking'' algorithm
	(section~\ref{sect-AH/algorithms/pretracking})
can greatly improve an elliptic-PDE algorithm's robustness,
by determining -- \emph{before} it appears -- approximately where
(in space) and when (in time) a new outermost apparent horizon will
appear.  Pretracking is implemented as a modification of an existing
elliptic-PDE algorithm, and is moderately slow: it typically has a
cost 5 to 10 times that of finding a single horizon with the
elliptic-PDE algorithm.

Finally, flow algorithms
	(section~\ref{sect-AH/algorithms/flow})
are generally quite slow (Metzger's algorithm~\cite{Metzger04} is a
notable exception), but can be very robust in their convergence.
They are moderately easy to program.  Flow algorithms are global
algorithms, in that their convergence does not depend on having
a good initial guess.

%%%%%%%%%%%%%%%%%%%%%%%%%%%%%%%%%%%%%%%%

\subsection{Summary of Publicly-Available Apparent-Horizon Finding Codes}
\label{sect-AH/summary/codes}

I know of 3~publicly-available apparent-horizon finding codes,
all implemented as modules (``thorns'') in the
\href{http://www.cactuscode.org}{\program{Cactus}} computational
toolkit:
\begin{description}
\item[\program{AHFinder}]\mbox{}\\
	Alcubierre's \program{AHFinder}~\cite{Alcubierre98b}
	includes both a ``fast flow'' algorithm based on the
	work of Gundlach~\cite{Gundlach98a}, and a minimization
	algorithm based on the work of Anninos \etal{}~\cite{Anninos98b}.
	\program{AHFinder} is part of the \program{CactusEinstein}
	set of thorns included with the standard
	\href{http://www.cactuscode.org}{\program{Cactus}}
	distribution, and has been used by many research groups.
\item[\program{AHFinderDirect}]\mbox{}\\
	Thornburg's \program{AHFinderDirect}~\cite{Thornburg2003:AH-finding}
	uses an elliptic-PDE algorithm, and has been used by many research
	groups, as well as ported to at least two
	non-\href{http://www.cactuscode.org}{\program{Cactus}}
	numerical relativity codes.
	Schnetter's pretracking algorithm~\cite{Schnetter04}
	is implemented as a set of modifications to \program{AHFinderDirect}.
\item[\program{TGRapparentHorizon2D}]\mbox{}\\
	Schnetter's \program{TGRapparentHorizon2D}~\cite{Schnetter02a,
	Schnetter03a} uses an elliptic-PDE algorithm.
	It's no longer maintained, but remains freely available.
\end{description}

%%%%%%%%%%%%%%%%%%%%%%%%%%%%%%%%%%%%%%%%%%%%%%%%%%%%%%%%%%%%%%%%%%%%%%%%%%%%%%%%

\section*{Acknowledgements}

I thank the many researchers who answered my E-mail queries on
various aspects of their work.
I thank Scott Caveny and Peter Diener for useful conversations
on event-horizon finders, and Badri Krishnan for useful conversations
on the properties of apparent, isolated, and dynamical horizons.
I thank Peter Diener, Luciano Rezzolla, and Virginia J.~Vitzthum
for helpful comments on various drafts of this paper.
I thank Peter Diener and Edward Seidel for providing unpublished figures.

I think the many authors named in this review for granting
permission to reprint figures from their published work.
I thank the American Astronomical Society,
the American Physical society,
and IOP Publishing
for granting permission to reprint figures published in their journals.
The American Physical Society requires the following disclaimer
regarding such reprinted material:
\begin{quote}
Readers may view, browse, and/or download material for
temporary copying purposes only, provided these uses are
for noncommercial personal purposes.  Except as provided
by law, this material may not be further reproduced,
distributed, transmitted, modified, adapted, performed,
displayed, published, or sold in whole or part, without
prior written permission from the publisher.
\end{quote}

I thank the Alexander von Humboldt foundation, the AEI visitors program,
and the AEI postdoctoral fellowship program for financial support.

%%%%%%%%%%%%%%%%%%%%%%%%%%%%%%%%%%%%%%%%%%%%%%%%%%%%%%%%%%%%%%%%%%%%%%%%%%%%%%%%
%%%%%%%%%%%%%%%%%%%%%%%%%%%%%%%%%%%%%%%%%%%%%%%%%%%%%%%%%%%%%%%%%%%%%%%%%%%%%%%%
\appendix
%%%%%%%%%%%%%%%%%%%%%%%%%%%%%%%%%%%%%%%%%%%%%%%%%%%%%%%%%%%%%%%%%%%%%%%%%%%%%%%%
%%%%%%%%%%%%%%%%%%%%%%%%%%%%%%%%%%%%%%%%%%%%%%%%%%%%%%%%%%%%%%%%%%%%%%%%%%%%%%%%

\chapter{Solving A Single Nonlinear Algebraic Equation}
\label{app-single-nonlinear-eqn}

In this appendix I briefly outline numerical algorithms and codes
for solving a single 1-dimensional nonlinear algebraic equation
$f(x) = 0$, where the function $f: \Re \to \Re$ is given.

The process generally begins by evaluating $f$ on a suitable grid of
points and looking for sign changes.  Assuming $f$~to be continuous, each
sign change must then bracket at least one root $x_\ast$:

Given a pair of ordinates $x_-$ and $x_+$ which bracket a root,
there are a variety of different algorithms available to accurately
and efficiently find the (a) root:

If $|x_+ - x_-|$ is small, say on the order of a finite-difference
grid spacing, then closed-form approximations are probably accurate enough:
\begin{itemize}
\item	The simplest approximation is a simple linear interpolation
	of $f$ between $x_-$ and $x_+$.
\item	A slightly more sophisticated algorithm,
	\defn{inverse quadratic interpolation}, is to use 3~ordinates,
	two of which bracket a root, and estimate the root as the
	root of the (unique) parabola which passes through the
	3~given $\bigl(x,f(x)\bigr)$ points.%%%
\footnote{%%%
	 The parabola generically has two roots, but normally
	 only one of them lies between $x_-$ and $x_+$.
	 }%%%
\end{itemize}

For larger $|x_+ - x_-|$, iterative algorithms are necessary to obtain
an accurate root:
\begin{itemize}
\item	Bisection (binary search on the sign of $f$), is a
	well-known iterative scheme which is very robust,
	but rather slow if high accuracy is desired.
\item	Newton's method can be used, but it requires that the
	the derivative~$f'$ be available.  Alternatively, the
	secant algorithm (similar to Newton's method but estimating
	$f'$ from the most recent pair of function evaluations)
	gives similarly fast convergence without requiring $f'$
	to be available.  Unfortunately, if $|f'|$ is small enough
	at any iteration point, both these algorithms can
	fail to converge, or more generally they can generate
	``wild'' trial ordinates.
\item	Probably the most sophisticated algorithm is that of
	van~Wijngaarden, Dekker, and Brent.  This is a carefully
	engineered hybrid of the bisection, secant, and inverse
	quadratic interpolation algorithms, and generally combines
	the rapid convergence of the secant algorithm with the robustness
	of bisection.  The van~Wijngaarden-Dekker-Brent algorithm is
	described by Forsythe, Malcolm, and Moler~\cite[chapter~7]{FMM77},
	Kahaner, Moler, and Nash~\cite[chapter~7]{Kahaner-Moler-Nash-1989},
	and Press \etal{}~\cite[section~9.3]{Press92}.
	An excellent implementation of this, the Fortran subroutine
	\program{ZEROIN}, is freely available from
	\href{http://www.netlib.org/fmm/}{\url{http://www.netlib.org/fmm/}}.
\end{itemize}

%%%%%%%%%%%%%%%%%%%%%%%%%%%%%%%%%%%%%%%%%%%%%%%%%%%%%%%%%%%%%%%%%%%%%%%%%%%%%%%%

\chapter{The Numerical Integration of Ordinary Differential Equations}
\label{app-ODEs}

\begingroup	% local latex macros for this appendix
\def\f{\mathsf{f}}
\def\y{\mathsf{y}}

The time-integration problem%%%
\footnote{%%%
	 The numerical-analysis literature usually refers to this
	 as the \defn{initial value problem}.  Unfortunately, in a
	 relativity context this terminology often causes confusion
	 with the \defn{initial data problem} of solving the ADM
	 constraint equations.  I use the term
	 \defn{time-integration problem for ODEs} to (try to)
	 avoid this confusion.
	 }%%%
$^,$%%%
\footnote{%%%
	 In this appendix sans-serif lower-case letters
	 $\mathsf{abc} \dots \mathsf{z}$ denote variables
	 and functions in $\Re^n$ (for some fixed dimension~$n$),
	 and sans-serif upper-case letters $\mathsf{ABC} \dots \mathsf{Z}$
	 denote $n \,{\times}\, n$ real-valued matrices.
	 }%%%
{} for ordinary differential equations (ODEs) is traditionally written as
follows:  We are given an integer $n > 0$ (the number of ODEs to integrate),
a \defn{right-hand-side} function $\f: \Re^n \times \Re \to \Re^n$,
and the value $\y(0)$ of a function $\y: \Re \to \Re^n$
satisfying the ODE
\begin{equation}
\frac{d\y}{dt} = f(\y,t)
\end{equation}
We wish to know (or approximate) $\y(t)$ for some finite interval
$t \in [0,t_{\max}]$.

This is a well-studied problem in numerical analysis.  See
Forsythe, Malcolm, and Moler~\cite[chapter~6]{FMM77}
or Kahaner, Moler, and Nash~\cite[chapter~8]{Kahaner-Moler-Nash-1989}
for a general overview of ODE integration algorithms and codes, or
Shampine and Gordon~\cite{Shampine-Gordon-1975},
Hindmarsh~\cite{Hindmarsh-1983}, or
Brankin, Gladwell, and Shampine~\cite{Brankin-Gladwell-Shampine-1992}
for detailed technical accounts.

\begin{sloppypar}
For our purposes, it suffices to note that highly accurate, efficient,
and robust ODE-integration codes are widely available.  In fact, there
is a strong tradition in numerical analysis of free availability of
such codes.  Notably,
the \program{RKF45} code described by
Forsythe, Malcolm, and Moler~\cite[chapter~6]{FMM77}
is freely available at
\href{http://www.netlib.org/ode/rkf45.f}%%%
     {\url{http://www.netlib.org/ode/rkf45.f}},
the \program{ODE} code described by
Shampine and Gordon~\cite{Shampine-Gordon-1975}
is freely available at
\href{http://www.netlib.org/ode/ode.f}{\url{http://www.netlib.org/ode/ode.f}},
the \program{ODEPACK}/\program{LSODE} suite of codes described by
Hindmarsh~\cite{Hindmarsh-1983}
are freely available at
\href{http://www.netlib.org/odepack/}%%%
     {\url{http://www.netlib.org/odepack/}},
and
the \program{RKSUITE} suite of codes described by
Brankin, Gladwell, and Shampine~\cite{Brankin-Gladwell-Shampine-1992}
are freely available at
\href{http://www.netlib.org/ode/rksuite/}%%%
     {\url{http://www.netlib.org/ode/rksuite/}}.
\end{sloppypar}

As well as being of high numerical quality, these codes are also very
easy to use, employing sophisticated adaptive algorithms to automatically
adjust step size and/or the precise integration scheme used.%%%
\footnote{%%%
	 \program{LSODE} can also automatically detect
	 and handle stiff systems of ODEs.
	 }%%%
{}  These codes can generally be relied upon to produce accurate results
both more efficiently and more easily than a hand-crafted integrator.
I have used the \program{LSODE} solver in several research projects
with excellent results.

\endgroup	% local latex macros for this appendix

%%%%%%%%%%%%%%%%%%%%%%%%%%%%%%%%%%%%%%%%%%%%%%%%%%%%%%%%%%%%%%%%%%%%%%%%%%%%%%%%
%%%%%%%%%%%%%%%%%%%%%%%%%%%%%%%%%%%%%%%%%%%%%%%%%%%%%%%%%%%%%%%%%%%%%%%%%%%%%%%%

\bibliography{references}

%%%%%%%%%%%%%%%%%%%%%%%%%%%%%%%%%%%%%%%%%%%%%%%%%%%%%%%%%%%%%%%%%%%%%%%%%%%%%%%%
%%%%%%%%%%%%%%%%%%%%%%%%%%%%%%%%%%%%%%%%%%%%%%%%%%%%%%%%%%%%%%%%%%%%%%%%%%%%%%%%

\end{document}